\documentclass[12pt]{article}

\usepackage{latexsym}
\usepackage{amsthm}
\usepackage{amsmath}
\usepackage[dvips]{graphicx}
\usepackage{amssymb}

\newcommand{\be}{\begin{eqnarray}}
\newcommand{\ee}{\end{eqnarray}}

\begin{document}

\baselineskip=18pt

\setcounter{footnote}{0}
\setcounter{figure}{0}
\setcounter{table}{0}

\begin{titlepage}

\begin{center}
\vspace{.1cm}

{\Large \bf What is the Simplest Quantum Field Theory?}

\vspace{0.1cm}

{\bf Nima Arkani-Hamed$^a$, Freddy Cachazo$^b$, Jared Kaplan$^{a,c}$}

\vspace{.1cm}

{\it $^{a}$ School of Natural Sciences, Institute for Advanced Study, Princeton, NJ 08540, USA}

{\it $^{b}$ Perimeter Institute for Theoretical Physics, Waterloo, Ontario N2J W29, CA}

{\it $^{c}$ Jefferson Laboratory of Physics, Harvard University, Cambridge, MA 02138, USA}

\end{center}

\begin{abstract}

Conventional wisdom says that the simpler the Lagrangian of a theory
the simpler its perturbation theory. An ever-increasing
understanding of the structure of scattering amplitudes has however
been pointing to the opposite conclusion. At tree level, the BCFW
recursion relations that completely determine the S-matrix are valid
not for scalar theories but for gauge theories and gravity, with
gravitational amplitudes exhibiting the best UV behavior at infinite
complex momentum. At 1-loop, amplitudes in ${\cal N} = 4$ SYM only
have scalar box integrals, and it was recently conjectured that the
same property holds for ${\cal N} = 8$ SUGRA, which plays an
important role in the suspicion that this theory may be finite. In
this paper we explore and extend the S-matrix paradigm, and suggest
that ${\cal N} = 8$ SUGRA has the simplest scattering amplitudes in
four dimensions. Labeling external states by supercharge
eigenstates--Grassmann coherent states--allows the amplitudes to be
exposed as completely smooth objects, with the action of SUSY
manifest. We show that under the natural supersymmetric extension of
the BCFW deformation of momenta, all tree amplitudes in ${\cal N}=4$
SYM and ${\cal N} = 8$ SUGRA vanish at infinite complex momentum,
and can therefore be determined by recursion relations. An important
difference between ${\cal N} = 8$ SUGRA and ${\cal N}=4$ SYM is that
the massless S-matrix is defined everywhere on moduli space, and is
acted on by a non-linearly realized $E_{7(7)}$ symmetry. We
elucidate how non-linearly realized symmetries are reflected in the
more familiar setting of pion scattering amplitudes, and go on to
identify the action of $E_{7(7)}$ on amplitudes in ${\cal N} = 8$
SUGRA. Moving beyond tree level, we give a simple general discussion
of the structure of 1-loop amplitudes in any QFT, in close parallel
to recent work of Forde, showing that the coefficients of scalar
``triangle" and ``bubble" integrals are determined by the ``pole at
infinite momentum" of products of tree amplitudes appearing in cuts.
In ${\cal N} = 4$ SYM and ${\cal N} = 8$ SUGRA, the on-shell
superspace makes it easy to compute the multiplet sums that arise in
these cuts by relating them to the best behaved tree amplitudes of
highest spin, leading to a straightforward proof of the absence of
triangles and bubbles at 1-loop. We also argue that rational terms
are absent. This establishes that 1-loop amplitudes in ${\cal N} =
8$ SUGRA only have scalar box integrals. We give an explicit
expression for 1-loop amplitudes for both ${\cal N}$=4 SYM and
${\cal N} = 8$ SUGRA in terms of tree amplitudes that can be
determined recursively. These amplitudes satisfy further relations
in ${\cal N} = 8$ SUGRA that are absent in ${\cal N} = 4$ SYM. Since
both tree and 1-loop amplitudes for maximally supersymmetric
theories can be completely determined by their leading
singularities, it is natural to conjecture that this property holds
to all orders of perturbation theory. This is the nicest analytic
structure amplitudes could possibly have, and if true, would
directly imply the perturbative finiteness of ${\cal N} = 8$ SUGRA.
All these remarkable properties of scattering amplitudes call for an
explanation in terms of a ``weak-weak" dual formulation of QFT, a
holographic dual of flat space.

\end{abstract}

\bigskip
\bigskip

\end{titlepage}

\section{Simple Lagrangians Versus Simple Amplitudes}

Scattering amplitudes are defined by evaluating the S matrix between
initial and final states of positive energy. Thinking of them as
functions of all incoming momenta, they can be written in a way that
is completely symmetric in external particles as
\begin{equation} M^{a_1 \cdots a_n}(p_1, \cdots, p_n)
\end{equation}
where the $a_i$ are the little group indices for whatever spin is
carried by particle $i$. Under Lorentz transformations, we must have
\begin{equation} M^{a_1 \cdots a_n}(p_1,\cdots,p_n) =
D[W(\Lambda,p_1)]^{a_1}_{\;\;b_1} \cdots D[W(\Lambda,p_n)]^{a_n}_{
\;\;b_n} M^{b_1 \cdots b_n}(\Lambda p_1, \cdots, \Lambda p_n)
\end{equation} where $D[W(\Lambda,p)]$ represents the action of the
little group~\cite{Weinberg:1995mt}.

It would naively appear that the field theories with the simplest amplitudes would be theories of scalar fields. There are no spin indices, and the Lagrangians for these theories take the simplest form. Indeed, in the usual formalism of field theory, there are additional complications for massless particles of spin $s \ge 1$.
For instance in the case of spin 1 particles, Feynman diagrams compute not amplitudes labeled by little group indices, but instead ``amplitudes" with Lorentz indices $M^{\mu_1 \cdots \mu_n}$. One would like to find polarization vectors $\epsilon_\mu^a(p)$ with both Lorentz and little group indices,transforming so that $\epsilon_{\mu_1}^{a_1} \cdots \epsilon_{\mu_n}^{a_n} M^{\mu_1 \cdots \mu_n}$ has the right transformation properties. However it is impossible to define such polarization vectors; the best that can be done is to find polarizaton vectors that transform in the proper way up to a  shift proportional to $p_\mu$. For the amplitudes to be properly Lorentz covariant, this additive piece must vanish when dotted into $M^{\mu_1 \cdots \mu_n}$, which requires the theory to have gauge redundancy. This already forces a relatively complicated Lagrangian structure with more intricate Feynman rules than simple scalar theories. For gravity, the redundancy balloons into diffeomorphism invariance, and perturbation theory becomes very complicated indeed--not only are there an infinite number of vertices, but even the simplest cubic vertex has $\sim 100$ terms!

Nevertheless, over the years, an increased understanding of scattering amplitudes in these theories has yielded a wonderful
surprise: the amplitudes for the naively most complicated theories exhibit beautiful simplicity and structure that is not present for the naively simpler theories~\cite{MHV}. To gather this data, it has been necessary to find ways of computing amplitudes other than the hopeless direct evaluation of Feynman diagrams~\cite{BG,reviews}.
The techniques for doing so revive the central ideas from the S-matrix program~\cite{Smatrix}. A particularly important fact is that amplitudes reveal their structure most transparently when studied for complex momenta~\cite{complex}. There is by now a well-developed industry for using these ideas to compute amplitudes to high orders in perturbation theory~\cite{Bern:2007dw}. But we will begin our discussion of the surprising inversion of what theories are simple in the simplest way, by considering tree amplitudes.

\subsection{BCFW Recursion Relations}

Consider the $n$-point amplitude $M(p_i,h_i)$ for massless particles with $h_i$ ``helicities" in a general number $D$ of spacetime dimensions. When we consider gauge theory, we will define $M(p_i,h_i)$ such that the color factors are already stripped away.
We will also suppress the trivial overall multiplicative coupling constant dependence.

One would like to study the amplitude for complex on-shell momenta in the simplest possible way. It is impossible to have only one momentum complex by momentum conservation. The key idea of BCFW~\cite{BCFW} is then to pick two external momenta $p_1$,$p_2$, and to analytically continue these momenta keeping them on-shell and maintaining momentum conservation. Specifically, BCFW take \be p_{1} \to p_1(z) = p_{1} + q z \ \ \ \mathrm{and} \ \ \ p_2 \to p_2(z) =
p_2 - q z \ee where to keep $p_1^2(z) = p_2^2(z) = 0$, we must have $q \cdot p_{1,2} = 0$ , $q^2 = 0$. This is impossible for real $q$, but possible for complex $q$. To be explicit, choose a Lorentz frame where $p_1,p_2$ are back to back with equal energy and use units where that energy is $1$. Then, we can choose \be p_1  = (1,1,0,0;0..,0), \ p_2 = (1,-1,0,0;0,..0), \ q = (0,0,1,i;0,..0) \ee We could keep all the momenta real but imagine that we are working in $SO(D-2,2)$ signature; however, this does not seem particularly fundamental, and it appears to be better to get used to complexifying all momenta in general.  Note that this deformation only makes sense for $D \ge 4$.

What about the polarization tensors? Note that for gauge theory in a
covariant gauge,  $q = \epsilon^+_1 = \epsilon^-_2$. This makes it
natural to use a $+,-,T$ basis for spin 1 polarization vectors where
\be \epsilon_1^+ = \epsilon_2^- = q, \ \ \ \epsilon_1^- =
\epsilon_2^+ = q^*, \ \ \ \epsilon_T = (0,0,0,0,...,1,...,0) \ee
with $D-4$ different $\epsilon_T$ forming a basis in the transverse
directions. When the momenta are deformed, the polarization vectors
must also change to stay orthogonal to their associated momenta and
maintain their inner products. This requires \be \epsilon_1^+(z) =
\epsilon_2^-(z) = q, \ \ \epsilon_1^-(z) = q^* - z p_2, \ \
\epsilon_2^+(z) = q^* + z p_1, \ \ \epsilon_T(z) =
(0,0,0,0,...,1,...,0) \ee Graviton polarization tensors are simply
symmetric, traceless products of these spin 1 polarization vectors.

With this deformation, $M(p_i,h_i) \to M(z)$ becomes a function of $z$. At tree level, $M(z)$ has an extremely simple analytic structure -- it only has simple poles.  This follows from a straightforward consideration of Feynman diagrams, as all singularities come from propagators, which are simply \be \frac{1}{P(z)^2} = \frac{1}{\left(\sum_{i \in L} p_i \right)^2} \ee where $L$ is some subset of the $n$ momenta. Since $p_1(z) + p_2(z)$ is independent of $z$, this only has non-trivial $z$ dependence when only one of $p_1(z)$ or $p_2(z)$ are included in $L$.  Without loss of generality we take $1 \in L$, in which case we have \be
\frac{1}{P(0)^2 + 2 z q \cdot P} . \ee This shows that all singularities are simple poles located at $z_P = - P(0)^2 / (2 q \cdot P)$. Furthermore, the residue at these poles has a very simple interpretation as a product of lower amplitudes: \be P^2(z) M(z) \stackrel{z\to z_P}{\longrightarrow} \sum_h M_L(\{p_1(z_P), h_1\}, \{-P(z_P), h\}, L) \times M_R(\{p_2(z_P), h_2\}, \{P(z_P), -h\}, R) \ee where we have a sum over helicities for the usual reason, guaranteed by unitarity, that the numerator of the propagator can be replaced by the polarization sum on shell.

So far everything has been kinematical and true for an arbitrary theory. What is remarkable is that for certain amplitudes in some theories, $M(z \to \infty)$ vanishes. Since meromorphic functions that vanish at infinity are completely characterized by their poles; $0 = \frac{1}{2 \pi i} \int_{\it \cal C} \frac{dz^\prime}{(z^\prime
- z)} M(z^\prime) = M(z) + \mathrm{residues}$ if ${\it \cal C}$ is a contour at infinity enclosing all the poles, we find the BCFW recursion relation~\cite{BCF,BCFW} for computing $M(z)$ \be M(z) = \sum_{L,h} M_L(\{p_1(z_P), h_1\}, \{-P(z_P), h\}, L) \frac{1}{P^2(z)} M_R(\{p_2(z_P), h_2\}, \{P(z_P), -h\}, R) \ee where $h$ indicates a possible internal helicity. These recursion relations produce a higher-point amplitude by sewing together lower-point on shell amplitudes. The lower amplitudes are on-shell (in complexified momentum space), because all the momenta are on shell though evaluated at a complex $z=z_P$. The original amplitude we are interested in is obtained by evaluating $M(z)$ at $z=0$.

\begin{figure}[h]
\begin{center} \label{FigBCFWRR}
\includegraphics[width=17cm]{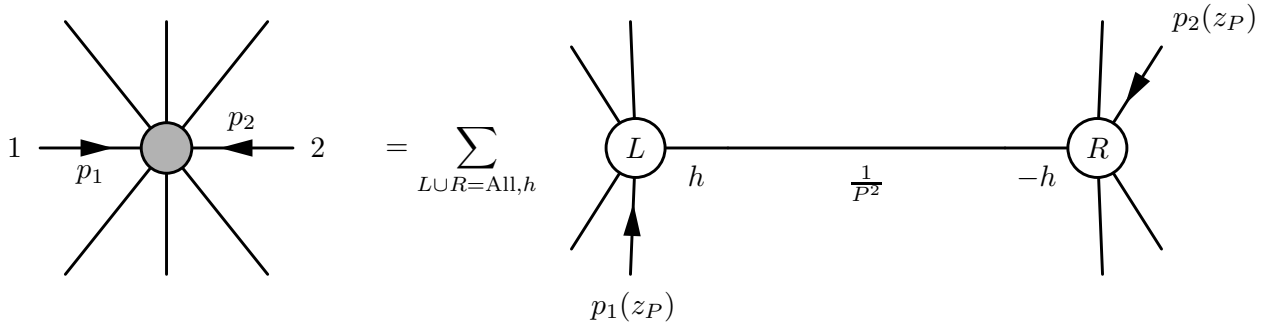}
\caption{\small{The BCFW recursion relation computes an $n$-point amplitude by sewing together lower-point amplitudes with complex on-shell momenta.}} \end{center} \end{figure}

The vanishing of $M(z)$ as $z \to \infty$ is far from obvious from inspection of Feynman diagrams-indeed it is naively never true! For instance, the amplitudes in $\phi^4$ theory go to a constant as $z \to \infty$. Naively, the situation is even worse for gauge theories and gravity, where momentum dependence in the vertices would make amplitudes blow up at infinite momentum increasingly badly.
Remarkably, however, certain amplitudes
\begin{equation}
M^{{\rm anything} -}_{{\rm YM}} \rightarrow \frac{1}{z}, \, M^{{\rm anything} -}_{{\rm Grav}} \rightarrow \frac{1}{z^2} \end{equation} do vanish at infinity~\cite{grav1, AK}. This is enough to get recursion relations, because for any amplitude, we can always adjust $q$ to correspond to the $-$ polarization of $2$.

A physical understanding of the behavior of amplitudes in the $z \to \infty$ limit has recently been given in \cite{AK}. The limit corresponds to a hard (complex) light-like particle blasting through a soft background, and can be conveniently studied using background field method and a background $q-$light-cone (or ``space-cone") gauge~\cite{ChS}. The particle spin plays an important role since there is an enhanced ``Spin-Lorentz" invariance at infinite momentum. This is easy to see in the case of Yang-Mills theory. We can expand the gauge field ${\cal A}_\mu = A_\mu + a_\mu$ where $A_\mu$ is the soft background; after standard gauge fixing the Lagrangian is \be L = -\frac{1}{4} \mathrm{tr} \, \eta^{a b} D_{\mu} a_{a} D^{\mu} a_{b} + \frac{i}{2} \mathrm{tr} [a_a, a_b] F^{a b}.
\ee The first term has the derivative coupling and dominates the large $z$ amplitude but is invariant under a ``Spin Lorentz"
symmetry which only rotates the spin indices, while the second term breaks the Lorentz symmetry as an antisymmetric tensor. This allows us to determine the form of the amplitude in $ab$ space, $M^{ab}$, to be \begin{equation} M_{s = 1}^{ab} = \left( c z \eta^{ab} + A^{ab}
+ \frac{B^{ab}}{z} + \cdots \right)
\end{equation}
where $A^{ab}$ is antisymmetric just as $F^{ab}$ is. Contracting this form of $M^{ab}$ with the polarization vectors and using the Ward identity gives the desired large $z$ scaling of the amplitudes.
The same analysis can be done for gravity. The ``Spin-Lorentz"
symmetry is here twice as big, with the graviton fluctuation having two ``Spin-Lorentz" indices $a,\bar{a}$. The large $z$ and tensor structure of the amplitude is of the form \begin{equation} M_{s = 2}^{ab \bar{a} \bar{b}} = M_{s = 1}^{ab} \times \bar M_{s=1}^{\bar{a} \bar{b}}.
\end{equation}
This is a concrete form of the ``Gravity = Gauge $\times$ Gauge"
connection which is so ubiquitous in perturbative gravity computations \cite{KLT}, and explains why the large $z$ scaling of gravity amplitudes is the square of the corresponding gauge amplitudes. This gives a physical understanding of why higher-spin amplitudes are better behaved at infinite momentum: they are governed by larger kinematical symmetries.

For YM and gravity, the recursion relations can be used to
systematically reduce the number of external legs until we reach the
three-point amplitude, which can't be recursed further. The
three-point amplitude is a fundamental object in field theory. While
it is impossible to have three particles on-shell for real momenta,
it is possible to do so for complex momenta. Since $p_1^2 = p_2^2 =
(p_1 + p_2)^2 = 0$, all the kinematic invariants $p_i \cdot p_j =
0$, so the structure of the amplitude is completely fixed by the
particle spins, (even non-perturbatively). Thus armed with the
recursion relations, all tree amplitudes for YM and gravity are seen
to be determined by Poincare invariance.

The analysis of the large $z$ scaling of amplitudes has recently
been extended by Cheung \cite{Cliff} to include general spin 1 and
spin 2 theories, including arbitrary matter fields with minimal
derivative couplings, finding that the pattern observed for pure
Yang-Mills and Gravity theories holds in general. For theories of
spin $s$, the $M^{{\rm anything}, -s }$ and $M^{{+s, \rm anything}}$
amplitudes vanish as $z \to \infty$.

We therefore see that tree amplitudes for gravity and gauge theory can be determined by BCFW recursion relations but that those of scalar theories are not. This can be said more invariantly as follows. An $n-$point tree amplitude has singularities--corresponding to its residues on its factorization channels. Can one recover the full amplitude from the knowledge of these singularities? The answer is no for scalar theories and yes for spin-1 and spin-2. In fact something stronger is true:
note that the BCFW construction makes factorization manifest for a subset of factorization channels, with $1,2$ on opposite sides. It is highly non-trivial that ensuring that these channels factorize correctly guarantees that {\it all} channels factorize correctly; this remarkable fact is encoded in the statement that
$M(z) \rightarrow 0$ as $z \to \infty$.

The physical understanding of the large $z$ behavior of scattering amplitudes in \cite{AK} relied heavily on the structure of the two-derivative Lagrangian, as well as gauge redundancy. It would be very enlightening to be able to derive this result directly from an amplitude-based argument, without any reference to the Lagrangian description. One would then have to directly prove that for gauge theories and gravity, the BCFW ansatz for the tree scattering amplitude automatically has the correct ``non-obvious"
factorizations.

Finally, note that already at tree-level we see hints of
 extra simplicity in gravitational amplitudes that are absent
 in gauge amplitudes. The BCFW constructible amplitudes all vanish manifestly as
 $\frac{1}{z}$ as $z \rightarrow \infty$. In Yang-Mills theory, \cite{AK}
 found that if the BCFW deformed legs are adjacent in color, the amplitude indeed scales
 as $\frac{1}{z}$, while for non-adjacent colors, it falls faster as $\frac{1}{z^2}$.
 For gravity, there is no color and thus no difference between different color orderings,
 and the amplitude vanishes as $\frac{1}{z^2}$. Explicitly, the cancelation of the $\frac{1}{z}$ term in $M(z)$ implies a non-trivial relation between tree amplitudes \begin{equation} \label{relations} \sum_{L,h} M_L(\{p_1(z_P), h_1\},\{-P(z_P), h\}, L) \frac{z_P}{P^2} M_R(\{p_2(z_P), h_2\}, \{P(z_P), -h\}, R) = 0.
\end{equation}
Another way of saying this is that $0=\int_{\it \cal C} dz M(z)$ where the contour ${\it \cal C}$ only encloses the point at infinity.

\subsection{Extra Kinematical Simplicity in 4D}

In four dimensions, it is especially simple to both solve for the on-shell kinematics and eschew any explicit reference to polarization vectors, by working with spinor helicity variables, with very well-known nice properties we quickly review \cite{spinor-helicity}. Since massless particles only have helicity in $4D$, the amplitudes are labeled by a string of helicities and the little group action is just multiplication by a phase. Recall that the states are defined by picking a reference light-like vector $k$, and for any other light-like vector $p$, choosing a particular Lorentz transformation $L(p)$ so that $p = L(p) k$. Then, states of momentum $p$ and helicity $h$ are defined as $|p,h \rangle = U(L(p))
|k, h \rangle$. Having defined the states in this way, a general
Lorentz transformation on $|p,h \rangle$ is \begin{equation}
U(\Lambda) |p, h \rangle = e^{i h \theta(\Lambda,p)} |\Lambda p, h \rangle \end{equation} where the phase is the little group rotation.

Now, the $2 \times 2$ matrix $(\sigma^\mu p_\mu)_{\alpha
\dot{\alpha}} = p_{\alpha \dot{\alpha}}$ associated with a massless
4-momentum has vanishing determinant and so has rank one; it can
thus be written as $p_{\alpha \dot{\alpha}} = \lambda^p_\alpha \bar
\lambda^p_{\dot{\alpha}}$. Note for real momenta this is ill-defined
up to a rephasing $\lambda^p \to e^{i \theta} \lambda^p, \bar
\lambda^p \to e^{-i \theta} \bar \lambda^p$. We can choose to write
$k_{\alpha \dot{\alpha}} = \lambda^k_\alpha \bar \lambda^k_{\dot
\alpha}$ for some fixed $\lambda^k, \bar \lambda^k$, and define the
spinors for other null momenta $p$ via $\lambda^p = ({\mathbf L}(p)
\lambda^k)$, where ${\mathbf L}(p)$ is the $SL(2,C)$ representation
of the Lorentz transformation $L(p)$. Having now picked a particular
way of assigning spinors to null momenta, it is easy to see that
\begin{equation} \lambda^{\Lambda p} = e^{\frac{i}{2}
\theta(\Lambda,p)} \left({\mathbf \Lambda} \lambda^p\right)
\end{equation} In other words, under a Lorentz transformation of
their defining momentum, the spinors transform according to the
corresponding $SL(2,C)$ Lorentz transformation {\it and}
multiplication by the little group phase associated with helicity
$\frac{1}{2}$. Thus, $SL(2,C)$ invariant objects built out of these
spinors with the appropriate powers are guaranteed to transform
properly as amplitudes under Lorentz transformation: in four
dimensions, amplitudes are directly functions of these spinors. We
will henceforth drop the superscript $``p"$ on the spinors
$\lambda$.

The invariants are built out of spinor products $\langle \lambda_1
\lambda_2 \rangle = \epsilon_{\alpha \beta} \lambda_1^\alpha \lambda_2^\beta$ and $[\bar \lambda_1 \bar \lambda_2] = \epsilon^{\dot{\alpha} \dot{\beta}} \bar \lambda_{1 \dot{\alpha}}
\bar\lambda_{2 \dot{\beta}}$. For instance, consider the 4-graviton scattering amplitude $M^{--++}_{4 \, {\rm grav}}$. Lorentz covariance alone fixes the structure of the amplitude to be \begin{equation}
M^{--++}_{4 \, {\rm grav}} = (\langle 1 2 \rangle [3 4])^4 F({\rm
s,t,u})
\end{equation}
with no reference to an auxiliary ``amplitude" $M^{\mu_1 \nu_1, \cdots, \mu_4 \nu_4}$ or to polarization tensors $\epsilon^{\mu \nu}$.  It is also particularly convenient that the spinor helicities are two dimensional vectors. This means that two generic spinors $\lambda_1,\lambda_2$ with $\langle 1 2 \rangle \neq 0$ give a basis for expanding any other spinor \begin{equation} \lambda = \frac{\langle \lambda 2 \rangle \lambda_1 - \langle \lambda 1 \rangle \lambda_2}{\langle 1 2 \rangle}.
\end{equation}
This also means that if we look at differences of product of the form $\langle a C \rangle \langle b D \rangle - \langle a D \rangle \langle b C \rangle$, that it would vanish if either $|a\rangle$ is proportional to $|b\rangle$ {\it or} $|C \rangle$ is proportional to $|D\rangle$. Thus we have \begin{equation} \langle a C \rangle \langle b D \rangle - \langle a D \rangle \langle b C \rangle = \langle a b \rangle \langle C D \rangle \end{equation} which is known as the Schouten identity.

As we have emphasized, amplitudes in 4D QFT's are directly a function of spinor helicities. One can put this knowledge to good use even using usual Feynman diagrams. Here polarization vectors are needed, and one can choose for spin 1 \begin{equation} \epsilon^+_{\alpha \dot{\alpha}} = \frac{\mu_\alpha \bar \lambda_{\dot{\alpha}}}{\langle \mu \lambda \rangle}, \, \epsilon^-_{\alpha \dot{\alpha}} = \frac{\lambda_\alpha \bar \mu_{\dot{\alpha}}}{[\bar \lambda \bar \mu]} \end{equation} where we have introduced arbitrary auxiliary spinors $\mu, \bar \mu$. Changing $\mu$ shifts the polarization vector by something proportional to $p_{\alpha \dot{\alpha}}$, which make vanishing contributions to the amplitude. Spin 2 polarization vectors are the obvious squares of the spin-1 ones.

A workable generalization of this formalism to higher dimensions is still lacking. However, since the discussion in this paper will be about ${\cal N} = 4$ SYM and ${\cal N} = 8$ SUGRA in four dimensions, we will make full use of the spinor helicity formalism. It is useful to state the BCFW deformation of momenta directly as a deformation of spinor helicities \begin{equation}
\lambda_1(z) = \lambda_1 + z \lambda_2,\; \bar \lambda_1(z) = \bar\lambda_1; \, \, \lambda_2(z) = \lambda_2,\; \bar \lambda_2(z) = \bar
\lambda_2 - z \bar \lambda_1.
\end{equation}

Beyond tree-level, scattering amplitudes in four dimensions can have both UV and IR divergences. The IR divergences in particular make a naive computation of ``S-matrix elements" ill-defined, unless they are combined into ``IR safe" observable quantities. This issue is side-stepped by imagining that the computations are done in $4 - 2 \epsilon$ dimensions for $\epsilon < 0$, with the external momenta fixed in 4D. This regulates the IR divergences, which appear as $1/\epsilon$ poles in the amplitudes.

\subsection{Why are ${\cal N}$=4 SYM and ${\cal N}=8$ SUGRA Special?}

The obvious answer to the question posed above is ``because they have so much symmetry"! Indeed, investigations of amplitudes in ${\cal N} = 4$ SYM and ${\cal N} = 8$ SUGRA at loop level over the past fifteen years have uncovered beautiful structures, with many of the computations made possible by the powerful on-shell ``unitarity based method" of Bern, Dixon and Kosower \cite{BDK}. In ${\cal N} = 4$ SYM, one loop amplitudes were found to contain only scalar ``box"
integrals. More recently, a remarkable connection between scattering amplitudes and Wilson line expectation values has been made, associated with a still mysterious ``dual conformal symmetry", beginning with the work of Alday and Maldacena who used AdS/CFT \cite{juan} to study amplitudes at strong coupling~\cite{Malda}, with confirming evidence from perturbative calculations \cite{perturbation}. In parallel with these developments, ${\cal N} = 8$ SUGRA amplitudes have also been intensively explored at loop level. The by now well-known surprise is that they are much better behaved in the UV than one would expect from power-counting, precisely mirroring that of ${\cal N} = 4$ SYM. Explicit computations for up to six external gravitons \cite{BohrNT} and for all MHV amplitudes \cite{Bern:1998sv} showed that ${\cal N} = 8$ SUGRA only has scalar box integrals just as ${\cal N} = 4$ SYM, leading to the ``no-triangle" hypothesis for all amplitudes in ${\cal N} = 8$ SUGRA \cite{Bern:2005bb}. For four external legs, a remarkable computation of Bern et. al. \cite{Bern2} has shown the divergence structure of ${\cal N} = 8$ to be the same as ${\cal N} = 4$ SYM up to three loops. This is in line with indirect string/M-theory arguments suggesting that the divergence structure of ${\cal N} = 4$ and ${\cal N} = 8$ SUGRA should be same to very high look order \cite{Pierre1,Pierre2,Nathan}. These results lead to the natural conjecture that ${\cal N} = 8$ SUGRA is perturbatively finite \cite{BernRoiban,Bern2,Pierre1}.

We wish to point out another elementary but important reason why the
amplitudes in these theories are likely to be simplest. As we have
seen, even beginning at tree level, amplitudes of particles with
spin are more nicely behaved than those of scalar theories. However,
this comes at an expense: for many external particles, the
amplitudes are labeled by an annoying discrete string of $+$'s and
$-$'s associated with the particle helicities. This is because in
Yang-Mills theory and Gravity, in order to have a CPT invariant
spectrum, one has to include separately positive and negative
helicity states. (Theories with only positive helicity particles
have a trivial S-matrix). This doubling of degrees of freedom is
un-natural and the discreteness adds significant complexity to the
amplitudes. A theory with the simplest amplitudes should somehow
have high spin particles without the additional discrete nature of
the scattering amplitudes associated with spin.

Theories with maximal SUSY uniquely accomplish this goal. For
representations with maximum spin $s$ =1 or 2, with ${\cal N} = 4s$
SUSY, the supersymmetries relate {\it all} the helicity states to
each other: the supermultiplet is CPT invariant all by itself, with
no need for doubling. This does not occur with less SUSY. This
remarkable feature of maximally supersymmetric theories allows us to
label the external states in a natural, smooth way, as Grassmann
coherent states $|\eta \rangle$ or $|\bar \eta \rangle$, built from
the ``ground state" of highest spin states $|-s \rangle$ and $+ s
\rangle$ respectively.  The scattering amplitudes then involve the
better behaved high-spin particles, but are also completely smooth
functions of the momenta and Grassmann parameters; for instance if
we choose to label all external states by $|\eta \rangle$ coherent
states, the amplitudes are of the form $M(\{\eta_i,\lambda_i,\bar
\lambda_i\})$.

Given that maximally supersymmetric theories likely have the simplest amplitudes, it behooves us to understand them in the simplest possible way. That is our goal in this paper.

\subsection{Outline of the Paper and Summary of Results}

We begin with a description of the on-shell superspace that makes the action of SUSY transparent. The states $|\eta \rangle$,$| \bar \eta \rangle$ diagonalize not only the momenta but also the $Q_I$ or $\bar Q^I$ supercharges, respectively; we can use one or other other to label external states. In this basis, the action of SUSY on amplitudes is simple and manifest. As we will see, maximal SUSY allows the good UV properties of the best-behaved high-spin amplitudes to be inherited by normally badly behaved lower-spin amplitudes.

The BCFW deformation $\lambda_1(z) = \lambda_1 + z \lambda_2, \bar
\lambda_2(z) = \bar \lambda_2 - z \bar \lambda_1$ has a natural
supersymmetric counterpart in deforming the corresponding Grassmann
parameters $\eta_1,\eta_2$, as $\eta_1(z) = \eta_1 + z \eta_2$.
Remarkably, with this SUSY generalization of the BCFW deformation of
momenta, we show that {\it all} amplitudes in maximally
supersymmetric theories vanish at infinity \begin{equation}
M(\{\eta_1(z),\lambda_1(z),\bar \lambda_1\},\{\eta_2,\lambda_2,\bar
\lambda_2(z)\}, \eta_i) \rightarrow \frac{1}{z^s}\,\; {\rm as}\; \,
z \rightarrow \infty
\end{equation} which implies that all tree amplitudes in these
theories can be obtained by recursion relations as \cite{wonders}
\begin{eqnarray} & M(\{\eta_1(z),\lambda_1(z),\bar
\lambda_1\},\{\eta_2,\lambda_2,\bar \lambda_2(z)\}, \eta_i) = &
\nonumber \\ & \sum_{L,R}\int d^{\cal N} \eta
M_L(\{\eta_1(z_P),\lambda_1(z_P),\bar\lambda_1\}, \eta, \eta_L) \,
\frac{1}{P^2(z)} \,
M_R(\{\eta_2,\lambda_2,\bar\lambda_2(z_P)\},\eta,\eta_R) &
\label{fullsusy}
\end{eqnarray}

\begin{figure}[h]
\begin{center} \label{FigBCFWRRforSUSY}
\includegraphics[width=16cm]{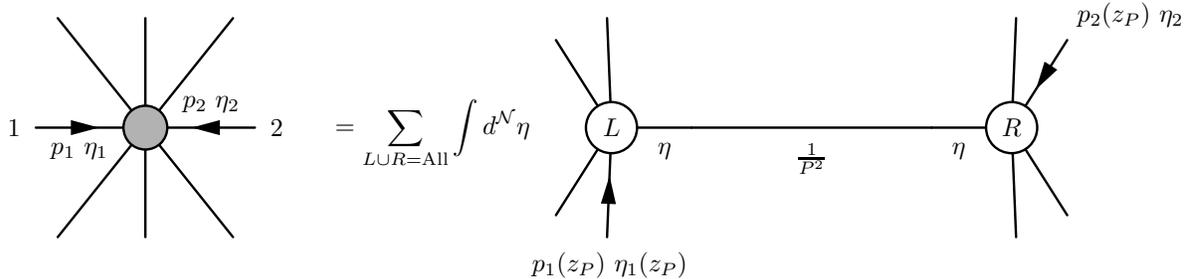}
\caption{\small{The BCFW recursion relations for ${\cal N} = 4$ SYM
and ${\cal N} = 8$ Supergravity.  Note that we must analytically
continue $\eta_1 \to \eta_1(z_P) = \eta_1 + z_P \eta_2$, but
$\eta_2$ is not continued.}} \end{center} \end{figure} Note that
this is not exactly a BCFW expression, since $M_L$ is evaluated at a
shifted value of $\eta_1(z_P)$. In components, this means that in
general, a given amplitude is determined by a recursion relation
involving lower-point amplitudes with different external states.

An important difference between ${\cal N} = 4$ SYM and ${\cal N} =
8$ SUGRA is reflected in their respective vacuum structures. Both
theories have moduli spaces of vacua, but for ${\cal N} = 4$, a
generic vacuum has mostly massive particles; the massless S-matrix
with its beautiful properties only exists at the origin of moduli
space. By contrast, the massless spectrum of ${\cal N} = 8$ is
unchanged along moduli space, and so the massless S-matrix is
defined everywhere. This makes is natural to try and relate the
theories at different points of moduli space, and as is well known,
the theory enjoys a non-linearly realized $E_{7(7)}$ symmetry, which
shifts the 70 scalar fields whose expectation values parametrize the
moduli space. The scalars are in the 4-index antisymmetric tensor of
$SU(8)$; the $E_{7(7)}$ algebra is obtained by appending generators
in this representation to the $SU(8)$ algebra. It is natural to ask
how the existence of the moduli space and the $E_{7(7)}$ symmetry is
reflected in the scattering amplitudes; not surprisingly, it is seen
in the behavior of amplitudes involving the emission of soft
scalars. However, the way this happens is interesting since the soft
limit is generally singular and must be taken with care. Working at
tree level, we first show how this works for the more elementary and
familiar example of the spontaneous breaking of a global symmetry
$G$ to a subgroup $H$; the presence of the non-linearly realized
$G/H$ generators is inferred from the ``anomalous" behavior of
amplitudes with double soft pion emission. We then move on to ${\cal
N} = 8$ SUGRA, using the recursion relation to find the soft
behavior for single and double emission. We prove that the amplitude
for single soft scalar emission vanishes -- this is how the
existence of a moduli space is reflected in the scattering
amplitudes. The $E_{7(7)}$ structure -- in particular the
commutation relations for the 70 ``broken" $E_{7(7)}$ generators
$X_{I_1,\cdots,I_4}$, where $[X,X]$ is an $SU(8)$ rotation -- is
revealed in the soft limit for double scalar emission. We prove that
the amplitude for emitting two soft scalars of momenta $p_1,p_2$ has
a universal form as $p_{1,2} \to 0$ \footnote{This result holds when
the two soft scalars 1,2 do not form an $SU(8)$ singlet. If they do
form an $SU(8)$ singlet, they can produce a soft graviton, and the
amplitude has the associated soft graviton singularity.}
\begin{equation} \label{E7} M_{n+2}(1,2;\eta_3,\ldots
\eta_{n+2})\longrightarrow \sum_{i=3}^{n+2} \frac{1}{2}
\frac{p_i\cdot (p_2-p_1)}{p_i\cdot (p_1+p_2)}
R_{[X_1,X_2]}(\eta_i)M_{n}(\eta_i,\ldots)
\end{equation}
where for a general $SU(8)$ generator $T^J_I$, \begin{equation}
R_{T^J_I}(\eta) = T^J_I \eta_J \partial_{\eta_I} \end{equation}
represents the action of $T^J_I$ on the $|\eta\rangle$ states. Said
in words, the amplitude for double soft emission is nothing but an
$SU(8)$ rotation of the amplitude with only hard momenta, where each
hard line is rotated by an amount dependent on its momentum. This
can be translated into a statement of the action of $E_{7(7)}$ on
the on-shell Hilbert space. Note that this double soft limit is
``anomalous" in the sense that the result is proportional to
$\frac{p_i\cdot (p_2-p_1)}{p_i\cdot (p_1+p_2)}$ and thus depends on
the directions in which the two momenta are sent to zero.
Physically, one is tempted to find a mapping between the external
particle states at differing points in moduli space; one can however
take different ``paths" between vacua, corresponding to taking the
soft momenta to zero in different ways, so there is no canonical
identification between states, but different paths can differ up to
a non-trivial $SU(8)$ rotation.

We then move on to discuss amplitudes at 1-loop. As a preamble, we give a simple general discussion of the structure of 1-loop amplitudes in any QFT. Our presentation is very closely related to the recent work of Forde \cite{Forde} and its generalizations~\cite{forde:extensions} (see also~\cite{OPP}), though our perspective is somewhat different.
The upshot is that the
usual real phase space integrals naturally become complex contour integrals, and the coefficients of scalar triangle and bubble integrals that appear at 1-loop are all determined by the ``pole at infinite momentum" of the products of tree amplitudes that appear in these cuts. In ${\cal N} = 4$ SYM and ${\cal N} = 8$ SUGRA, the on-shell superspace makes it straightforward to compute the multiplet sums that arise in these cuts and relate them to the best behaved tree amplitudes with the highest spin, giving a simple proof for the absence of triangles and bubbles. We also give an argument for the absence of rational terms, thereby establishing the validity of ``no-triangle" conjecture.
This conjecture has also recently been
proven by Bjerrum-Bohr and Vanhove in \cite{BV}, using very different ideas. Having established that only boxes occur in both theories, we can write down an explicit expression for all 1-loop amplitudes \be M^{1\hbox{-}{\rm
loop}}_n(\{\eta_i,\lambda_i,\tilde\lambda_i\}) = \!\!\!\!\!
\sum_{L_t, R_t, L_b, R_b} \!\!\! C_{L_b, R_b}^{L_t, R_t} \left(\{ \eta_1, \lambda_1, \bar \lambda_1 \}, \ldots, \{ \eta_n, \lambda_n, \bar \lambda_n \} \right) I(P_{L_t}, P_{R_t}, P_{L_b}, P_{R_b}) \ee
where $L_t,R_t,L_b,R_b$ label the sets of particles in the top left, top right, bottom left and bottom right of the box respectively, the $I$'s are the standard scalar box integrals, and the coefficients $C_{L_b, R_b}^{L_t, R_t}$ are determined as products of four tree amplitudes \begin{eqnarray} && C_{L_b, R_b}^{L_t, R_t} \left(\{ \eta_1, \lambda_1, \bar \lambda_1 \}, \ldots, \{ \eta_n, \lambda_n, \bar \lambda_n \}
\right) =
 \\ && \sum_{\ell_*} \int \Pi_i d^{\mathcal{N}} \eta_i M_{L_t}(\ell_1^*, \ell_2^*;\eta_1,\eta_2)M_{R_t}(\ell_2^*, \ell_3^*;\eta_2,\eta_3) M_{R_b}(\ell_3^*; \ell_4^*,\eta_3,\eta_4)M_{L_b}(\ell_4^*,
\ell_1^*;\eta_4,\eta_1) \nonumber
\end{eqnarray}
Here the $\ell_i^*$ are the complex on-shell loop momenta which put
all four loop momenta on shell, and the sum refers to the (in
general) two solutions for the frozen momenta. Having shown that all
tree amplitudes with maximal SUSY can be computed via recursion
relations, this gives an algebraic procedure for determining all the
1-loop amplitudes in these theories. Needless to say, these are the
only theories for which such a simple and explicit form of 1-loop
amplitudes can be presented!

Given that both tree and 1-loop amplitudes for maximally supersymmetric theories can be completely determined by their leading singularities, it is natural to conjecture that this property holds to all orders of perturbation theory. This is the nicest analytic structure field theory amplitudes could possibly have, and if true, would directly imply perturbative finiteness for both ${\cal N} = 8$ SUGRA (and of course ${\cal N} = 4$ SYM).

Our investigations suggest that the amplitudes in ${\cal N} = 8$ SUGRA may be even simpler and better behaved than those in ${\cal N} = 4$ SYM. Already at tree level SUGRA amplitudes die off at infinity faster than in SYM, implying relations between tree amplitudes for ${\cal N} = 8$ that are absent for ${\cal N} = 4$. This is inherited at 1-loop, since the ``box" scalar integral coefficients are determined by tree amplitudes, and therefore also satisfy further relations in ${\cal N} = 8$ SUGRA that are absent in ${\cal N} = 4$ SYM. The persistence of the massless S-matrix everywhere on moduli space and the associated $E_{7(7)}$ symmetry is another hint in this direction. We close with some further speculations about the existence and utility of a ``weak-weak" dual formulation of QFT, with ${\cal N} = 8$ SUGRA as its simplest case.

\section{On-Shell Supersymmetry}

We have argued that amplitudes in maximally supersymmetric theories
should be labeled by smooth Grassmann parameters, since all the
states in the supermultiplet are related by SUSY and the mutiplets
are self-CPT conjugate. In this section we will introduce and gain
some familiarity with a very natural formalism that makes this
manifest, using Grassmann coherent states labeled by Grasmann
parameters $|\eta \rangle$ or $|\bar \eta \rangle$. Just as it is
convenient to work with momentum eigenstates, these states are
further eigenstates of the $Q$,$\bar Q$ supercharges; since $Q, \bar
Q$ have a non-vanishing anticommutator, we can diagonalize one or
the other but not both. Using these variables allows us to
diagonalize as many of the operators in the SUSY algebra as we can.
In this basis the amplitudes become completely smooth functions of
the spinor helicities $\lambda_i,\bar \lambda_i$, and $\eta_i$ (or
$\bar \eta_i$). The action of SUSY on the states and the amplitudes
is simple and manifest.

In much of the literature on ${\cal N} = 4$ SYM and ${\cal N}=8$ to
date,  the consequences of SUSY were derived in components, using
SUSY Ward identities, for instance in the case of MHV amplitudes
where they relate pairs of amplitudes. However, this has had
limitations; for instance Ward identities are not very useful for
more general amplitudes, including the next simplest case, {\it
i.e.} NMHV amplitudes, since there are not enough equations to get
relations between pairs of amplitudes.

On the other hand the $\eta, \bar \eta$ variables furnish an
on-shell superspace that allows us to harness the power of
supersymmetry in a more powerful and transparent way than previously
appreciated. The construction is very natural and appears in the
early paper of Nair \cite{Nair}, with a close relationship to the
light-cone superspace of Mandelstam \cite{Mandelstam}. The $\eta$
variables were also used by Witten in \cite{Twistor} to give a
compact form for MHV amplitudes in ${\cal N} = 4$ SYM. Similar ideas
have recently been used in \cite{Dan}, and this basis was recently
used in the last two papers of \cite{perturbation} to study dual
super-conformal invariance for ${\cal N} = 4$ SYM.

This simple formalism will play an essential role in everything we
do in the rest of this paper. It will allow us to control the large
BCFW deformations of general amplitudes, showing that with a
suitable supersymmetric generalization of the BCFW deformation of
momenta, {\it all} amplitudes in ${\cal N} =4$ SYM and ${\cal N} =
8$ SUGRA vanish at infinity, and can therefore be determined by
recursion relations. It will let us easily study the soft limits of
amplitudes with the emission of one and two particles, allowing us
to expose the action of the $E_{7(7)}$ symmetry of ${\cal N} = 8$
SUGRA on amplitudes in a transparent way.  And it will allow us to
replace discrete multiplet sums by smooth Grasmann integrals, which
will be critical in allowing us to understand the multiplet sums
that arise in cuts of loop diagrams, leading to a proof of the
``no-triangle" hypothesis.

\subsection{Coherent States and SUSY Transformations}

Our discussion will be completely parallel for ${\cal N} = 4$ SYM
and ${\cal N} = 8$ SUGRA; we can write ${\cal N} = 4 s$ where $s$ is
the highest spin in the theory, $s=1$ for SYM and $s=2$ for SUGRA.
Let us denote the superchages by $Q_{I\alpha}$ and $\bar{Q}^{I\dot
\alpha}$, where $\alpha$ and $\dot\alpha$ are right and left handed
spinor indices while $I$ is the $SU({\cal N})$ R-symmetry index. As
usual, an object with an (upper) lower $I$ index is in the (anti-)
fundamental representation of $SU({\cal N})$.

As mentioned in the previous section, the main simplification with
maximal SUSY arises from the fact that all helicity states are
related by supersymmetry. The nicest way to make this manifest is to
introduce Grassmann variables $\eta_I$ or $\bar\eta^I$. Then, for a
given massless momentum $p_{\alpha \dot{\alpha}} = \lambda_\alpha
\bar{\lambda}_{\dot{\alpha}}$,
 one can represent all states in the theory in terms of the analog of
coherent states as follows
\begin{equation}
|\bar{\eta}, \lambda, \bar{\lambda} \rangle = e^{\bar{Q}^{I
\dot{\alpha}}\bar{w}_{\dot{\alpha}} \bar{\eta}_I} |+s ,\lambda,
\bar{\lambda} \rangle, \quad |\eta,\lambda, \bar\lambda\rangle =
e^{Q_{I \alpha} w^{\alpha} \eta^I}|-s,\lambda, \bar\lambda\rangle
\end{equation}
where $w_\alpha$ and $\tilde w_{\dot a}$ are spinors such that
$\langle w,\lambda\rangle = 1$ and $[\bar{w},\bar{\lambda}] = 1$.
Here we have chosen a convention for the helicities where
$Q|+\rangle = \bar{Q}|- \rangle = 0$, and e.g.
\begin{equation}
Q_{\alpha I} |-s\rangle = \lambda_\alpha |-s + \frac{1}{2}
\rangle_I, \, \, Q^{\dot{\alpha} I} |+s \rangle = \bar
\lambda^{\dot{\alpha}} |+s - \frac{1}{2} \rangle^I
\end{equation}
Note that e.g. $w_\alpha$ is not uniquely defined, but is fixed  up
to an additive shift $w_\alpha \sim w_\alpha + c \lambda_\alpha$;
however, the state $|\eta \rangle$ is the same for all these choices
of $w_\alpha$ since $\lambda^\alpha Q_{\alpha I} |- \rangle = 0$.
Indeed we could have defined states somewhat more symmetrically,
labeled by $\eta_{I \alpha}$, identifying states $|\eta_{I \alpha}
\rangle \sim |\eta_{I \alpha} + c_I \lambda_\alpha \rangle$. We have
fixed this redundancy by writing $\eta_{I \alpha} = w_\alpha
\eta_I$.

Note that the $\eta$ and $\bar\eta$ representations are equally
valid but complementary descriptions of the same object; the
complete supermultiplet. One diagonalizes $Q$, the other $\bar{Q}$:
\begin{equation}
Q_{I \alpha} |\bar \eta \rangle = \bar \eta_I \lambda_\alpha |\bar
\eta\rangle, \quad \bar{Q}^{I \dot{\alpha}} |\eta \rangle =
\bar{\lambda}^{\dot{\alpha}} \eta^I |\eta \rangle.
\end{equation}
Labeling the states by e.g. $\eta$'s and momenta then diagonalizes
as many of the operators in the SUSY algebra as we can. The
$\eta,\bar \eta$ representations are related via a Grassmann Fourier
transform
\begin{equation}
|\bar{\eta}\rangle = \int d^{\cal N} \eta e^{\eta \bar{\eta}} |\eta\rangle,
\quad |\eta \rangle = \int d^{\cal N} \bar{\eta} e^{\bar{\eta} \eta}
|\bar{\eta}\rangle
\end{equation}
where here and in the rest of the paper, we will typically suppress
the $\lambda$ and $\bar\lambda$ information unless it is needed.

These states clearly transform nicely under SUSY. Under the $Q$
supersymmetric transformation with parameter $\zeta^{I \alpha}$ one
has
\begin{equation}
e^{Q_{I \alpha} \zeta^{I \alpha}} |\eta
\rangle = |\eta + \langle \zeta \lambda \rangle \rangle, \quad e^{Q_{I
\alpha}\zeta^{I \alpha}} |\bar{\eta} \rangle =
e^{\bar{\eta}_J \langle \lambda \zeta^J \rangle} |\bar{\eta}\rangle
\end{equation}
and analogously for the $\bar{Q}$ supersymmetries. Note that $Q$
shifts $\eta$ and rephases the $\bar{\eta}$ state, while $\bar{Q}$
does the opposite.

In general, a given external state can be labeled with $\eta$ or
$\bar \eta$; scattering amplitudes are smooth functions
\begin{equation}
M(\{\eta_i,\lambda_i,\bar \lambda_i\};\{\bar \eta_{\bar
i},\lambda_{\bar i},\bar \lambda_{\bar i}\})
\end{equation}
The correct little group transformations require that
\begin{equation}
M(\{t_i \eta_i,t_i \lambda_i,t^{-1}_i \bar \lambda_i\};\{
t^{-1}_{\bar i} \eta_{\bar i},t_{\bar i} \lambda_{\bar
i},t^{-1}_{\bar i} \bar \lambda_{\bar i}\}) = \prod_{i,\bar i}
t_i^{2s} t_{\bar i}^{-2s} M(\{\eta_i,\lambda_i,\bar
\lambda_i\};\{\bar \eta_{\bar i},\lambda_{\bar i},\bar \lambda_{\bar
i}\})
\end{equation}

SUSY is reflected in the scattering amplitudes following from the
SUSY transformation properties of the states:
\begin{equation}
M(\eta_i;\bar \eta_{\bar{i}})= e^{\sum_j [\bar\lambda_j \bar \zeta]
\eta_j + \sum_{\bar{j}} \langle \lambda_{\bar{j}} \zeta \rangle \bar{\eta}_{\bar{j}}} M(\eta_i +
\langle \lambda_i\zeta \rangle,\bar{\eta}_{\bar{i}} + [\lambda_{\bar{i}} \bar \zeta])
\label{deda}
\end{equation}
where the index $i$ ($\bar{i}$) labels all external states expressed
in term of the $\eta$ ($\bar\eta$) representation.

A special case which will be repeatedly useful in the next sections
is to use the $\eta$ basis for all external states; the action of
the $Q$ SUSY's alone reduces to a simple translation of the $\eta$
co-ordinate
\begin{equation}
M(\eta_i) = M(\eta_i +
\langle \lambda_i, \zeta \rangle)
\label{tras}
\end{equation}
Another fact that we will use frequently is that, because $\zeta_{I
\alpha}$ has the extra $\alpha$ index, in general we can use
(\ref{tras}) to translate up to two $\eta$'s to zero. For instance,
if we write
\begin{equation}
\zeta_{I} = a_I \lambda_1 + b_I \lambda_2
\end{equation}
then
\begin{equation}
\eta_1 \to \eta_1 + \langle 12 \rangle  b, \, \eta_2 \to \eta_2  - \langle 1 2 \rangle a
\end{equation}
so by choosing
\begin{equation}
\zeta_{I \alpha} = \frac{\eta_{2 I} \lambda_{1
\alpha} - \eta_{1 I} \lambda_{2 \alpha}}{\langle 12 \rangle}
\label{gaki}
\end{equation}
we can send $\eta_{1,2} \to 0$. Note this can only be done if
$\langle 1 2 \rangle \neq 0$. We can also use the $\bar{Q}$ SUSY's
to set two $\bar{\eta}$'s to zero, so all in all, we can set two
$\eta$'s and two $\bar{\eta}$'s to zero using the full power of
$Q,\bar{Q}$ SUSY's.

Under the $\bar{Q}$ SUSY's alone, we have
\begin{equation}
M(\eta_i) = e^{\bar \zeta \sum_j \bar \lambda_j \eta_j} M(\eta_i)
\end{equation}
This is in complete parallel to the action of translations $M(p_i) =
e^{i x \cdot \sum_j p_j}M(p_i)$ which tells us that all amplitudes
must be proportional to the momentum conserving delta function
$M(p_i) \propto \delta(\sum_i p_i)$. For the $\bar Q$ SUSY, this
tell us that all amplitudes must be proportional to
\begin{equation}
M(\eta_i) = \delta^{2 {\cal N}} \left(\sum_i \bar \lambda_i \eta_i
\right) \hat{M}(\eta_i)
\end{equation}
Indeed, the object
\begin{equation}
\Delta(\{\eta_i, \bar \lambda_i \}) = \delta^{2 {\cal N}}
\left(\sum_i \bar{\lambda}_i \eta_i\right)
\end{equation}
with a similar definition for
$\bar{\Delta}(\{\bar{\eta}_{\bar{i}},\lambda_i\})$, is
supersymmetric.  It is manifestly so under the $\bar{Q}$ SUSY; under
the $Q$ SUSY the argument of the delta function shifts by $\sum_i
\bar \lambda_i \lambda_i \zeta$ which vanishes due to momentum
conservation $\sum_i \lambda_i \bar \lambda_i = 0$. The function
$\hat{M}(\eta_i)$ then only transforms under the $Q$ SUSY by $\eta$
translations. This form is useful when we wish to commit to using
only $\eta$'s or $\bar \eta$'s to label the external states; in
general we will find it useful to be able to label the states by
either $\eta$ or $\bar \eta$ as needs dictate. We will therefore
work generally with $M(\eta_i)$ rather than $\hat{M}(\eta_i)$.

Finally, we note that while ${\cal N} = 4$ SYM and ${\cal N} = 8$
SUGRA are both C,P,T invariant theories, our choice of labeling
states by $\eta$'s or $\bar \eta$'s makes either negative or
positive helicities special, and therefore does not make these
discrete symmetries manifest. The PT invariance of the theory is
reflected in the fact that the amplitude in the $\{\bar
\eta,\lambda,\bar \lambda\}$ representation, is the same function as
if we worked with the $\eta$ representation, but replaced $\eta \to
\bar \eta$ and flipped $\lambda \leftrightarrow \bar \lambda$.
Explicitly, we have
\begin{equation}
\label{PT} \int \prod_i d^{\cal N} \eta_i e^{\bar \eta_i \eta_i}
M(\{\eta_i,\lambda_i,\bar \lambda_i\}) = M(\{\bar \eta_i,\bar
\lambda_i,\lambda_i\})
\end{equation}
where on the RHS, $M(\bar \eta_i)$ is not the amplitude in the $\bar
\eta$ representation, but rather the amplitude in the $\eta$
representation evaluated with $\eta_i \to \bar \eta_i$.

\subsection{Simple Applications}

Let us see how we can use this formalism to reproduce classic
results following from SUSY, such as the vanishing of $M^{++ \cdots
+}$ and $M^{++ \cdots +-}$ for the highest spin $s$ particles. Take
the first
\begin{equation}
M^{++ \cdots +} = \int d^{\cal N} \eta_1 \cdots d^{\cal N} \eta_n
M(\eta_1,\cdots,\eta_n)
\end{equation}
This vanishes because we can do a $Q$ SUSY translation to set e.g.
$\eta_1$ to zero; the rest of the $\eta$'s translate but we are
integrating over them, so we get
\begin{equation}
M^{++ \cdots +} = \int d^{\cal N} \eta_1 d^{\cal N} \eta_2 \cdots
d^{\cal N} \eta_n M(0,\eta_2,\cdots,\eta_n) = 0
\end{equation}
because there are no $\eta_1$'s to soak up the $d^{\cal N} \eta_1$
integral. Similarly,
\begin{equation}
M^{++ \cdots +-} = \int d^{\cal N} \eta_1 d^{\cal N} \eta_2 \cdots
d^{\cal N} \eta_{n-1}d^{\cal N} \bar{\eta}_n
M(\eta_1,\eta_2,\cdots,\eta_{n-1},\bar{\eta}_n)
\end{equation}
Here we can use the $Q$ SUSY transformation to the maximum, by
translating both $\eta_{1,2} \to 0$. This translates the other
$\eta$'s but again this is irrelevant since we integrate over them,
and also gives a phase involving $\bar{\eta}_n$:
\begin{equation}
\int d^{\cal N} \eta_1 d^{\cal N} \eta_2 \cdots d^{\cal N}
\bar{\eta}_n e^{\bar{\eta}_n (A \eta_1 + B \eta_2)}
M(0,0,\eta_3,\cdots,\eta_{n-1},\bar{\eta}_n) = 0
\end{equation}
because the only dependence on $\eta_{1,2}$ is through $A \eta_1 + B
\eta_2$ and we get $0$ by the Grassmann integral over the orthogonal
combination. Here $A$ and $B$ can be explicitly computed using
(\ref{gaki}) but their form is not important for this argument
except to note that we can do this only when $\langle 1 2 \rangle
\neq 0$; this is good, because as we will see in a moment, for the
three-point amplitude, when $\langle 1 2 \rangle = 0$, $M^{++-}$ is
indeed non-vanishing.

Next consider the case of the Maximally Helicity Violating (MHV)
amplitude for spin $s$, $M^{++\cdots +--}$, which we choose to write
as
\begin{equation}
M^{++\cdots +--} = \int d^{\cal N} \eta_1 d^{\cal N} \eta_2 \ldots
d^{\cal N} \eta_{n-2}d^{\cal N} \bar\eta_{n-1}d^{\cal N} \bar\eta_n
M(\eta_1,\eta_2,\eta_3,\ldots, \eta_{n-2},\bar\eta_{n-1},\bar\eta_n)
\end{equation}
and use the $Q$ SUSY to translate, say, $\eta_1$ and $\eta_2$ to
zero at the expense of picking up phases and translating all other
$\eta$'s,
\begin{equation}
M(\eta_1,\eta_2,\eta_3,\ldots, \eta_{n-2},\bar\eta_{n-1},\bar\eta_n) = e^{\sum_{i=n-1}^n \bar\eta_i(A_i\eta_1+B_i\eta_2)}M(0,0,\eta_3',\ldots, \eta_{n-2}',\bar\eta_{n-1},\bar\eta_n)
\end{equation}
where $A_i$ and $B_i$ are computed using (\ref{gaki}). Performing a
change of variables from $\eta_1, \eta_2$ into two new variables
$\eta_{n-1}$ and $\eta_n$ as $\eta_{n-1} = A_1\eta_1+B_1\eta_2$ and
$\eta_{n} = A_2\eta_1+B_2\eta_2$ one picks up a jacobian which is
given by
\begin{equation}
{\cal J} = \left(\frac{\langle n 1 \rangle \langle (n-1) 2 \rangle
-\langle (n-1) 2 \rangle \langle n 1 \rangle}{\langle 1 2
\rangle^2}\right)^{\cal N} = \left(\frac{ \langle (n-1)  n
\rangle}{\langle 1 2 \rangle}\right)^{\cal N}.
\end{equation}
where in the last equality we used the Schouten identity.

Using this in the equation one finds
\begin{eqnarray}
\label{mhv}
M^{++\cdots +--} & = & \left(\frac{\langle (n-1) n \rangle}{\langle
1 2 \rangle}\right)^{\cal N} \int d^{\cal N} \eta_{n-1} d^{\cal N}
\eta_n d^{\cal N} \eta_3\ldots
\\
& & \int d^{\cal N} \eta_{n-2}\int d^{\cal N}
\bar\eta_{n-1}d^{\cal N} \bar\eta_n
e^{\bar\eta_{n-1}\eta_{n-1}}e^{\bar\eta_{n}\eta_{n}}M(0,0,\eta_3,\ldots,
\eta_{n-2},\bar\eta_{n-1},\bar\eta_n).\nonumber
\end{eqnarray}
Now we can perform the $\bar\eta$ integrations which simply produce the corresponding $\eta$ state representations for $\eta_{n-1}$ and $\eta_n$. This gives the $\eta$ representation of the $M^{--++\ldots +}$ amplitude. Therefore we conclude that
\begin{equation}
M^{++\cdots +--} = \left(\frac{ \langle (n-1) n \rangle}{\langle 12
\rangle}\right)^{\cal N} M^{--++\ldots +}
\end{equation}
This is the well known form of the Ward identities for MHV
amplitudes. Note that this implies
\begin{equation}
\label{hatMHV}
M^{+ \cdots -_j \cdots -_k \cdots +} = \langle j k \rangle^{\cal N}
\hat{M}_{MHV}(\lambda_i,\bar \lambda_i)
\end{equation}
where the function $\hat{M}_{MHV}$ is independent of the $-$
helicity states $j,k$. For ${\cal N} = 4$ SYM, $\hat{M}$ is the
simply the denominator of the famous Parke-Taylor amplitude
\begin{equation}
\hat{M}_{MHV} = \frac{1}{\langle 1 2 \rangle \langle 2 3 \rangle \cdots \langle n 1 \rangle}
\end{equation}
and is only a function of the $\lambda_i$. For gravity there is no
similarly explicit for of $\hat{M}_{MHV}$, and it is not
holomorphic, depending on both $\lambda_i,\bar \lambda_i$.

\subsection{The 3 and 4 Particle Amplitudes}

As mentioned in the introduction, the three-point amplitude is a
fundamental object. While it is impossible to have three particles
on-shell for real momenta, it is possible to do so for complex
momenta. Since $p_1^2 = p_2^2 = (p_1 + p_2)^2 = 0$, all the
kinematic invariants $p_i \cdot p_j = 0 \to \langle i j \rangle [i
j] = 0$. In fact one can easily see by momentum conservation that
either $\langle i j \rangle = 0$ or $[i j] = 0$, so the 3-point
amplitude is purely holomorphic or anti-holomorphic. Together with
the physical requirement that the amplitude vanish for real momenta,
the 3-point amplitude can be completely determined by the external
particle spins. For instance, for particles of spin $s$, the
amplitudes for $(--+)$ and $(++-)$ helicities must be of the form
\begin{equation}
M_3^{--+}  =  \left(\frac{\langle 12\rangle^4}{\langle 12 \rangle
\langle 23\rangle\langle 31 \rangle} \right)^s, \, M_3^{++-} =
\left( \frac{[12]^4}{[12] [23][31]} \right)^s
\end{equation}
while for the $(+++)$ and $(---)$ amplitudes are of the form
\begin{equation}
M_3^{+++} = ([12][23][31])^s, M_3^{---} = (\langle 12 \rangle
\langle 23 \rangle \langle 31 \rangle)^s
\end{equation}
These forms are exact, even non-perturbatively, and are determined
up to an overall coupling constant, (which might be zero). The
leading two-derivative terms in the action for Yang-Mills and
Gravity give the $(--+)$ and $(++-)$ amplitudes, while the $(+++)$
and $(---)$ arise from higher-derivative $F^3$ and $R^3$
interactions that may be present in the low-energy effective action,
and in the case of gravity, can be generated as the classic leading
counterterm at two loops~\cite{Goroff:1985th}. With maximal SUSY, we
know that $M^{+++}$ vanishes, and hence it is impossible to
supersymmetrize the $R^3$ term as is well known.

Given the $M^{++-}$ and $M^{--+}$ amplitudes for gluons and
gravitons, it is trivial to use SUSY to determine the entire 3-point
function. If $[ij] = 0$ but $\langle i j \rangle \neq 0$, expressing
the 3-point amplitude as $M(\eta_1,\eta_2,\bar{\eta}_3)$, we can use
SUSY to send $\eta_{1,2}$ and $\bar{\eta}_3$ to zero, relating it to
the $(--+)$ gluon and graviton amplitude for ${\cal N} = 4$ and
${\cal N} = 8$ respectively. We do the opposite operation for the
case with $[ij] \neq 0$. This fixes the amplitude to be, say in the
$\eta_{1,2,3}$ basis,
\begin{equation}
M_3(\eta_i) = \frac{\Delta(\eta_i)}{([12][23][31])^s} +
\frac{\bar{\Delta}(\eta_i)}{(\langle 1 2 \rangle \langle 23 \rangle
\langle 31 \rangle)^s}
\end{equation}
Note that the structure of the denominator is fixed by the required
little group transformation of the amplitude.  Note also that here
$\bar{\Delta}$ is in the $\eta$ representation, i.e.
\begin{equation}
\bar{\Delta}(\eta_i) = \int d^{\cal N} \bar \eta e^{\bar \eta \eta}
\bar{\Delta}(\bar \eta) = \int d^{\cal N} \bar \eta e^{\bar \eta
\eta} \delta^{2 {\cal N}}(\sum_i \lambda_i \bar \eta_i)
\end{equation}
It is easy to verify this form is correct,  for instance
\begin{eqnarray}
M_3^{++-} &=& \int d^{\cal N} \eta_1 d^{\cal N} \eta_2
M_3(\eta_1,\eta_2,0) \nonumber \\ &=&  \frac{1}{([12][23][31])^s}
\int d^{\cal N} \eta_1 d^{\cal N} \eta_2 \delta^{2 {\cal N}}(\bar
\lambda_1 \eta_1 + \bar \lambda_2 \eta_2)=
\frac{1}{([12][23][31])^s} \times [12]^{4s}
\end{eqnarray}

It is also easy to determine the structure of the full 4-pt
amplitude by SUSY. Here we can look at
$M(\eta_1,\eta_2,\bar{\eta}_3,\bar{\eta}_4)$, and use all the SUSY's
to translate all the $\eta,\bar{\eta}$ to zero, yielding the
$(++--)$ amplitude whose form is fixed by Lorentz invariance. For
gravity, it is given by
\begin{equation}
M^{--++} = (\langle 1 2 \rangle [34])^4 \times \left(\frac{1}{{\rm s
\, t \, u}} \, + {\rm polyn.(s, \, t, \, u)} \right)
\end{equation}
where the 1/(s t u) term arises from graviton exchange in the
two-derivative theory, and higher polynomial terms arise from
possible $R^4$ corrections to the effective action. Restricting
purely to the two-derivative theory, this gives us the 4-particle
amplitude for ${\cal N} = 8$ SUGRA
\begin{equation}
\label{M4}
M_4(\eta_1,\eta_2;\bar \eta_3,\bar \eta_4) = \frac{(\langle 12
\rangle [34])^4}{{\rm s \, t \, u}}
\rm{exp}\left[\left(\begin{array}{cc}\eta_1 & \eta_2
\end{array} \right) \left(\begin{array}{cc} \frac{\langle 2 3 \rangle}{\langle 1 2 \rangle} &
\frac{\langle 2 4 \rangle}{\langle 1 2 \rangle} \\ \frac{\langle 3 1
\rangle}{\langle 1 2 \rangle} & \frac{\langle 4 1 \rangle}{\langle 1
2 \rangle}
\end{array} \right) \left(\begin{array}{c} \bar \eta_3 \\ \bar
\eta_4 \end{array} \right) \right]
\end{equation}
Note that in deriving this form, we first translated $\eta_{1,2}$ to
the origin picking up the phase factor in the action on the
$\bar{\eta}$ states, and then shifted $\bar \eta_{3,4}$ to zero
incurring no additional phase (since the $\eta$'s have already been
set to zero). This is the origin of the asymmetry between $\langle
\rangle$ and $[ \, ]$ brackets in this expression for $M_4$; our
form can be seen to be equivalent to the one we would have obtained
translating the $\bar \eta$'s to zero first by using momentum
conservation in the form $\langle 1 2 \rangle [2 4] + \langle 1 3
\rangle [3 4] = 0$.

Another observation \cite{Twistor, Dan} is that there is a simple
generating function for all MHV amplitudes:
\begin{equation}
M_{MHV}(\bar \eta_1,\cdots, \bar \eta_n) = \hat{M}_{MHV} \delta^{2 {\cal N}}
\left(\sum_i  \lambda_i \bar \eta_i \right)
\end{equation}
where $\hat{M}_{MHV}$ is as defined in eqn.(\ref{hatMHV}). This can
be seen from the fact that the delta function, which has to be there
in any amplitude, already gives the correct behavior to satisfy all
Ward identities.

\subsection{Supermultiplet Sums}

One last simple observation which will be useful when discussing
recursion relations and loops is that this formalism is especially
well suited to compute sums over helicity in intermediate states.
Suppose one has a tree amplitude near a multi-particle singularity.
Then one would expect that the amplitude factorizes as the product
of two amplitudes with a new physical state. One has to add
contributions from all the particles in the spectrum of the theory.
This is usually written as
\begin{equation}
\sum_{h\in {\rm Multiplet}} M_L(\{h,\ell\})M_R(\{-h,-\ell\})
\end{equation}
where $M_L$ and $M_R$ are the Left and Right factors that come out
of the limit. We will find it very useful to write this in a
manifestly supersymmetric manner as follows
\begin{equation}
\int d^{\cal N} \eta M_L(\{\eta,\lambda,\bar\lambda\})M_R(\{\eta,
\lambda,-\bar\lambda\}).
\end{equation}
Note the minus sign judiciously chosen on the $\bar \lambda$, which
ensures that the second momentum has the opposite sign as the first.
It also ensures that the amplitude is supersymmetric: under $Q$
SUSY, the $\eta$ shifts equally in both terms since the $\lambda$'s
are the same, while under the $\bar{Q}$ SUSY, the minus sign on the
second $\bar{\lambda}$ ensures that the phases cancel out in the
product of the two terms. The advantages of working with amplitudes
labeled by continuous rather than discrete parameters is very clear
here, with smooth, translationally invariant integrals replacing
discrete sums.

\section{Recursion Relation for General Tree Amplitudes}

In this section we will see that maximally supersymmetric theories
are special even at tree-level. The main result of this section is
to show that {\it all} amplitudes vanish at infinite complex
momentum, provided that ``infinite momentum" is defined in a
naturally supersymmetric way. This immediately implies that BCFW
recursion relations can be applied to any amplitude and to any
particles, regardless of their helicities, in ${\cal N}=8$
supergravity and ${\cal N}=4$ super Yang-Mills.

Naively this is impossible, since e.g. in ${\cal N}=8$ SUGRA,
the amplitudes involving spin 0,1, particles all diverge as $z \to \infty$. In other words, if we BCFW deform the momenta 1 and 2, then in general \begin{equation} M(\{\eta_1,\lambda_1(z),\bar \lambda_1\},\{\eta_2,\lambda_2,\bar
\lambda_2(z)\}, \eta_i) \, {\rm does \,  not \, vanish \, as\, } \, z \to \infty \end{equation} However, this deformation, which only changes the momenta without touching the $\eta$'s,  is clearly un-natural from a supersymmetric perspective. Indeed, recall that we are forced to deform both $\lambda_1$ and $\bar \lambda_2$ in order to conserve momentum. But recall also that e.g. for all $\eta$ labeled amplitudes, there is also the super-delta function $\delta^{2 {\cal N}} (\sum_i \bar \lambda_i \eta_i)$. Under the BCFW shift of $\bar \lambda_2 \to \bar
\lambda_2 - z \bar \lambda_1$, the argument of the delta function changes, unless we also shift \begin{equation}
\eta_1 \to \eta_1(z) = \eta_1 + z \eta_2 \end{equation} which can be seen as enforcing the supersymmetric counterpart of momentum conservation.

We will now prove that with this natural supersymmetric extension of the BCFW deformation, {\it all} amplitudes in maximally supersymmetric theories vanish at infinity \begin{equation} M(\{\eta_1(z),\lambda_1(z),\bar \lambda_1\},\{\eta_2,\lambda_2,\bar
\lambda_2(z)\}, \eta_i) \rightarrow \frac{1}{z^s}\,\; {\rm as}\; \, z \rightarrow \infty \end{equation} The proof is very simple. We can use  $Q$ SUSY's to send
$\eta_1(z),\eta_2 \to 0$, with translation \begin{equation} \zeta = \frac{\lambda_2 \eta_1(z) - \lambda_1(z) \eta_2}{\langle
1(z) 2 \rangle} = \frac{\lambda_2 \eta_1 - \lambda_1 \eta_2}{\langle
1 2 \rangle}
\end{equation}
Note that $\zeta$ is manifestly $z$ independent. We therefore find \begin{eqnarray} M(\{\eta_1(z),\lambda_1(z),\bar \lambda_1\},\{\eta_2,\lambda_2,\bar
\lambda_2(z)\}, \eta_i) &=& M(\{0,\lambda_1(z),\bar \lambda_1\},\{0,\lambda_2,\bar \lambda_2(z)\},\eta_i + \langle \zeta i \rangle) \nonumber \\ & \rightarrow & \frac{1}{z^s} \,\; {\rm as } \,\; z \to \infty \end{eqnarray} where we have used the fact that the translated amplitude is that of two $(-s)$ particles in some general ($z$-independent) background, which as we mentioned in our review of BCFW, are known to vanish at large $z$ as $1/z^s$.

We can therefore conclude that any amplitude can be determined by
recursion as \begin{eqnarray} & M(\{\eta_1(z),\lambda_1(z),\bar
\lambda_1\},\{\eta_2,\lambda_2,\bar \lambda_2(z)\}, \eta_i) =
\nonumber \\ \sum_{L,R}\int d^{\cal N} \eta &
M_L(\{\eta_1(z_P),\lambda_1(z_P),\bar\lambda_1\}, \eta, \eta_L) \,
\frac{1}{P^2(z)} \,
M_R(\{\eta_2,\lambda_2,\bar\lambda_2(z_P)\},\eta,\eta_R)
\label{fullsusy}
\end{eqnarray}
Note again that this is not precisely a BCFW expression, since $M_L$ is evaluated at a shifted value of $\eta(z_P)$. In components, this means that a given amplitude is determined by a recursion relation involving lower-point amplitudes with different external states.


We can also write our recursion relation by redefining $\eta_1(z) \to \eta_1$ as follows \begin{eqnarray} & M(\{\eta_1,\lambda_1(z),\bar \lambda_1\},\{\eta_2,\lambda_2,\bar
\lambda_2(z)\}, \eta_i) = \nonumber \\ \sum_{L,R}\int d^{\cal N} \eta & M_L(\{\eta_1(z_P - z),\lambda_1(z_P),\bar\lambda_1\}, \eta,
\eta_L) \, \frac{1}{P^2(z)} \,
M_R(\{\eta_2,\lambda_2,\bar\lambda_2(z_P)\},\eta,\eta_R)
\end{eqnarray}
Note that, with $\eta_1(z_p - z) = \eta_1 + (z_P - z) \eta_2$, as $z
\to \infty$, there can be large positive powers of $z$ in the
expansion of the right hand side that can overwhelm the $1/z$ from
the $P^2(z)$ factor. This is of course to be expected, since as we
remarked, it is true that many individual amplitudes diverge as $z
\to \infty$. Note, however, that with $\eta_2 = 0$, the amplitude
manifestly vanishes at infinity; so the (anything -) amplitude
vanishes as $z \to \infty$, as also proven by Cheung~\cite{Cliff}.

Our derivation shows very clearly how the notion of what amplitudes are ``simplest" is completely reversed relative to naive expectations. The way we succeeded in proving the good behavior at infinity was by using supersymmetry to relate general amplitudes to that of gluons/gravitons and not to very naively ``simpler" scalar amplitudes.

We also see vividly the extra simplicity of maximally supersymmetric theories. For pure gauge theory and pure gravity, only some of the amplitudes vanish at infinity, while others do not. By contrast, we see that {\it all} amplitudes vanish at infinity in maximally supersymmetric theories, provided that ``infinity" is approached in a naturally supersymmetric way.

Note that for the usual BCFW recursion relations in YM and Gravity,
there is a natural asymmetry between particles $1,2$, since e.g.
$2$, for which $\bar \lambda_2$ is deformed, has to have negative
helicity. On the other hand, with maximal SUSY, we can deform either
$\lambda_1$ or $\bar \lambda_1$ for particle 1, and we get a
recursion relation either way; in the first case, $\eta_1$ is
deformed and $\eta_2$ isn't, in the second case $\eta_2$ is deformed
and $\eta_1$ isn't. Working in components, this gives us two
different recursion relations for the same amplitude. For instance,
consider an amplitude with all gravitons (or all gluons), with
particle 2 having $-$ helicity. The first form of the recursion
relation corresponds to the ``usual" BCFW formula, since $\eta_2 =
0$ and $\eta_1$ is undeformed, and so we write an all-graviton or
all-gluon amplitude in terms of lower-point all-graviton and
all-gluon amplitudes. But we can also do the ``wrong" deformation,
which does not have a usual BCFW formula, but which is written in
terms of a sum over lower-point amplitudes involving all the
particles in the multiplet. It is in fact easy to show that,
starting with the recursion relation in one form, applying the PT
transformation as defined in eqn.(\ref{PT}) and Fourier-transforming
both sides, we end up with the recursion relation in the other form,
so that the two forms should be thought of as being related by PT
invariance. However, the equality of the two forms of the recursion
relation for the same amplitude gives us a relation between
amplitudes \begin{eqnarray} \label{equality} & & \sum_{L,R}\int
d^{\cal N} \eta
M_L(\{\eta_1(z_{P_L}),\lambda_1(z_{P_L}),\bar\lambda_1\}, \eta,
\eta_L) \, \frac{1}{P_L^2}
M_R(\{\eta_2,\lambda_2,\bar\lambda_2(z_{P_L})\},\eta,\eta_R) =
\nonumber
\\ && \sum_{L,R}\int d^{\cal N} \eta
M_L(\{\eta_1,\lambda_1,\bar\lambda_1(z_{P_R}) \}, \eta, \eta_L) \, \frac{1}{P_R^2}
M_R(\{\eta_2(z_{P_R}),\lambda_2(z_{P_R}),\bar\lambda_2\},\eta,\eta_R)
\end{eqnarray}
which is {\it not} directly a consequence of PT invariance. One can write the recursion relation for the amplitude in a symmetrized form between deforming $\eta_1$ and $\eta_2$, to be manifestly PT invariant, but then eqn.(\ref{equality}) must be taken as an additional non-trivial relation.

While the recursion relations have the same form for ${\cal N} =8$ SUGRA and ${\cal N} = 4$ SYM,  just as for pure YM and gravity, the fact that ${\cal N} = 8$ amplitudes vanish as $\frac{1}{z^2}$ rather than $\frac{1}{z}$ at infinity implies a further non-trivial relation between tree-amplitudes for ${\cal N} = 8$ SUGRA, that is absent for ${\cal N} = 4$ SYM:
\begin{equation} 0 = \sum_{L} \int d^8 \eta M_L(\{p_1(z_P),
\eta_1(z_P) \},\{-P(z_P), \eta \}, L) \frac{z_P}{P^2} M_R(\{p_2(z_P), \eta_2\}, \{P(z_P), \eta\}, R) \label{relation} \end{equation}


Before closing this section, it is worth revisiting the relation
between BCFW recursion relations and the infrared singular behavior
in the maximally supersymmetric theories. The original BCF recursion
relations for gluons were inspired by the IR singular behavior of
${\cal N}=4$ SYM. The tree amplitude for the emission of a soft
gluon from a given $n$ particle process is given by a universal soft
factor multiplying $M_n$; the phase space integral over the emitted
soft gluon is IR divergent and this divergence is canceled by IR
divergences from soft gluons in the 1-loop correction to $M_n$. This
fixes the leading IR divergent part of the loop amplitude to have
the form
\begin{equation} \left. M^{\rm 1\hbox{-} loop}_{IR} \right. = -\frac{1}{\epsilon^2}\sum_{i=1}^n (-s_{i,i+1})^{-\epsilon} M^{\rm tree}. \label{known}
\end{equation}
It turns out that the left hand side of this equation can be
computed in terms of linear combinations of products of four tree
level amplitudes. The four amplitudes are ``connected" by an
internal on-shell line which carries all particles in the multiplet.
For a particular linear combination which was found in~\cite{RSV},
one can simplify this further~\cite{BCF} to make it look like a
quadratic recursion relation. Given that our on-shell formalism is
especially suited to transform sums over the supermultiplet into
smooth integrals over superspace, we re-examined this problem. We
have found that the IR equations directly imply the manifestly PT
invariant, supersymmetric form of the recursion relations, i.e.
symmetrized between $\eta_1$ and $\eta_2$ deformations. A detailed
derivation is given in appendix A. However, the IR equations
themselves do {\it not} tell us the additional equality
eqn.(\ref{equality}), reflecting the remarkable fact that each of
the two terms in the symmetrized expression gives the correct
amplitude by itself. This is an additional non-trivial property  not
following from IR behavior or SUSY, but instead following from the
good large $z$ behavior of the amplitudes that allowed a direct
derivation of the SUSY form of the recursion relations to begin
with.

A natural question is whether the analogous IR equations for same ${\cal N}=8$ SUGRA imply a PT symmetrized form of the SUSY recursion relations. Somewhat surprisingly, we have found by explicitly studying the IR singular behavior of 1-loop five-particle amplitudes that the recursion relation does not follow from it! This means that the IR equations and the recursion relations are independent relations among tree-amplitudes which in turn translate into relations among the coefficient of scalar boxes in ${\cal N}=8$ SUGRA. This is one more indication that ${\cal N}=8$ SUGRA amplitudes are more special than their ${\cal N}=4$ SYM counterparts.

\section{Vacuum Structure, Soft Emission and $E_{7(7)}$}

As already mentioned in the introduction, an important difference
between ${\cal N} = 8$ SUGRA and ${\cal N} = 4$ SYM is their
respective vacuum structures. In both theories, there is a moduli
space of vacua, but there are massless states only at the origin of
moduli space for ${\cal N} = 4$, while they are there in all the
${\cal N} = 8$ vacua. We move around in moduli space by giving
expectation values to the 70 scalar fields of the theory, which
transform under the $SU(8)$ symmetry as a four-index antisymmetric
tensor. One might naively think that giving the scalars a vev breaks
the $SU(8)$; but given that the spectrum and supersymmetries are
exactly the same, it must be that a different $SU(8)$ is realized at
a different point in moduli space. This happens because the $SU(8)$
sits inside a larger non-linearly realized symmetry, the famous
$E_{7(7)}$ symmetry of ${\cal N} = 8$ SUGRA in four
dimensions~\cite{E7(7)}. We can think of $E_{7(7)}$ as the group
that we get by taking $SU(8)$ with generators $T_I^J$ and adding
generators $X_{I_1,\cdots,I_4}$ which transform as a four-index
antisymmetric tensor representation under $SU(8)$, associated with
shifting the 70 scalars. Two $X$'s commute into an $SU(8)$ rotation
as
\begin{equation} -i \left[X_{I_1,\cdots,I_4},X_{I_5,\cdots,I_8}\right] =  \epsilon_{J
I_2 \cdots I_8} T^J_{I_1} + \cdots + \epsilon_{I_1, \cdots, I_7 J}
T^J_{I_8} \end{equation} and one can check that the Jacobi identity
is satisfied (this is non-trivial, which is why such constructions
only work in exceptional cases associated with exceptional groups).
As a matter of nomenclature, the ``(7)" in $E_{7(7)}$ refers to the
fact that the $X$ generators are multiplied by an $i$ relative to
the analogous construction for $E_7$, so that they are non-compact
directions, as they clearly must be since the scalar fields can take
arbitrary values. The number in brackets is the number of
non-compact generators minus the number of compact generators, which
in this case is 7 = 70 non-compact - 63 compact generators of
$SU(8)$. The usual compact $E_7$ would be $E_{7(-133)}$ in this
funny notation.

This symmetry is realized in a highly non-trivial way in the
Lagrangian of ${\cal N} = 8$ SUGRA; indeed, its complete action on
all the SUGRA fields was only found in~\cite{Kallosh:2008ic}
explicitly earlier this year! It is therefore very natural to ask
how this non-trivial structure--both the presence of the moduli
space as well as the non-linearly realized $E_{7(7)}$ symmetry--is
reflected in the scattering amplitudes. We aim to clarify this issue
in this section.

\subsection{Pion Preamble}

It is useful to get some intuition from a more familiar example of
non-linearly realized (compact) symmetries associated with ordinary
Goldstone bosons. Consider spontaneous symmetry breaking of a global
symmetry $G$ to a subgroup $H$. Let the unbroken symmetry generators
be $T_i$, and the broken generators be $X_\alpha$; the Goldstone
bosons (pions) are in one-to-one correspondence with the $X_\alpha$.
The $G$ commutation relations are of the schematic form
\begin{equation} [T,T] \sim T, \, [X,T] \sim X, \, [X,X] \sim T
\end{equation} When we have a Lagrangian description, the non-linear
realization of $G$ is manifest as an action on the pion fields. But
how does the presence of the other vacua and the non-linearly
realized $G$ manifest itself in the structure of the scattering
amplitudes? Clearly, since the other vacua can be thought of as
giving expectation values to the pions, we need to look at the
amplitude for soft emission. While this was an intensively explored
subject in the 60's \cite{Adler:1964um,Weinberg:1966kf},to our
knowledge the precise question we are asking has not been addressed
in the literature, and as we will see, it has a very pretty but
somewhat subtle answer.

Let us begin with what one might naively expect the answer to look
like, which will quickly be seen to be incorrect, but will guide us
to the right answer.  Given that the theories at two different vacua
are related by $G$, one is tempted to try and identify states in the
Hilbert spaces of the two vacua. For instance, giving an expectation
value $\langle \pi^\alpha \rangle = \theta^\alpha$, one would be
tempted to define \begin{equation} |\psi\rangle_\theta = e^{i
Q^\alpha \theta^\alpha} |\psi \rangle
\end{equation}
Of course the vacuum is not invariant, and so the $|vac
\rangle_\theta$ state--the $\theta$ vacuum--is a coherent state of
zero momentum goldstone bosons. But the symmetry should tell us that
for scattering amplitudes \begin{equation} M^{\psi_\theta} =
M^{\psi} \end{equation} and if we expand \begin{equation} |\psi
\rangle_\theta = |\psi \rangle + \theta_\alpha |\psi\rangle^\alpha +
\frac{1}{2} \theta_\alpha \theta_\beta |\psi \rangle^{\alpha \beta}
+ \cdots = |\psi \rangle + |\psi^{(1)}\rangle + |\psi^{(2)}\rangle +
\cdots
\end{equation}
the invariance tells us that
\begin{equation}
M^{\psi^{(1)}} = 0 , \, \, M^{\psi^{(2)}} = 0 , \, \cdots \end{equation} Clearly, since the state $\psi^\alpha$ includes a part with one zero momentum pion, $\psi^{\alpha \beta}$ contains two zero-momentum pions, the structure of $G$ is revealed in the behavior of scattering amplitudes with additional pions of zero momentum. It is a famous fact that the amplitude for the emission of a single soft pion vanishes--this ``Adler zero" is the first-order reflection of the presence of the other degenerate vacua~\cite{Adler:1964um}. The first non-trivial information about the group structure should then involve amplitudes with two soft pions.

\begin{figure}[h]
\begin{center} \label{FigAdlerZero}
\includegraphics[width=7cm]{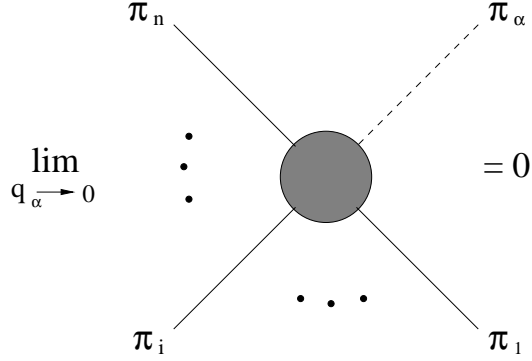}
\caption{\small{There is an Adler zero when a single pion becomes soft.}}
\end{center}
\end{figure}

However, this picture is not quite correct, as can be seen in a
number of ways. For instance, it is clear that the amplitude for the
soft emission of two pions $\alpha,\beta$ should know that
$[X_\alpha,X_\beta] = f_{\alpha \beta}^j T_j$. But there is a little
paradox. Bose statistics tells us that $M^{\alpha
\beta}(p_\alpha,p_\beta, \cdots) = M^{\beta
\alpha}(p_\beta,p_\alpha, \cdots)$; as $p_{\alpha},p_\beta \to 0$,
it seems that $M^{\alpha \beta}(0,0, \cdots) = M^{\beta,
\alpha}(0,0,\cdots)$. But then how can this object, symmetric in
$(\alpha,\beta)$, know about $[X_\alpha,X_\beta]$, which is
antisymmetric in $(\alpha, \beta)$? Related to this, we have assumed
that there is a canonical way to relate the states in the theory
expanded around $\langle \pi^\alpha \rangle = \theta^\alpha$ with
those of $\langle \pi^\alpha \rangle = 0$; but this is too much to
expect for non-Abelian $G$. Abstractly, we should expect that there
can be different ``paths" from $\langle \pi^\alpha \rangle = 0$ to
$\langle \pi^\alpha \rangle = \theta^\alpha$, and that there is some
$G$ rotation associated with each path, but that different paths can
differ up to a $G$ rotation; of course for infinitesimal paths at
quadratic order, this is how the $G$ commutation relations arise.
And yet we don't seem to have the required  notion of a ``path"
connecting two different vacua.

There is a simple physical reason behind these difficulties that
also points to their resolution. We have been pretending that we can
think of the states of one theory expanded around one vacuum as some
excitation around another vacuum. Of course we know this is formally
incorrect-- the charge operator $Q$ for broken symmetries does not
exist as it creates states whose norm diverges with the volume of
space. This may seem like a technicality: after all, it just
reflects the fact that $Q_\alpha |vac\rangle = |\pi^\alpha(q = 0)
\rangle$ creates a zero momentum pion. While we should in principle
carefully regulate zero momentum pions by giving them some tiny
momentum and take the momentum to zero, this seems physically
irrelevant: by turning on sufficiently long wavelength pion
excitations in one vacuum, we can simulate another one over an
arbitrarily large region of space. For instance, suppose we want to
give pions pointing in two different directions $\alpha, \beta$ a
tiny expectation value. We could probe what amplitudes in this
vacuum look like by considering amplitudes that contain in addition
to some hard pions of momenta $p_i$, the two soft pions
$\alpha,\beta$ with momenta $q_\alpha,q_\beta$ and eventually send
$q_{\alpha,\beta} \to 0$.

As we will see, however, while this double soft limit is indeed
non-singular, {\it it is ambiguous}: the amplitude depends on ratios
such as  $\frac{q_\alpha \cdot p_i}{q_\beta \cdot p_i}$. This is how
the natural notion of ``taking different paths from one vacuum to
another" is physically realized, by taking the double soft pion
limit in different ways. This observation invalidates our naive
paradoxes above: for instance, we will find that due to this
``anomalous" ambiguity in the soft limit,  the limit of $M^{\alpha
\beta \cdots}(q_\alpha,q_\beta)$ as $q_{\alpha,\beta} \to 0$ is in
fact completely {\it antisymmetric} in $(\alpha \beta)$, and nicely
encodes the $[X,X] \sim T$ commutation relations of $G$. We should
find that the amplitude in the double soft limit is the same as the
amplitude without the soft pions, up to the $H$ rotation on the hard
goldstones lines corresponding to the $[X^\alpha,X^\beta]$
generator; in fact, we will see that the hard goldstone states with
momenta $p_i$ are $H$ rotated in this way by an amount that depends
on $p_i$ as well as  $q_{\alpha,\beta}$.

It is best to see this at work in a concrete example, working at
tree level.  Let us take the most familiar case of pions arising
from $SU(2) \times SU(2) \to SU(2)$; we will actually think of them
more conveniently as being Goldstone bosons of $SO(4) \rightarrow
SO(3)$, and label them by a vector $SO(3)$ ``flavor" index $a$. The
four pion amplitude is
\begin{equation} M^{abcd} = s_{12} \delta^{ab} \delta^{cd} + s_{13}
\delta^{ac} \delta^{bd} + s_{14} \delta^{ad} \delta^{bc}
\end{equation} where we define in general $s_{i_1, \cdots, i_n} =
(p_{i_1} + \cdots + p_{i_n})^2$. Clearly the single and double soft
limits vanish here; this is consistent because the amplitude with
the two soft pions removed is a 2 point amplitude that vanishes
manifestly. In order to see a non-trivial double soft limit, we have
to start with a 6 point amplitude $M^{a_1 \cdots a_6}$. The are 15
different pairs that can be made from the 6 external legs and hence
15 different flavor $\delta$ contractions for the amplitude, one of
which is e.g. $\delta^{i_1 i_2} \delta^{i_3 i_4} \delta^{i_5 i_6}$,
which is the only one that contributes to the $M^{aabbcc}$ amplitude
for $a,b,c$ distinct. The six-point amplitude was first computed by
Frye and Susskind~\cite{Susskind:1970gf}, with result
\begin{eqnarray} M^{aabbcc}&=& s_{12} s_{34} \left(\frac{1}{s_{125}}
+ \frac{1}{s_{126}}\right) + s_{12} s_{56} \left(\frac{1}{s_{123}} +
\frac{1}{s_{124}}\right) + s_{34} s_{56}
\left(\frac{1}{s_{341}} + \frac{1}{s_{342}}\right) \nonumber \\
&-&(s_{12} + s_{34} + s_{56}) \end{eqnarray} It is easy to see that
in the limit as any one pion becomes soft the amplitude vanishes as
expected. We can then examine the limit when two pions become soft.
It is easy to see that if the pions with the same flavor index are
taken to be soft the amplitude vanishes. This is consistent with our
expectation, since the $H$ rotation in this case is $[X^a,X^a] = 0$.

Consider then the case where e.g. $p_1,p_3 \rightarrow 0$. Note that
the pre-factors of the first three terms vanish in this limit, so to
get a finite result we need to get a singularity from the
denominators; since $s_{abc} = s_{ab} + s_{bc} + s_{ac}$, to find a
singular term, we must have an $(abc)$ that contains both soft
momenta $1,3$. We find
\begin{equation} M \rightarrow s_{56} \left(\frac{s_{12}}{s_{12} +
s_{32}} + \frac{s_{34}}{s_{34} + s_{14}} - 1 \right) \end{equation}
Note that as advertised, this has a finite but ambiguous soft limit!
Furthermore, writing the $-1$ as $-1/2 - 1/2$ we can rewrite this as
\begin{equation} M \rightarrow \frac{1}{2} s_{56} \left(\frac{s_{12}
- s_{32}}{s_{12} + s_{32}} - \frac{s_{14} - s_{34}}{s_{14} + s_{34}}
\right)
\end{equation}
which is manifestly antisymmetric in $p_1,p_3$! Note that our
original partial amplitude was invariant only under under the
exchange of e.g. $(12) \leftrightarrow(34)$, with no special
properties for simply exchanging $1,3$. Of course, in the double
soft limit, the amplitude is still symmetric under $(12)
\leftrightarrow (34)$, but it is also separately antisymmetric under
$1 \leftrightarrow 3$ and $2 \leftrightarrow 4$.

\begin{figure}[h]
\begin{center} \label{FigPionDoubleSoft}
\includegraphics[width=15cm]{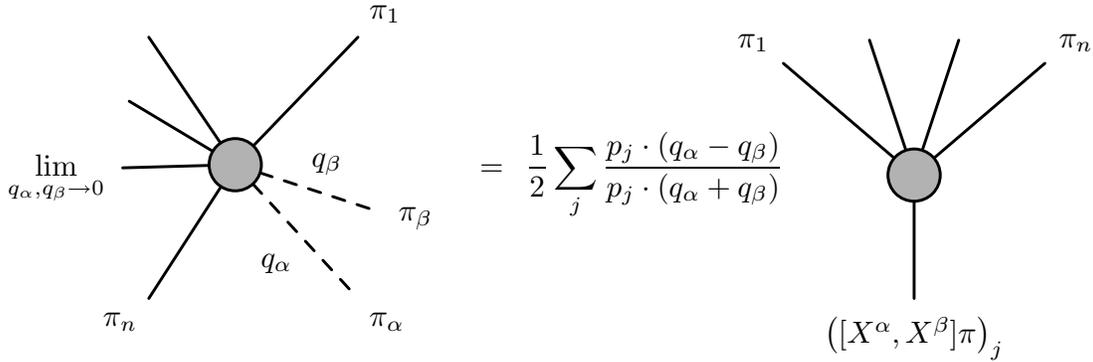}
\caption{\small{When we take the double soft limit of two pions, the
result is a momentum-dependent $H$ rotation on each of the remaining
hard states.}}
\end{center}
\end{figure}

To complete the interpretation of this double soft limit as reflecting the $[X,X] \sim T$ commutation relations of $G$, note that $s_{56}$ is nothing but the 4 point amplitude made out of the hard momenta $p_2,p_4,p_5,p_6$ for some choice of flavors:
\begin{equation}
M^{ddcc}(p_2,p_4,p_5,p_6) = s_{56}
\end{equation}
The action of $[X^a,X^b]$ on any vector index is \begin{equation} ([X^a,X^b]V)^d = V^a \delta^{bd} - V^b \delta^{da} \end{equation}
 Stripping off the soft particles 1,3 from $M^{aabbcc}$,  and acting on the flavor index of particle 2 with $[X^a,X^b]$
 we get \begin{equation} M^{bbcc} = s_{56} \end{equation} while acting on the flavor index of particle
 4 we get \begin{equation} -M^{aacc} = - s_{56} \end{equation}
 Acting on the flavor indices of $5,6$ gives zero. We conclude that the amplitude for
 emitting two soft pions of flavors $a,b$ and momenta $p_1,p_3$ is an $[X^a,X^b]$
 transformation of the amplitude with only hard pions, where a pion of momentum $p_i$
 is rotated by an amount proportional to $\theta_i$ given by \begin{equation}
 \theta_i = \frac{1}{2} \frac{(p_1 - p_3) \cdot p_i}{(p_1 + p_3) \cdot p_i} \end{equation}
 Note that this result is consistent with the vanishing of the
 amplitude as any one of the soft momentum $p_1,p_3$ are sent to zero: in this limit,
 $\theta_i \to \pm 1$, and so we have nothing but a common $H$ rotation on all the external particles,
 which vanishes by the $H$ invariance of the theory.

We can summarize what we have seen in the following way. There was something missing from our attempt to construct states in the different vacua by simply acting with $e^{i Q \theta}$; that is, the understanding that this operator does not really exist, and that we instead have to add soft pions and take the zero momentum limit carefully. This means that in expanding \begin{equation}
|\psi\rangle_ \theta = |\psi\rangle + |\psi^{(1)} \rangle +
|\psi^{(2)}\rangle + \cdots
\end{equation}
there is more ``data" labeling the variations $|\psi^{(n)}\rangle$,
associated with the directions of the $n$ soft momenta
$q_1,\cdots,q_n$; we will write explicitly \begin{equation} q_i = t
\hat{q}_i \end{equation} and take the soft limit as $t \to 0$. Lets
take the state $|\psi\rangle$ to consist of many hard pions
\begin{equation} |\psi \rangle = |\pi_{\sigma_1}(p_1) \cdots
\pi_{\sigma_n}(p_n) \rangle
\end{equation}
Then for the first order variation we have simply \begin{equation}
|\psi \rangle^{(1)} = {\rm lim}_{t \to 0}  |\psi + \pi_\alpha(t
q_\alpha) \rangle
\end{equation}
and the statement that single soft pion emission vanishes at zero momentum is
just \begin{equation} M^{\psi^{(1)}} = 0 \end{equation} But for the second order variation,
 our expression for the double soft pion emission can be interpreted as one of $G$ invariance
  \begin{equation} M^{\psi^{(2)}} = 0 \end{equation} if we take
  the second order variation to include the addition of two soft pions together with the momentum-dependent $[X^\alpha,X^\beta]$ rotation on the hard pions:
\begin{eqnarray}
|\psi\rangle^{(2)} &=& {\rm lim}_{t \to 0} \, |\psi + \pi_\alpha(t
\hat{q}_\alpha) + \pi_\beta(t \hat{q}_\beta) \rangle \nonumber \\
&+& \sum_j \frac{1}{2}\frac{(\hat{q}_\alpha - \hat{q}_\beta) \cdot
p_j}{(\hat{q}_\alpha + \hat{q}_\beta) \cdot p_j} \, \,
|\pi_{\sigma_1}(p_1) \cdots ([X^\alpha,X^\beta] \pi)_{\sigma_j}(p_j)
\cdots \pi_{\sigma_n}(p_n) \rangle
\end{eqnarray}
We have thus explicitly constructed the appropriately IR regulated
action of the non-linearly realized $G$ on the states of the theory,
up to the second order variation.

It should be clear that this is giving a very interesting
representation of $G$! While we have arrived at this result by a
direct inspection of the six-point amplitude, it must be possible to
derive it in generality using current algebra techniques. It would
also be very interesting to push this further and understand how the
finite transformation is properly built up, but we will stop our
discussion here and move on to see how this structure arises in
${\cal N} = 8$ SUGRA with $G=E_{7(7)}$ and $H= SU(8)$.

\subsection{Single Soft Emission in Gauge Theory and Gravity}

In order to see the $E_{7(7)}$ structure in ${\cal N} = 8$ SUGRA, we
need to study the soft limit for scalar emission. Unlike our
analysis of the pion case, where we relied on the explicit form of
the 6 point scattering amplitude to infer the behavior of the soft
limit, for ${\cal N} = 8$ the story is more systematic. Indeed, the
amplitude for soft scalar emission should be related by SUSY to that
of soft gluon/graviton emission, and these have a famous universal
form first given by Weinberg \cite{Weinberg:1964ew}. We begin by
recalling this form, as a warm-up to the problem we are interested
in.

Weinberg considered attaching a single soft particle to some
amplitude, working at tree level.  In terms of Feynman diagrams,
those diagrams where the soft particle of momentum $q$ is attached
to an external line of momentum $p_j$ have a $\frac{1}{p_j \cdot q}$
singularity from the extra nearly on-shell propagator, while those
diagrams where the soft particle attaches to off-shell intermediate
lines will have no such singularity. The soft limit is then
dominated by the singular behavior as $q \to 0$. The soft factor for
photons is
\begin{equation} \sum_j e_j \frac{p_{j\mu}\epsilon^\mu_j}{p_j \cdot
q} \end{equation} The factor is the same for YM, except for a given
color ordering there is only a single term in the sum, with the soft
gluon in the appropriate place according to its color. For gravity
the soft factor is
\begin{equation} \sum_j \frac{p_{j \mu} p_{j \nu}\epsilon^{\mu \nu}}{p_j \cdot q} \end{equation}
Using the explicit form of polarization vectors written in terms of
spinor helicities, we find the YM soft factor for $+$ polarization
is \begin{equation} \frac{\langle \mu j \rangle}{\langle \mu s
\rangle \langle s j \rangle} \end{equation} while for a graviton
with the same $+$ polarization we have \begin{equation} \sum_j
\left(\frac{\langle \mu j \rangle}{\langle \mu s \rangle} [sj]
\right)^2 \frac{1}{[sj] \langle s j \rangle} \end{equation} here the
soft momentum $q = \lambda_s \bar \lambda_s$, and $\mu$ is a
reference spinor entering in the definition of the polarization
tensors. For $-$ helicity particles, we simply interchange $\langle
\rangle$ and $[\,]$ brackets.

Can we re-derive this result using purely on-shell methods?
Obviously this is not important for gauge/gravity emission where we
have long known the answer, but apart from its intrinsic interest
this will shortly help us understand the behavior of soft scalar
emission in ${\cal N} = 8$ SUGRA. We are interested in isolating the
leading singularity as some momentum becomes soft; actually, for
complex momentum, we could do this by examining what happens when
either $\lambda_s$ or $\bar \lambda_s$ are made small, and we can
probe for each singularity separately.

\begin{figure}[h]
\begin{center} \label{FigSugraSoftLimit}
\includegraphics[width=7cm]{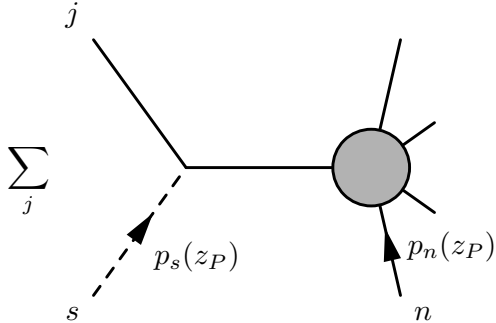}
\caption{\small{Here we consider the amplitude for emitting a single
soft graviton at tree level. The figure shows the only terms that
contribute in the soft limit when we compute the amplitude using the
BCFW Recursion Relations, analytically continuing the soft graviton
momentum and some other hard leg. Terms with more than one hard leg
associated with the soft graviton do not contribute directly because
they do not have a soft emission pole. They do not contribute
through an additional pole in the soft graviton sub-amplitude
because in such a sub-amplitude, $p_s$ is analytically continued to
a value $p_s(z_P)$ that does not vanish as $p_s \to 0$.}}
\end{center}
\end{figure}

Suppose we are considering the soft limit for an $(n+1)$ point amplitude with a $+$ helicity graviton. We can compute this amplitude using BCFW recursion, by choosing this soft particle, and one of the other hard particles, say particle $n$, as the ones undergoing BCFW deformation, $\lambda_2(z) = \lambda_s + z \lambda_n, \, \bar \lambda_n(z) = \bar \lambda_n - z \bar \lambda_s$. Now, of all the terms that appear in the recursion relation, there is a special one, where the soft particle and a single other hard particle $j$ appear together in a a three-point vertex. By the BCFW shift of $\lambda_s$, we make $\lambda_s(z),\lambda_j$ and the internal line $\lambda_I$ all proportional to put the three-point vertex on shell; we can always choose $\lambda_I = \lambda_j$ and using momentum conservation dotted into $\lambda_n$ to solve for $\bar \lambda_I$ as \begin{equation} \bar \lambda_I = - \frac{\langle n s \rangle}{\langle n j \rangle} \bar \lambda_s - \bar \lambda_j \end{equation} note that this of course goes to $-\bar \lambda_j$ in the limit as the soft momentum goes to zero. Taking $j$ also to have positive helicity (the soft limit result is the same for negative helicity) the three-point vertex is \begin{equation} \left(\frac{[s j]^3}{[sI][jI]}\right)^2 = \left(\frac{\langle n j \rangle}{\langle n s \rangle} [sj] \right)^2 \end{equation} As the momentum is taken to zero, we get a singularity from the BCFW propagator; summing over the allowed $j$'s (all the hard lines excepting the BCFW deformed line $n$) we find from these special diagrams \begin{equation} M_{n+1} \to \sum_{j \neq n} \left(\frac{\langle n j \rangle}{\langle n s \rangle} [sj] \right)^2 \times \frac{1}{[sj]\langle s j \rangle} \times M_n \end{equation} But it is easy to see that these are the only terms in the recursion relation for $M_{n+1}$ that have any singularity in the soft limit.
Any term with two or more lines with total momentum $P$ will have
$P^2 \neq 0$ and so combined together with $s$ will clearly not have a singular propagator factor. One might think that the presence of an amplitude with a soft line on one side might itself give a singularity in the soft limit, but recall that the amplitudes are evaluated at the BCFW shifted momentum, which is necessarily a hard momentum, since it must combine with the remaining momentum $P$ to become lightlike! Thus, there are no singularities in any of these terms in the recursion relation in the limit as $\lambda_s \to 0$; all the singularities are isolated in the special case we first discussed.

This form of the soft limit agrees perfectly with Weinberg's result;
if we take the reference spinor $\mu = \lambda_n$, the term with $j
= n$ in Weinberg's sum vanishes and we are left with our form of the
result. For the negative helicity soft graviton, repeat the argument
using instead the $\lambda_n(z),\bar \lambda_s(z)$ BCFW shift.

\subsection{Single Emission in ${\cal N} = 8$ SUGRA}

Turning to ${\cal N} = 8$ SUGRA, what is the soft limit for the
emission of any of the particles? Following the basic logic we have
been repeatedly using in this paper, our first instinct is to use
SUSY to relate the amplitude for emitting any soft particle to that
of a graviton, and indeed we can in principle do this. For instance,
we can send the $\eta$ of a soft particle $\eta_s \to 0$ by using
the SUSY translation $\zeta_\alpha = - \mu_\alpha \eta_s/\langle \mu
s \rangle$. The problem is that this transformation blows up in the
soft limit $\lambda_s \to 0$, so taking the soft limit is not
transparent.

Let us instead start by seeing what the soft limit of the
three-particle amplitude $123$ looks like. Say all the $\lambda$'s
are parallel so that the amplitude is a function of the $\bar
\lambda$'s. Suppose we keep $\lambda_{1,2,3}$ large while making
particle 3 soft by taking $\bar \lambda_3$ to zero. Concretely, let
us put $\lambda_1 = \lambda_2 = \lambda_3$; then using momentum
conservation we can solve for $\bar \lambda_2 = - \bar \lambda_1 -
\bar \lambda_3$ and
\begin{equation}
M_3 = [13]^2 \delta^8(\eta_1 - \eta_2) \delta^8(\eta_3 -\eta_2)
\end{equation}
which all vanish as $\bar \lambda_3 \to 0$. Now suppose we instead
keep $\bar \lambda_3$ large and make $3$ soft by sending $\lambda_3
\to 0$: $\lambda_1 = \lambda_2 = \lambda$, $\lambda_3 = \epsilon
\lambda$. Then, by momentum conservation, $\bar \lambda_2 = - \bar
\lambda_1 - \epsilon \bar \lambda_3$, and we find the 3 particle
amplitude to be
\begin{equation} M_3 = \frac{[13]^2}{\epsilon^2} \delta^8(\eta_1 -
\eta_2) \delta^8(\eta_3 - \epsilon \eta_2)
\end{equation}
We see from here that the soft limit $\epsilon \to 0$ of a positive helicity
graviton and gravitino is singular, that of a positive helicity graviphoton goes to
 a constant, and of the rest, including soft scalar emission, vanish. Making the opposite choice
 of $\bar \lambda$'s be parallel, the behavior will be conjugate to this one, but note that the
 3 point amplitude with a soft scalar vanishes in all cases.

It is very easy to see that this is a general result: the amplitude
for the emission of a single soft scalar vanishes in ${\cal N} = 8$
SUGRA.  Imagine computing the amplitude $M_{n+1}$ with the
additional soft scalar using BCFW, this time {\it not} deforming the
soft momentum. Then we can continue to recurse down to the
three-point amplitude involving the soft scalar, which vanishes.
Thus, just as is very familiar for pions, this vanishing amplitude
is the first-order indication of the moduli space of vacua.

It is interesting that the single emission of a graviphoton of the
appropriate helicity goes to a constant in the soft limit. This
hints at an enlarged symmetry which acting on the vacuum would
produce not a scalar but some graviphoton flux. Such a symmetry has
very recently been discussed by Berkovits and Maldacena
\cite{nathanjuan}, and it would be interesting to pursue this
possible connection further.

\subsection{Double Soft Emission and $E_{7(7)}$}

By analogy to our discussion for pions,  in order to see the
$E_{7(7)}$ group structure, we have to examine double scalar
emission. Fortunately, given the vanishing of the soft limit for the
emission of a single scalar, it is straightforward to understand the
double soft limit. Consider the amplitude $M_{n+2}$ with two soft
scalars. We would like to compute this via BCFW deforming one of the
soft legs $\lambda_{s_1}$; one might think it most natural to deform
the other soft particle as the second BCFW leg but this will not be
convenient for our purposes, since in a typical term of the
recursion relation the deformed $s_1,s_2$ will both be hard and we
can't exploit the known vanishing of the amplitude for single soft
emission easily. Thus we need to choose one of the hard lines as the
second BCFW line, but this unpleasantly breaks the symmetry between
all the hard particles. We are therefore motivated to use a useful
trick: we will add an additional negative helicity soft graviton $g$
to the amplitude, and use it as the second leg for the BCFW
procedure. This is nice because it allows us to treat all of the
hard particles symmetrically, and also because it means that we do
not need to analytically continue any of the $\eta$'s, for the
simple reason that $\eta_g = 0$.  We know that the addition of a
soft graviton simply multiplies the original amplitude by a
universal factor. Furthermore, because the soft scalar momenta are
going to zero, the graviton soft factor is precisely the same for
the amplitude with and without the two soft scalars. Therefore after
we perform the calculation we can simply strip off this factor to
obtain our result!

We therefore begin with the BCFW formula in the form
\begin{equation}
M_{n+3}(\eta_1,\ldots, \eta_{n+2}, g) = \sum_{L,R} \int d^8\eta
M_L(\{\eta_1,\lambda_1(z_P),\bar\lambda_1\},\eta,\eta_L)
\frac{1}{P^2}M_R(\{0,\lambda_g,\bar\lambda_g(z_P)\},\eta,\eta_R)
\end{equation}
As usual, $z_P$ is found by requiring $P(z_P)^2=0$,
$\lambda_1(z_P)=\lambda_1+z_P\lambda_g$,
$\bar\lambda_g(z_P)=\bar\lambda_g-z_P\bar\lambda_1$.  Note again
that although generically we would have $\eta_1(z_P)$, in this case
$\eta_1(z_P) = \eta_1 + z_P \eta_g = \eta_1$.  We will take the soft
limit $p_1, p_2 \to 0$ before taking $p_g \to 0$, so until the end
of the calculation we will view $g$ as a hard particle.

Now, there are terms in the recursion relation where $s_2$ is not on
the same side as $s_1$. If $s_1$ is together with more than one hard
momentum, then its deformation will make it hard and there is no
propagator singularity, while $s_2$ remains soft on the other side,
and since the single soft scalar amplitude vanishes, these terms
don't contribute to the soft limit. This is even true if $s_1$ is
together with a single hard line; while there is a propagator
singularity, this is overcompensated by the vanishing of the
3-vertex and again it vanishes. So we only have to consider cases
where $s_1,s_2$ are on the same side. But if they are together with
2 or more hard lines, once again under the deformation the $s_1$
becomes hard, and we are left with a single soft scalar amplitude
for $s_2$ which vanishes. We are therefore left with the two cases
where $s_1,s_2$ and a single hard line $j$ are joined to the
intermediate line via the four-point amplitude $M_4$, and where
$s_1,s_2$ join with a 3-point amplitude $M_3$.

\begin{figure}[h]
\begin{center} \label{FigSugraDoubleSoftLimit}
\includegraphics[width=13cm]{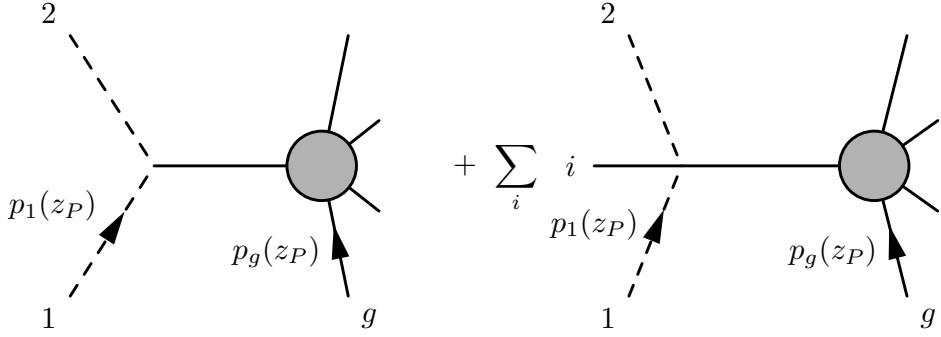}
\caption{\small{The double scalar soft limit in ${\cal N} = 8$
Supergravity can be obtained from the BCFW Recursion Relations.  In
the figure we display all of the terms that contribute to this limit
when we analytically continue one of the soft scalar momenta and the
auxiliary additional graviton. The term on the left only contributes when the
two scalars form an $SU(8)$ singlet, and in the soft limit, this
term gives rise to graviton emission.  The term on the right
contributes when the two scalars form an $SU(8)$ singlet and also
when their $\eta$'s share one index.  In the former case, this term
also leads to graviton emission, while in the latter case it leads
to a sum of terms giving a momentum dependent $SU(8)$ rotation on
each of the hard states.}}
\end{center}
\end{figure}

Note that the $M_3$ term can only arise in the case where the two
soft scalars can combine into an $SU(8)$ singlet; the intermediate
state in this case is a graviton. This contribution will have a soft
singularity which is nothing other than the one associated with the
emission of a single graviton. There is a similar singular
contribution when the two scalars form an $SU(8)$ singlet from
$M_4$. We will not consider these terms in detail here, since we are
interested in the $SU(8)$ rotation associated with $[X,X]$ which
vanishes when the $X$ indices combine into a singlet.

The interesting case that reveals the $E_{7(7)}$ structure is when
the two soft scalars cannot form a singlet (and therefore they
cannot produce a graviton). In this case we expect ${\cal N}=8$
SUGRA to mirror the story for pions. Labeling the state with hard
particles by $|\eta_1,\cdots,\eta_n\rangle$ we should find for the
second order variation under $E_{7(7)}$
\begin{eqnarray} \label{sugravacuumstructure}
|\{\eta_i\}\rangle^{(2)} &=& {\rm lim}_{t \to 0} \, |\{\eta_i\} +
\phi_{I_1 \cdots I_4}(t \hat{q}_1) + \phi_{I_5 \cdots I_8}(t
\hat{q}_2) \rangle \\ &+& \sum_j \frac{1}{2} \frac{(\hat{q}_1 -
\hat{q}_2) \cdot p_j}{(\hat{q_1} + \hat{q_2}) \cdot p_j}
R(\eta_j)_{(I_1,\cdots,I_4),(I_5,\cdots,I_8)} |\{\eta_i\}\rangle,
\nonumber
\end{eqnarray}
here $R$ is the differential operator
\begin{equation}
R(\eta)_{(I_1,\cdots,I_4),(I_5,\cdots,I_8)} =
\left(\left[X_{I_1,\cdots,I_4},X_{I_5,\cdots,I_8} \right] \eta
\right)_J \partial_{\eta_J}
\end{equation}
that generates the $SU(8)$ rotation on $\eta$ associated with the
$[X,X]$ commutator. This has exactly the same structure as for the
pion case: the addition of two soft scalars, together with a
momentum-dependent $SU(8)$ rotation. Just as for the pion case, with
this action of $E_{7(7)}$ on the states, the behavior of the double
soft scalar amplitude is equivalent to the statement of $E_{7(7)}$
invariance of the amplitudes \begin{equation} M^{\psi^{(2)}} = 0
\end{equation}

Let us see how the structure of equation
(\ref{sugravacuumstructure}) arises explicitly in the amplitudes of
${\cal N}=8$ SUGRA. Let particles $1$ and $2$ be the two scalars
that become soft. The corresponding $SU(8)$ indices are
$\phi_{1}^{abcd}$ and $\phi_{2;efgh}$; note that we have chosen to
express the indices in the second scalar as antifundamental indices.
In the coherent state representation it is convenient to introduce
the notation $\eta_1^{(abcd)}$ and $\bar\eta_{2;(efgh)}$ to denote
$\eta_1^a\eta_1^b\eta_1^c\eta_1^d$ and
$(\bar\eta_2)_e(\bar\eta_2)_f(\bar\eta_2)_g(\bar\eta_2)_h$
respectively. Then we consider the amplitude
\begin{equation}
\int d^8\eta_1 \eta_1^{(abcd)}\int d^8\bar\eta_2 \bar\eta_{2;(efgh)}
M_{n+3}(\eta_1,\bar\eta_2,\eta_3,\ldots, \eta_{n+2}, \eta_g = 0)
\end{equation}
We will analyze the behavior of this amplitude as $p_1, p_2$ vanish
by taking the $\lambda$'s and $\bar \lambda$'s of the soft particles
to be proportional to a parameter $t$ and look at $t \rightarrow 0$
limit. Recall that the BCFW deformation is for the soft line $1$ and
the auxilliary graviton $g$. As we discussed above, when we BCFW
deform the particles $1$ and $g$, the only terms that contribute in
the limit $p_1, p_2 \to 0$ in the case where $1$,$2$ do not form an
$SU(8)$ singlet, are those with particles $1$ and $2$ with a hard
particle $i$.  A single term from the sum will take the form
\begin{equation}
\label{doso} \int d^8\eta
M_4(\{\eta_1,\lambda_1(z_P),\bar\lambda_1\},\eta,\eta_2,\bar\eta_i)
\frac{1}{P^2_{12i}}
M_{n+1}(\{0,\lambda_g,\bar\lambda_g(z_P)\},\ldots)
\end{equation}
Note that in this case, $z_P$ is of order $t$ and so all of
$\lambda_1(z_P)$ is of order $t$ just as $\lambda_1$ is.

Now let us use the explicit representation of the four particle
amplitude from equation (\ref{M4}).
%
%
We will represent the spinor for the intermediate particle with
$\hat P$, and we will choose to represent the hard particle $i$
using the $\bar \eta$ representation. Then the four particle
amplitude with two scalar emission is

%
\begin{eqnarray}
\label{sami} && M^{(abcd)}_{4 (efgh)}(\eta,\bar \eta_i) =
\frac{(\langle \hat 1 ~ \hat P \rangle [2~i])^4}{{\rm s_{\hat 1,
\hat P} \, s_{\hat 1, 2} \, s_{\hat 1,i}}} \times \nonumber \\ & &
\int d^8\eta_1\eta_1^{(abcd)}\int
d^8\bar\eta_2\bar\eta_{2;(efgh)}\rm{exp}\left[\left(\begin{array}{cc}\eta_1
& \eta
\end{array} \right) \left(\begin{array}{cc} \frac{\langle \hat P 2 \rangle}{\langle 1 \hat P\rangle} &
\frac{\langle \hat P i \rangle}{\langle 1 \hat P \rangle} \\
\frac{\langle 2 1 \rangle}{\langle 1 \hat P \rangle} & \frac{\langle
i 1 \rangle}{\langle 1 \hat P \rangle}
\end{array} \right) \left(\begin{array}{c} \bar \eta_2 \\ \bar
\eta_i \end{array} \right) \right]
\end{eqnarray}
where $s_{ij}=(p_i+p_j)^2$ and $\hat p_1 = p_1(z_P)$.

The matrix in the exponential has a very interesting structure. In
the double soft limit $t\to 0$, the off-diagonal terms are of order
$t$ while the diagonal components are of order $t^0$. Since the only
source of divergences is the propagator $1/P_{12i}^2$ which scales
as $1/t^2$, any terms in the integral which are of order $t^3$ or
higher are irrelevant.

Expanding out the exponential, the terms that depend on $\eta_1$ and
$\bar\eta_2$ are of the form
\begin{equation}
\label{m1m2} \left(\frac{\langle \hat P 2 \rangle}{\langle \hat 1
\hat P\rangle}\eta_1\bar\eta_2\right)^{m_1}\left(\frac{\langle 2
\hat 1 \rangle}{\langle \hat 1 \hat P
\rangle}\eta\bar\eta_2\right)^{m_2} \left(\frac{\langle \hat P
i\rangle}{\langle \hat 1 \hat P
\rangle}\eta_1\bar\eta_i\right)^{m_3}.
\end{equation}
These terms are of order $t^{m_2+m_3}$ and their contribution
to~(\ref{doso}) is
\begin{eqnarray}
\label{expi} && \int d^8\eta \int d^8\eta_1 \;\eta_1^{(abdc)}\int
d^8\bar\eta_2\;\bar\eta_{2;(efgh)}\left(\prod_{i=1}^{m_1}\eta_1^{I_i}\bar\eta_{2;I_i}\right)\left(\eta^J\bar\eta_{2;J}\right)^{m_2}
\left(\eta_1^K\bar\eta_{i;K}\right)^{m_3}\times \\
&& {\rm exp}\left(\frac{\langle i\hat 1\rangle}{\langle \hat 1\hat
P\rangle}\eta\bar\eta_i\right)
M_{n+1}(\{0,\lambda_g,\bar\lambda_g(z_P)\},\eta,\ldots). \nonumber
\end{eqnarray}
where we have stripped out a factor that contains the four-graviton
amplitude in~(\ref{sami}) and the momentum factors
from~(\ref{m1m2}).

Note that the integrals over the $\eta_1$ and $\bar\eta_2$ Grassmann
variables can be non-zero only if $m_1+m_2=4$ and $m_1+m_3=4$. Now,
if $(abcd)$ and $(efgh)$ are the same, i.e., the scalars can form an
$SU(8)$ singlet then $(m_1,m_2,m_3)=(4,0,0)$ contributes and it is
of order $t^0$. in this case, the divergence from the propagator is
not canceled and reflects the presence of the soft graviton factor
mentioned above. Consider now the case when $(abcd)$ and $(efgh)$
differ in two or more indices. This means that $m_1 \leq 2$.
Therefore $m_2+m_3\geq 4$ which means that these terms vanish in the
$t\to 0$ limit. Therefore, the only case of interest is when
$(abcd)$ and $(efgh)$ differ in exactly one index. Therefore
$(m_1,m_2,m_3)=(3,1,1)$. In this case the contribution is non zero
and finite as expected.

The argument of the exponential in eqn.(\ref{expi}) contains a
factor $\langle i\hat 1\rangle/\langle \hat 1\hat P\rangle$ which is
equal to $1+{\cal O}(t)$. Since the rest of the integral is finite,
the ${\cal O}(t)$ terms can be dropped giving
\begin{equation}
\int d^8\eta \int d^8\eta_1 \;\eta_1^{(abdc)}\int
d^8\bar\eta_2\;\bar\eta_{2;(efgh)}\left(\prod_{i=1}^3\eta_1^{I_i}\bar\eta_{2;I_i}\right)\left(\eta^J\bar\eta_{2;J}\right)
\left(\eta_1^K\bar\eta_{i;K}\right) {\rm
exp}\left(\eta\bar\eta_i\right) M_{n+1}(\eta,\ldots).
\end{equation}
We note that $\eta^J \bar{\eta}_{i;K}$ exp($\eta \bar \eta_i$) =
$\bar \eta_{i;K} \partial_{\bar \eta_{i;J}}$ exp$(\eta \bar
\eta_i)$, so upon doing the $\eta$ integral we get simply
\begin{equation}
\int d^8\eta_1 \;\eta_1^{(abdc)}\int
d^8\bar\eta_2\;\bar\eta_{2;(efgh)}\left(\prod_{i=1}^3\eta_1^{I_i}\bar\eta_{2;I_i}\right)
\left(\eta_1^K\bar\eta_{i;K}\right)\left(\bar\eta_{2;J}
\partial_{\bar\eta_{i:J}}\right)
M_{n+1}(\bar\eta_i,\ldots)
\end{equation}
where the effect of the $\eta$ integration is to switch from the
$\eta$ to the $\bar\eta$ representation, i.e., from $M_{n+1}(\eta,
\ldots)$ to $M_{n+1}(\bar\eta_i,\ldots)$.

Further rewriting this we get
\begin{equation}
\int d^8\eta_1 \;\eta_1^{(abdc)}\int
d^8\bar\eta_2\;\bar\eta_{2;(efgh)}\left(\prod_{i=1}^3\eta_1^{I_i}\bar\eta_{2;I_i}\right)
\left(\eta_1^K\bar\eta_{2;J}\right)\times \left(\bar\eta_{i;K}
\partial_{\bar\eta_{i:J}}\right)
M_{n+1}(\bar\eta_i,\ldots)
\end{equation}
Performing the integrals over $\eta_1$ and $\bar\eta_2$ we get
\begin{equation}
\epsilon^{abcdI_1I_2I_3K}\epsilon_{efghI_1I_2I_3J}\times
\left(\bar\eta_{i;K}
\partial_{\bar\eta_{i:J}}\right)
M_{n+1}(\bar\eta_i,\ldots)
\end{equation}
where there is a sum for each repeated index. This formulas has to
be multiplied by all the momentum factors we left out, i.e.,
\begin{equation}
\frac{(\langle \hat 1 ~ \hat P \rangle [2~i])^4}{{\rm s_{\hat 1,
\hat P} \, s_{\hat 1, 2} \, s_{\hat 1,i}}}\times\left(\frac{\langle
\hat P 2 \rangle}{\langle \hat 1 \hat
P\rangle}\right)^{3}\frac{\langle 2 \hat 1 \rangle}{\langle \hat 1
\hat P \rangle} \frac{\langle \hat P i\rangle}{\langle \hat 1 \hat P
\rangle}\times\frac{1}{p_i\cdot (p_1+p_2)}
\end{equation}
where the last factor comes from the propagators $1/P^2_{12i}$ which
in the soft limit can be written as $\frac{1}{2 p_i\cdot
(p_1+p_2)}$. After a simple computation and recalling that $\hat 1$
means $\lambda_1(z_P)$ one finds
\begin{equation}
\frac{p_i\cdot p_2}{p_i\cdot (p_1+p_2)}
\epsilon^{abcdI_1I_2I_3K}\epsilon_{efghI_1I_2I_3J}\times
\left(\bar\eta_{i;K}
\partial_{\bar\eta_{i:J}}\right)
M_{n+1}(\bar\eta_i,\ldots).
\end{equation}
Since there is a sum over repeated indices one can replace the
product of the two $\epsilon$ tensors by
\begin{equation}
\frac{p_i\cdot p_2}{p_i\cdot (p_1+p_2)}
\epsilon^{abcdK}_{efghJ}\times \left(\bar\eta_{i;K}
\partial_{\bar\eta_{i:J}}\right)
M_{n+1}(\bar\eta_i,\ldots)
\end{equation}
where $\epsilon^{abcdK}_{efghJ}$ vanishes if the set on top is not
the same as the one on the bottom and it gives $\pm 1$ depending on
the order of the permutation needed to bring one set to the other.

Let us define the operator
\begin{equation}
R(\bar\eta_i) = \epsilon^{abcdK}_{efghJ}\times \left(\bar\eta_{i;K}
\partial_{\bar\eta_{i:J}}\right).
\end{equation}
This operator is the one that implements an infinitesimal $SU(8)$
rotation on the $i^{\rm th}$ particle associated with
$[X_{(abcd)},X^{(efgh)}]$, which is non-vanishing only in the case at
hand, where $(abcd)$ and $(efgh)$ share a single common index.

We are now ready to add up all the contributions from the BCFW
recursion relation,
\begin{equation}
M_{n+3}(1,2,\ldots, \{0,\lambda_g,\bar\lambda_g\})\longrightarrow
\sum_{i=3}^{n+2} \frac{p_i\cdot p_2}{p_i\cdot (p_1+p_2)}
R(\bar\eta_i)M_{n+1}(\bar\eta_i,\ldots,
\{0,\lambda_g,\bar\lambda_g\})
\end{equation}
where on the right hand side we used that in the soft limit
$\bar\lambda_g(z_P)$ can be replaced by $\bar\lambda_g$ up to order
$t$ terms which do not contribute.

Both amplitudes depend on the auxiliary graviton
$\{0,\lambda_g,\bar\lambda_g\}$ we added in by hand. Using that the
soft graviton factor is universal and that the contribution from the
scalars vanish we conclude that the same factor is produced on both
sides of the equation and hence it can be removed leaving
\begin{equation}
\label{sima} M_{n+2}(1,2,\ldots)\longrightarrow \sum_{i=3}^{n+2}
\frac{p_i\cdot p_2}{p_i\cdot (p_1+p_2)}
R(\bar\eta_i)M_{n}(\bar\eta_i,\ldots).
\end{equation}

Finally, note that by simple rewriting $p_2 = \frac{1}{2}(p_2+p_1) -
\frac{1}{2}(p_2-p_1)$ the double soft limit (\ref{sima}) becomes
\begin{equation}
M_{n+2}(1,2,\ldots)\longrightarrow \sum_{i=3}^{n+2} \frac{1}{2}
\frac{p_i\cdot (p_2-p_1)}{p_i\cdot (p_1+p_2)}
R(\bar\eta_i)M_{n}(\bar\eta_i,\ldots) + \frac{1}{2}\sum_{i=3}^{n+2}
R(\bar\eta_i)M_{n}(\bar\eta_i,\ldots).
\end{equation}
The second term is nothing but a global $SU(8)$ rotation, where are
states are rotated by the {\it same} $SU(8)$ operator, which
vanishes by $SU(8)$ invariance,  leaving the result we wanted to
prove
\begin{equation}
M_{n+2}(1,2,\ldots)\longrightarrow \sum_{i=3}^{n+2} \frac{1}{2}
\frac{p_i\cdot (p_2-p_1)}{p_i\cdot (p_1+p_2)}
R(\bar\eta_i)M_{n}(\bar\eta_i,\ldots)
\end{equation}

\subsection{Discussion}

Before concluding this section let us point out precisely where the
difference with ${\cal N}=4$ SYM lies in this discussion.  In SYM the
3-point amplitude vanishes with the momentum of the soft scalar scalar, but it does so more weakly than in ${\cal N} = 8$.  This means that the leading single soft emission behavior in ${\cal N}=4$ SYM is a constant, because the vanishing 3-point vertex is canceled by a vanishing propagator.  This cancelation of the Adler zero is a well-known phenomenon in theories with spontaneously broken global symmetries that have 3-pt vertices involving scalars.  What was special in ${\cal N}=8$ is that the zero is of second order in the soft momentum, so it was not canceled by the propagator.

The 4-point amplitude of ${\cal N}=4$ SYM can be related to the 4-point gluon amplitude by SUSY. In ${\cal N}=4$ SYM, however, we have to consider color-ordered amplitudes, and each partial amplitude has to be taken into account independently of the others. For the color ordering where the two
scalars are not color adjacent, this 4-gluon amplitude pre-factor is
\begin{equation}
\frac{\langle \hat 1\hat P\rangle^3}{\langle \hat P 2\rangle\langle
2 i \rangle\langle i \hat 1\rangle} \sim t^0
\end{equation}
which is finite in the double soft limit just as in the case of
gravity. However, consider now the case of the color-ordering where
the two scalars are color adjacent; the same factor becomes
\begin{equation}
\frac{\langle \hat 1\hat P\rangle^4}{\langle \hat P 2\rangle\langle
2 \hat 1 \rangle\langle \hat 1 i\rangle\langle i\hat P\rangle} \sim
1/t
\end{equation}
which diverges as $t\to 0$. This divergence reflects the fact that
for this color ordering of scalars, the remaining two lines in $M_4$
are becoming massive; the amplitudes with massless external states
will therefore necessarily be singular. Therefore the double soft
limit is telling us that the origin of the moduli space is special
and there is no massless S-matrix away from it. The fact that some
color-orderings give a finite answer while others are divergent show
that once again the presence of the color structure obstructs the
softer behavior that is seen in gravity amplitudes.

We have clearly only begun the exploration of the implications of
$E_{7(7)}$ for the scattering amplitudes of ${\cal N}=8$. Just as
for the story of pions, it is important to understand how to
exponentiate the state variations to the full finite group action on
the Hilbert space; it would also be illuminating to make precise the
connection between the differing ways of taking the soft pion limit
and different paths connecting vacua in the moduli space. More
specifically for ${\cal N} = 8$, we have already seen hints for more
simplicity and structure in tree-level scattering amplitudes,
following from the fact that they vanish as $1/z^2$ rather than
$1/z$ at infinity, and it would be interesting to see if this is
connected to the presence of $E_{7(7)}$ in any direct way. Another
interesting question is related to the KLT-type ``Gravity =
Gauge$\times$Gauge" relationship between ${\cal N} = 8$ and ${\cal
N} = 4$ amplitudes. The ``Gauge $\times$ Gauge" part only has a
manifest $SU(4) \times SU(4)$ invariance, while the ``Gravity" side
should have at least an $SU(8)$ invariance; the way this arises has
recently been studied by \cite{Dan}. But while the $SU(4)$
associated with ${\cal N} = 4$ is ``rigid", in the sense that is
well-defined and unbroken only at the origin of moduli space, there
is a different $SU(8)$ embedded in different ways inside $E_{7(7)}$
at different points in moduli space. This should be reflected in the
nature of the ``Gravity = Gauge $\times$ Gauge" relationship in an
interesting way. It is worth mentioning in this connection an
inspiring recent clarification of the KLT relations by Bern et. al.
\cite{ZviKLT},which relies on extra redundancy in the description of
the scattering amplitude. Finally, while we have focused here on the
case of soft scalar emission, as we already mentioned, it is
intriguing that the amplitude for emission of a single soft
graviphoton of appropriate helicity can go to a constant, instead of
diverging as for gravitons and vanishing as for scalars. It would be
interesting to see if this is related to any enlarged symmetry of
the theory.

\section{Structure of One Loop Amplitudes in any QFT}

We now turn to an analysis of ${\cal N} = 4$ SYM and ${\cal N} = 8$
SUGRA at 1-loop. As pre-amble, we will discuss the general structure
of 1-loop amplitudes in {\it any} QFT, and its relation to tree
amplitudes. This discussion has very significant overlap with the
beautiful recent work of Forde~\cite{Forde} and
others~\cite{forde:extensions} (see also~\cite{OPP} and
\cite{Mastrolia1}); we arrived at these constructions independently
and our perspective is therefore somewhat different.

Beyond tree-level, scattering amplitudes possess singularities which
can have branch cuts in addition to poles. Unitarity imposes strong
constraints on what the discontinuity across the singularities must
be in terms of lower loop and/or lower point on-shell amplitudes.
The textbook use of unitarity relates double cuts of 1-loop
amplitudes, which are given by the product of two tree amplitudes
integrated over a Lorentz invariant phase space, to the imaginary
part of the loop amplitude. This in turn is the discontinuity across
a branch cut once the amplitude is considered an analytic function
of the kinematical invariants. A notion of generalized unitarity,
where more propagators are cut, was developed in the 60's in order
to deal with more intricate situations where singularities could be
located away from the physical region~\cite{Smatrix}. In the
original literature most discussions were carried out for massive
particles (at the time QCD had not been established as the theory of
strong interactions and photons were considered peculiar objects!).
More recently, the power of generalized unitarity for massless
particles has been employed forcefully by Bern, Dixon and Kosower
under the name of the ``unitarity based method"~\cite{BDK}. In a
nutshell, the idea is to start from an ansatz using a basis of
simple scalar integrals, in terms of which any amplitude can be
expanded with coefficients which are rational functions of the
external kinematical invariants. The appropriate basis can be
determined using techniques like the Passarino-Veltman reduction
procedure~\cite{reduction}. The coefficients are determined by using
unitarity to write the integrand of cut amplitudes as tree
amplitudes and then expand the integrand to turn it into an
expression that resembles the cut of the basis of integrals.
Sophisticated analysis to ensure that double counting does not
happen, to exclude pieces that integrate to zero and to incorporate
dimensional regularization in $D=4-2\epsilon$ are all part of the
impressive machinery.

Here we will follow a similar philosophy, but will emphasize that
the phase space integrals themselves turn into contour integrals
naturally. These contour integrals give rise to a canonical
decomposition where pieces that integrate to zero never appear.
Another byproduct of our analysis is that recent techniques which
rely on complexifying momenta, like quadruple cuts~\cite{quadruple}
and BCFW~\cite{BCFW}, come out from completely standard (real
momentum) phase space integrals.

\subsection{Brief Review of Unitarity and Discontinuity Arguments}

At 1-loop it is known that any amplitude can be written as a linear
combination of scalar integrals. As mentioned above, this can be
achieved using reduction procedures~\cite{reduction} (in appendix F
we review on such procedure, and define the corresponding integrals
in four dimensions in appendix C). Near $d$ (integer) dimensions, we
need to include up to $d-$ gon scalar integrals with $d$
propagators--a ``box" in four dimensions, a ``pentagon" in five
dimensions and so on. Schematically, we write
\begin{equation}
\label{unity}
M^{\rm 1\hbox{-}loop} = \sum_i C_d^i I^i_d + \sum_j C_{d-1}^j I_{d-1}^j+ \cdots + \sum_k
C_2^k I_2^k+R
\end{equation}
where $I_m$ denotes a scalar integral with $m$ propagators that will
be defined in dimensional regularization, while $C's$ and $R$ are
rational functions of the kinematical invariants which are kept in
$d$ integer dimensions.

It is not entirely obvious that all 1-loop amplitudes can be
decomposed in this way. For instance, the Feynman diagrams
contributing to a 1-loop amplitude with 100 external particles in
four dimensions will have 100 propagators, so how are they reduced
to scalar integrals with at most 4 propagators? Loosely speaking,
the reduction procedures proceed by using partial fractions to break
apart the denominator factors, using the fact that with enough
external legs, the loop momentum can be expanded in a basis made
from some a subset of external momenta. This is a technical way of
describing the decomposition eqn.(\ref{unity}); one may wonder
whether there is any physical meaning to using this basis of scalar
integrals. As we will mention, there is a simple reason that 1-loop
amplitudes can be written in this way that makes it clear that this
decomposition has real physical significance.

The left hand side of (\ref{unity}) is defined by the sum over all
1-loop Feynman diagrams determined by the lagrangian of the theory.
Using dimensional regularization to render the integrals well
defined, it is clear that the amplitude is an analytic function of
the kinematical invariants. The right hand side of (\ref{unity}) is
obtained from Feynman diagrams by reduction procedures which give
rise to an alternative representation of the same analytic function.
Since the two sides of the equation represent the same analytic
function (at least in the physical sheet) they must have the same
analytic continuation. This means that when continued as analytic
functions of the kinematical invariants, which are allowed to be
complex variables, the two functions must have the same branch cuts
and the same discontinuities across them. The unitarity based method
uses that the discontinuities across given cuts on both sides are
the same to get information about the coefficients $C_i$. Of course,
for this method to be useful one has to have a systematic way of
computing discontinuities. As is well known, this method was first
provided by Cutkosky \cite{Cutkosky} and fully developed in
\cite{Smatrix}.

Consider all integrals with measure $d^d\ell$ and split it as
$d^{d-1}\vec{\ell}d\ell^0$. The $\ell^0$ integral is then taken to
be a contour integral along the real axis in the complex $\ell^0$
plane. The Feynman $i\epsilon$ prescription for propagators splits
all poles off the real axis sending half of them to the upper half
plane and the other half to the lower half plane. Deforming the
contour integral in the lower half plane gives rise to a sum of
contour integrals enclosing each of the poles. The residues of each
of these poles (taking $\epsilon$ to be infinitesimal) can also be
computed by taking the original integral and replacing the
propagator, which contains the pole in consideration, say,
$1/(\ell^2+i\epsilon)$, by $\delta^+(\ell^2)$, where the $(+)$
indicates choosing the solution which has $\ell^0>0$ (these are the
ones that lie on the lower half plane). This replacement is what is
called cutting a propagator. Now we are left with a sum over
``single cut" terms. Using the delta functions to perform the
$\ell^0$ integral one finds integrals over $d^{d-1}\vec{\ell}$. The
radial part of these integrals is of the form
\begin{equation}
\label{conti}
I = \int_0^\infty dE_{\vec{\ell}}~F(E_{\vec{\ell}}, \theta_i)
\end{equation}
where $\theta_i$ are the angular variables. The function $F$ has
poles (the remaining propagators) in the complex $E_{\vec{\ell}}$
space, whose location is a function of the external kinematical
invariants (see figure 7). If we choose the kinematical invariants
to vary in such a way that a single pole moves in the complex
$E_{\vec{\ell}}$ plane, travels in a circle around the origin and
back to its original position, one finds that the pole drags the
contour of integration of (\ref{conti}) with it. In order to get
something related to the original integral one has to force the
contour back into its original position but there is a price; as the
contour goes back it leaves behind a contour enclosing the pole.
This means that under the operation we performed (a monodromy), the
integral $I$ transforms as
\begin{equation}
I \longrightarrow I + \int_{C} dE_{\vec{\ell}}~F(E_{\vec{\ell}},
\theta_i)
\end{equation}
where $C$ is a small circle around the pole, see figure 7. This
means that $I$ is not a singled valued function and has branch cuts.
The particular branch cut under consideration has as branch point
the value of the kinematical invariants that make the pole coincide
with the origin $E_{\vec{\ell}}=0$. The discontinuity across the
branch cut is given by this contour integral.

\begin{figure}[h]
\begin{center} \label{fig:Loopy}
\includegraphics[width=16.5cm]{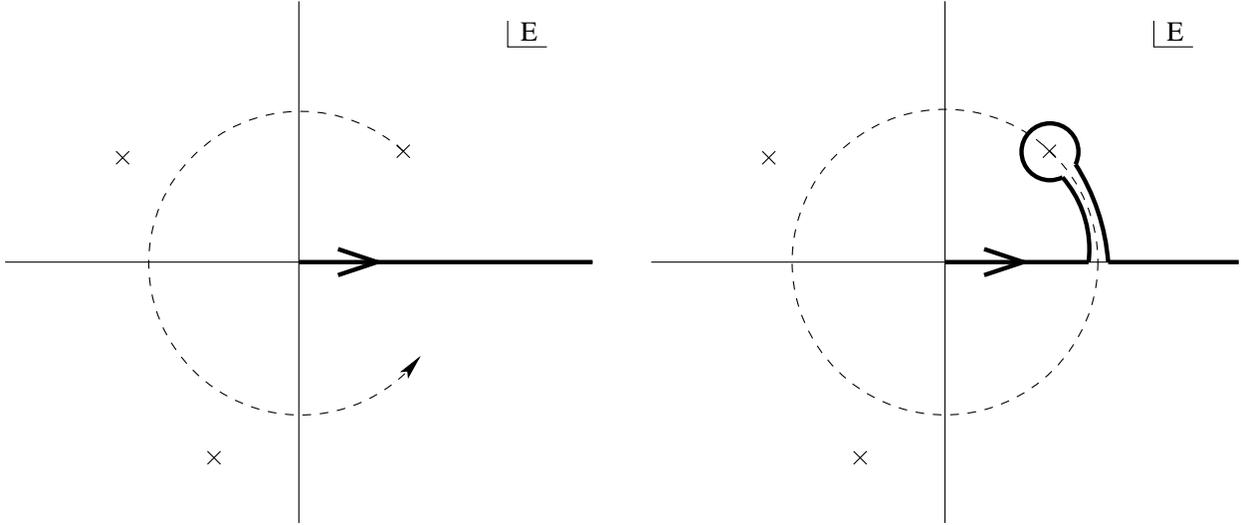}
\caption{\small{On the left, complex $E_{\vec{\ell}}$ plane with the
contour integral along the positive real axis and a pole moving in a
circle around the origin.
On the right, the result after moving the pole back into its original
position. The contour is dragged by the pole which indicates the
presence of a branch cut. At the end, there is an additional contour
integral around the pole which is the discontinuity across the branch
cut.}}
\end{center}
\end{figure}

Once again, the contour integral that computes the residue can be
computed by cutting the corresponding propagator in the original
integral if the pole is infinitesimally close to the real positive
axis. In other words, a second propagator has been cut. This means
that this discontinuity has the usual meaning of two physical
tree-level amplitudes; one emitting two on-shell particles with
positive energy that become the in-states of the other. The leftover
integrations make up the Lorentz invariant phase space integral of
two on-shell particles satisfying a momentum conservation condition.
If, on the other hand, the pole is not located close to the real
positive axis the discontinuity of the integral, which is still
computed by the residue on the pole, does not have such a physical
meaning.

Consider for example the simple bubble integral in four dimensions
\begin{equation}
I(P^2) = \int d^4\ell
\frac{1}{(\ell^2+i\epsilon)((\ell-P)^2+i\epsilon)}.
\end{equation}
Following the steps outlined above one finds that the pole in the
$E_{\vec{\ell}}$ complex plane is located close to the real positive
axis only if $P^2>0$ and hence the discontinuity (which is also in
this case the imaginary part) of the integral is computed by a
double cut integral
\begin{equation}
\Delta I(P^2) = \int d^4\ell \delta^+(\ell^2)\delta^+((\ell-P)^2).
\end{equation}
This is the same as the residue of the pole obtained by taking
$P^2\to {\rm exp}(2\pi i)P^2$ in a circle around the origin and
picking up the contour in figure 7. If $P^2<0$ then the integral is
real on the principal branch. On a different sheet, which is
reached, e.g., by taking $P^2\to {\rm exp}(2\pi i)P^2$, the integral
has an imaginary part coming from the pole contribution. So, there
is a discontinuity for both signs of $P^2$, but only for $P^2 > 0$
can it be expressed as the double-cut phase space integral. Note
also that to explicitly see the contour integrals associated with a
discontinuity we chose a frame, performed the $l_0$ integration,
deformed the $E_l$ contour an so on, but of course the final result
is Lorentz-invariant; an advantage of being able to intepret the
discontinuity as a cut integral is that this Lorentz invariance is
obvious from the outset. We can then compute the discontinuity for
$P^2 > 0$ in this convenient way, and continue the result to $P^2 <
0$.

\begin{figure}[h]
\begin{center} \label{fig:Loopy}
\includegraphics[width=16.5cm]{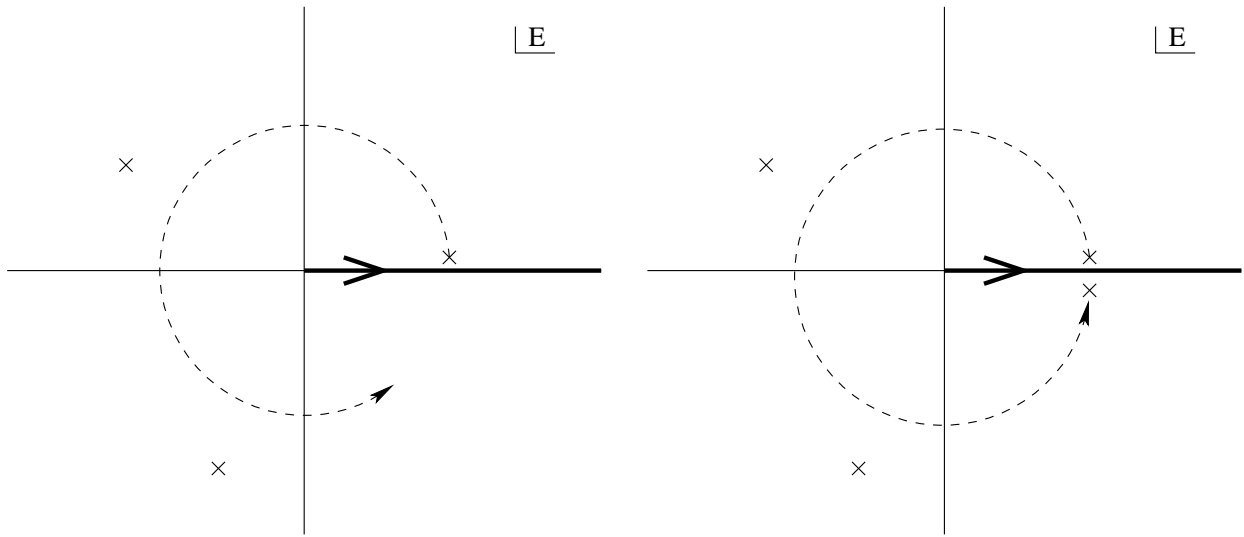}
\caption{\small{In cases when the pole is infinitesimally close to
the real axis then after almost completing the rotations about the
origin, it is easy to see that the computation of the residue is the
same as taking $1/((\ell-P)^2+i\epsilon) - 1/((\ell-P)^2-i\epsilon)$
which in the limit as $\epsilon\to 0$ becomes
$\delta((\ell-P)^2)$.}}
\end{center}
\end{figure}

Thus, for kinematical invariants in an appropriate range, the
discontinuity across branch cuts (of lowest codimension in the space
of kinematical invariants) is computed by cutting two propagators,
i.e., a double cut. In general the remaining angular integrals also
possess branch cuts (of higher codimension). One can continue the
study of monodromies by making poles travel in a path that goes back
to their original location but crosses the contour of integration.
This leads to triple cuts, etc. At some point, the location of the
poles will not be infinitesimally close to the real axis for
kinematical invariants in the physical region. This means that it is
not possible to think about them as taking the original integral and
replacing propagators by delta functions, i.e. as cutting
propagators. As we will see shortly, this is what happens for the
quadruple cut in four dimensions, where the four delta functions
localize the loop integral to a complex value. In the 60's this
problem was addressed by analytically continuing masses to force all
singularities into the physical region. In massless theories this
does not work and the singularities associated to ``would-be"
quadruple cuts do not have an interpretation for real momenta. We
will however shortly see how they arise by studying real triple
cuts, but we need first to establish our conventions.

\begin{figure}[h]
\begin{center} \label{FigGeneralCuts}
\includegraphics[width=8cm]{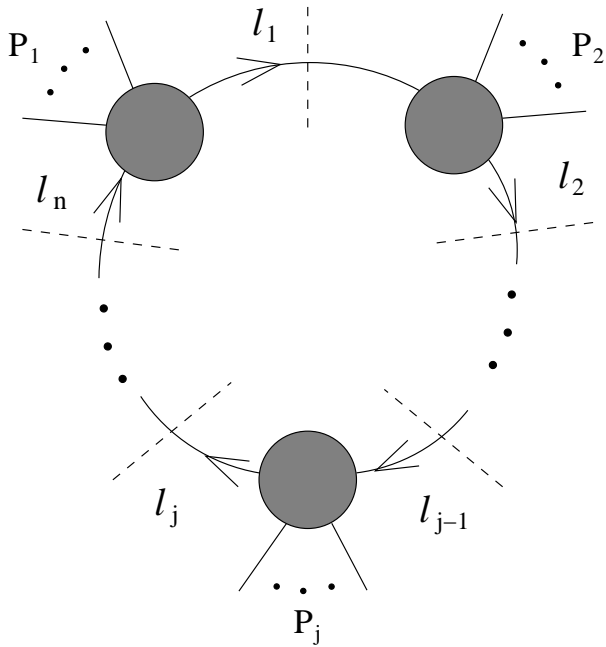}
\caption{\small{The structure of an $n$-cut in a general theory in
any number of dimensions.}}
\end{center}
\end{figure}

\subsection{General Cut Analysis}

Let us continue our analysis by defining a general $n-$cut 1-loop
amplitude in the following way.  Let us divide the external states
into $n$ different groupings, $\{1\},\{2\},\cdots,\{n\}$, where we
will cyclically define $\{n+1\} \equiv \{1\}$. Let the sum of the
momenta in these groupings be $P_1,\cdots,P_n$. We denote the loop
momentum flowing from $\{j\} \to \{j + 1 \}$ as $l_{j}$. Then, the
$n$ cut of the amplitude, associated with the grouping
$\{1\},\cdots,\{n\}$ is
\begin{equation}
\left[{\rm Cut}^{(n)}_{\{1\},\cdots,\{n\}}\right]M =\int \prod_{i = 1}^n \frac{d^D
l_{i}}{(2 \pi)^D} \, \delta^+(l_{i}^2) \sum_{{\rm species,spins}}
\prod_{j =1}^n M_{{\rm
tree}}^{h_{j-1},-h_j}\left(l_{j-1},\{j\},-l_j)\right)
\delta^D(l_{j-1} + P_j - l_j)
\end{equation}
Note that since cutting the loop momenta puts various internal lines
on shell, the loop ingtegrand directly becomes a product of the
appropriate tree amplitudes. The numerators of the propagators can
be replaced by helicity sums on shell. Also note that there is no
need to talk about ghosts--their role is to cancel unphysical
polarizations and they indeed guarantee that the cut is reproduced
by the physical helicity sum.

Note that since there is only a single free loop momentum, in $d$
integer dimensions we cannot put more than $d$ lines on
shell\footnote{For fixed external momenta. Also in the 60's, more
cuts were considered~\cite{Smatrix} which impose polynomial
relations among the external momenta and those computed the
discontinuity across poles!}, and hence the $n$ cut manifestly
vanishes unless $n \le d$.

As we have seen, these cut integrals correspond to discontinuities
of the full 1-loop amplitude, and therefore, we can equate then to
the corresponding cut of the RHS of equation(\ref{unity}). This
relates the coefficients $C_d,C_{d-1},\cdots,C_2$ to cuts as
\begin{equation}
\left[{\rm Cut}^{(n)}_{\{1\},\cdots,\{n\}}\right]M =
\sum_i C^i_d \left[{\rm Cut}^{(n)}_{\{1\},\cdots,\{n\}}\right] I_d^i + \cdots +
\sum_i C^k_2 \left[{\rm Cut}^{(n)}_{\{1\},\cdots,\{n\}}\right] I_2^i
\end{equation}
Here, the $n-$ cut of the scalar integrals replaces the appropriate
propagators in the loop integrand by $\delta^+$ functions, and are
thus fixed known functions of the external momenta. Since the LHS is
determined by tree amplitudes, these ``cut equations" can be used to
try and determine the coefficients $C_{d},\cdots, C_2$ from tree
amplitudes. Indeed as we will see, we can explicitly determine the
$C$'s, and find that they have a very nice physical interpretation:
the box coefficients are determined by products of four tree
amplitudes at the frozen (complex) value of the loop momenta
determined by the quadruple cut, while the coefficients of the
triangles and bubbles are determined by the ``pole at infinite"
momentum of the product of three and two amplitudes appearing in the
triple and double cuts respectively. We will return at the end to
discuss the rational pieces.

\subsection{Quadruple Cuts and Boxes}

Let us specialize for simplicity to four dimensions. Then, the
maximal cut we can consider is the quadruple cut. It is easy to see
that in general, the four lines can only be put on shell for complex
momenta, but the integral as defined is for real momenta. However,
one is very tempted to deform the contour of integration to capture
these cuts. Alternately, if we work in (2,2) signature, the momenta
would be real and this would happen automatically; since the scalar
integral coefficients are rational functions, we can freely continue
them from (2,2) back to the physical (3,1) signature.

Now, of the box, triangle and bubble scalar integrals, obviously
only the box has a non-vanishing quadruple cut. Thus, the quadruple
cut of the amplitude associated with a given partition of the
external momenta into four sets precisely determines the coefficient
of the corresponding scalar box integral as

\begin{equation}
C_{L_b, R_b}^{L_t, R_t}  =  \sum_{\ell^*} \sum_{{\rm species,spins}} M_{L_t}(\ell_1^*,
-\ell_2)M_{R_t}(\ell_2^*, -\ell_3^*)
M_{R_b}(\ell_3^*; -\ell_4^*)M_{L_b}(\ell_4^*,
-\ell_1^*)
\end{equation}
where $L_t, R_t, L_b,$ and $R_b$ are sets of momentum that partition
the external states.  We use these labels because we think of the
sets as the top-left, top-right, bottom-left, and bottom-right
corners of a scalar box integral. Also, since we are solving
quadratic equations there are two solutions for $\ell^*$, and we are
adding the contribution from both of them.

One might naively hope that the coefficients of the triangle and the
bubble are determined by the triple and double cuts of the
amplitude; however, note that the scalar box integrals have
non-vanishing triple and double cuts in addition to the
non-vanishing quadruple cut. The triangle and bubble coefficients
are non-vanishing only if the triple and double cuts that we inherit
from the boxes do not exhaust all triple and double cuts.

\subsection{Triple Cuts and Triangles}

Let us then move on to consider the triple cut.  
To take a concrete example, suppose
that we have $P_1^2 > 0$. We can go to a frame and choose energy
units so that $P_1 = (1,1,0,0),P_3 = (E_+,E_-,0,0)$, where for
convenience we are using light-cone co-ordinates for the first two
entries. Let us work out all the kinematics to determine the form of
the on-shell cut momenta. It is straightforward to solve the
constraints $l_3^2 = 0, l_2^2 = (l_3 - P_3)^2 = 0, l_1^2 = (l_3 +
P_1)^2 = 0$. The three momenta $l_i$ are then of the form
\begin{equation}
(\sigma \cdot  l)_i = \left( \begin{array}{cc} \alpha^+_i & \ell
\\\bar \ell& \alpha^-_i \end{array} \right)
\end{equation}
with
\begin{equation}
\ell \bar{\ell} = \alpha^+_i \alpha^-_i \equiv r^2 = \frac{-E_+ E_-
(1 + E_+) (1 + E_-)}{(E_+ - E_-)^2}
\end{equation}
The precise form of the $\alpha^\pm_i$ will not be particularly
important to us, we give them here for completeness
\begin{eqnarray}
\alpha_1^+ = \frac{E_-(1 + E_+)}{E_- - E_+} &,& \alpha_1^- =
\frac{E_+ (1 + E_-)}{E_+ - E_-} \nonumber \\ \alpha_2^+ =
\frac{E_+(1 + E_+)}{E_-
- E_+} &,& \alpha_2^- = \frac{E_- (1 + E_-)}{E_+ - E_-} \nonumber \\
\alpha_3^+ = \frac{E_+(1 + E_-)}{E_- - E_+} &,& \alpha_3^- =
\frac{E_- (1 + E_+)}{E_+ - E_-}
\end{eqnarray}
Note that there is in general a 1-parameter set of complex momentum
solutions to these equations, with $\ell = r t, \bar \ell = r
t^{-1}$. The $r=0$ case corresponds to a cut where $P_3^2=0$ and
there is a {\it unique} solution, i.e., the integral localizes. For
$r^2> 0$, we have that the on-shell momentum lies on a circle with
$\ell = e^{i \theta}, \bar \ell = r e^{- i \theta}$. Note that if
$r^2 < 0$ there are no real solutions; this is precisely analagous
to our the discussion of discontinuity of the bubble integral, which
only has a cut interpretation for $P^2 > 0$; for $P^2 < 0$ there is
still a discontinuity which can't be computed as a cut integral, but
can be obtained from continuing the result from $P^2 > 0$. Similarly
in what follows we will everywhere put $r^2 > 0$; we will end up
with an expression for the coefficient of a triangle integral which
is a rational function, that can then be freely continues to other
values for which $r^2$ might not be positive.

The triple cut is
\begin{equation}
\left[{\rm Cut}^{(3)}\right] M = \int d \theta F_3(r {\rm cos} \theta,r {\rm sin} \theta)
\label{oris}
\end{equation}
here
\begin{equation}
F_3 \sim  M_1 M_2 M_3
\end{equation}
is the product of the tree amplitudes in the definition of the
triple cut.

\begin{figure}[h]
\begin{center} \label{FigContourIntegral}
\includegraphics[width=16cm]{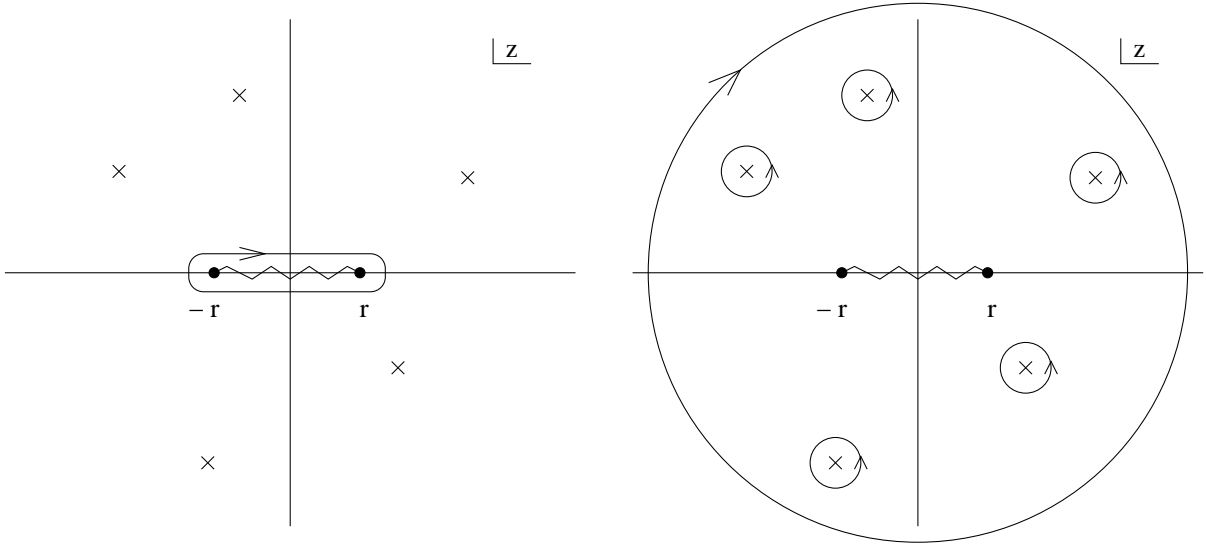}
\caption{\small{The real phase space integral derived from a triple
cut can be represented as a complex contour integral.  The contour
of integration shown on the left is derived from the phase space
integral, but it can be deformed as shown on the right.  The poles
we pick up correspond to additional propagators that can go on-shell
-- these are box contributions.  A possible pole at infinity is due
to the presence of a genuine triangle.}}
\end{center}
\end{figure}

Since we know that the tree amplitudes are rational functions of the
momenta, it is convenient to write $\cos \theta = z$ and
\begin{equation}
\ell = z + i \sqrt{r^2 - z^2}, \bar \ell = z - i \sqrt{r^2 - z^2}
\end{equation}
Then we can re-write
\begin{equation}
\left[{\rm Cut}^{(3)}\right]M = \int_{-r}^r \frac{d z}{\sqrt{r^2 - z^2}} G(z)
\end{equation}
where
\begin{equation}
G(z) = \left(F_3(z,\sqrt{r^2 - z^2}) + F_3(z,-\sqrt{r^2 - z^2})
\right)
\end{equation}
We now view the integrand for complex $z$. The function $\sqrt{r^2 -
z^2}$ has a branch cut that can be taken to run on the real axis
from $(-r,r)$. Note that while each term in the definition of $G(z)$
has a branch cut, $G(z)$ itself is a rational function of
$z$. We can therefore write the triple cut as a contour integral
\begin{equation}
\left[{\rm Cut}^{(3)}\right]M = \int_{{\cal C}} \frac{d z}{\sqrt{r^2 - z^2}} G(z)
\label{somi}
\end{equation}
where the contour encircles the cut from $[-r,r]$. Now, the rational
function $G(z)$ has simple poles in the complex $z$ plane. Since $G$
is a product of three tree amplitudes, the poles of $G$ correspond
to another line going on shell at some complex momentum. This is
precisely the condition for fixing momentum in the quadruple cut,
and the residue of $G$ is nothing but the product of the appropriate
four tree amplitudes associated with that quadruple cut! If we
perform the contour integral for the triple cut by deforming the
pole around infinity, therefore, we get a contribution from poles
which are nothing other than the triple cut of scalar box integrals,
whose coefficients are determined by the complex quadruple cut. The
remaining term comes from the pole of $G(z)$ at infinity, and this
determines the coefficient of the scalar triangle integral
associated with the triple cut!

Let us see this very explicitly, beginning with the derivation of
the quadruple cut from the triple cut. In the triple cut, we have a
product of amplitudes that we will call $M_t(z) M_{L_b}(z)
M_{R_b}(z)$; as we approach a pole of $M_t(z)$, it factorizes as \be
M_t(z) \rightarrow M_{L_t}(z) \frac{1}{(P_{L_t} + \ell(z))^2}
M_{R_t}(z) = M_{L_t}(z) \frac{1}{A+B z + C \sqrt{1-z^2}} M_{R_t}(z)
\ee where for brevity we have omitted the sum over helicites and
species in this expression. If we take $M_{L_t}(z) M_{R_t}(z)
M_{L_b}(z) M_{R_b}(z) = f(z, \sqrt{1-z^2})$ then we can write the
triple cut integral as \be \left[{\rm Cut}^{(3)}\right]M = \int
\frac{dz}{\sqrt{1-z^2}} \left(\frac{f(z,\sqrt{1-z^2})}{A + B z + C
\sqrt{1-z^2}} +\frac{f(z,-\sqrt{1-z^2})}{A + B z - C \sqrt{1-z^2}}
\right) \ee Here we only wish to evaluate the residues from the two
propagators that we have explicitly displayed, because we want to
make contact with a particular box coefficient.  If we call the two
poles $z_1$ and $z_2$, then the result is \be \left. \left[{\rm
Cut}^{(3)}\right]M \right|_{z_1, z_2} & = & \frac{2 \pi i
C}{(B^2+C^2)(z_1 - z_2)}
\left[M_{L_t}(z_1) M_{R_t}(z_1) M_{L_b}(z_1) M_{R_b}(z_1) \right. \nonumber \\
& &  \left. + M_{L_t}(z_2) M_{R_t}(z_2) M_{L_b}(z_2)
M_{R_b}(z_2)\right] \ee As we know the sum of the product of the
four tree amplitudes inside the square brackets is the coefficient
of the associated box function. Furthermore, the overall coefficient
is nothing but the value of the triple cut of the associated box
integral! This can be seen trivially from the fact that the triple
cut of the box is the integral we would get by replacing
$M_{L_t},M_{R_t}$ in the above with $1$. Thus we see that the
contribution to the triple cut coming from the poles at finite $z$
are simply the triplet cuts of the scalar box integrals whose
coefficients have already been fixed by the quadruplet cut.

If there is no pole at infinity, all triple cuts are determined by
quadruple cuts -- but if there is a pole at infinity, then the
scalar triangle coefficients are non-vanishing, and are determined
to be
\begin{equation}
C_3 = \int_{{\cal C}} \frac{d z}{\sqrt{r^2 - z^2}} \left[ F_3(z,\sqrt{r^2 - z^2}) + F_3(z,-\sqrt{r^2 - z^2})\right]
\end{equation}
where the contour ${\cal C}$ encircles the pole at infinity; equivalently we
can change variables to $z = 1/w$ and encircle the pole at $w = 0$.

Note that this construction (\ref{somi}) has a smooth limit as $r\to
0$. This is to be contrasted with the original integral (\ref{oris})
which localizes and does not have a parameter to expand in or to
integrate over. In (\ref{somi}), as $ r \to 0$ and the branch cut
shrinks to zero one can keep the contour of integration unchanged.
In this limit this construction has a very nice interpretation, the
complex loop momentum is $l = (l_+,l_-,z,iz)$, while the real
momentum appearing in (\ref{oris}) is fixed at the value $z=0$. Note
that the complex momentum is nothing but a BCFW deformation, and the
contour integral reduces to the integral of a contour surrounding
the origin $\int dz G(z)/z$, which precisely evaluates the value of
$G(z)$ at the unique real momentum $z=0$.

It is remarkable that merely evaluating the real phase space
integral associated with the triple cut naturally exposes many of
the new ingredients that have been used to shed light on amplitudes
in recent years, including the quadruple cut and BCFW deformation of
momenta! It also gives us a lovely picture for the coefficient of
the scalar triangle and box integrals. The box integral coefficients
are given by the species/helicity sum of the product of four tree
amplitudes with the internal momenta summed over the pair of points
in momentum space picked out by the quadruple cut. Similarly, we
interpret the triangle coefficient as the product of three tree
amplitudes evaluated at infinite $z$ -- the pole at infinity after
we perform the triple cut -- and summed over internal helicities.

\subsection{Double Cuts and Bubbles}

A similar argument relates the coefficient of the scalar bubble
diagram to the pole at infinity of the product of two tree
amplitudes. The phase space integral can be deformed so that poles
at finite locations in the complex plane correspond precisely to
phase space integrals of triple cuts. Since scalar bubbles do not
contribute to triple cuts and triple cuts account for all
contributions from triangles and boxes, then the pole at infinity
gives the information of the scalar bubble directly. The essential
idea is exactly the same: the phase space integral in this case
involves an integration over a sphere, which we can parametrize with
$(\theta,\phi)$ variables, and the integration over the $\phi$
circle can be complexified as above. The deformation to complex
momenta is here even easier to describe: the product of the two tree
amplitudes is a function of the cut loop momenta $\ell_{1,2}$,
$F_2(\ell_1,\ell_2)$. Since these are both lightlike, it is natural
to consider, for a fixed $\ell_{1,2}$, the BCFW deformation under
$\ell_1 \to \ell_1 + q z, \ell_2 \to \ell_2 - qz$, under which
$F_2(l_1,l_2) \to F_2(z)$; (note that $q$ also depends on
$\ell_{1,2}$). Then, the bubble coefficients are determined by the
pole at infinity of $F_2(z)$. The details are somewhat are more
involved and we postpone its discussion to future work \cite{WIP},
but the result for the coefficient of the bubble is very natural and
can be written as
\begin{equation}
C_2 = \int d {\rm LIPS}[\ell_1,\ell_2] \int_{{\cal C}} \frac{dz}{z}
M_L(\ell_1(z),\ell_2(z)) M_R(\ell_1(z),\ell_2(z))
\end{equation}
where the $d$LIPS$[\ell_1,\ell_2]$ is the Lorentz-invariant phase
space associated with the two cut momenta $\ell_1,\ell_2$, and the
contour ${\cal C}$ is taken to enclose the pole at infinity. This
way of obtaining bubble coefficients is very reminiscent of the
discussion of Bjerrum-Bohr et.al. in~\cite{BohrNT}. There is
however, an important difference: in the argument of~\cite{BohrNT}
the double-cut integrand is studied and manipulations are done at
the level of this integrand to try and isolate the contribution from
bubbles (this also applies to the technique by Forde~\cite{Forde}).
Whenever manipulations are done at the level of the integrand in
such arguments, one has to argue that pieces that integrate to zero
do not affect the result. This was assumed in~\cite{BohrNT} and
proven by Forde~\cite{Forde}. Our analysis instead directly examines
the full cut integral, and so we don't encounter issues having to do
with terms that may or may not integrate to zero.

In four dimensions, the bubble integrals have a special
significance, as they are the only ones with UV (logarithmic)
divergences. Suppose we have a renormalizable theory like Yang-Mills
or Yang-Mills minimally coupled to matter, where there is only a
single coupling constant. Then, logarithmic divergences in
an $n-$ particle amplitude must either reflect the logarithmic
running of the coupling or IR divergences. This tells us that we can compute
$b_0$, the coefficient of the one loop beta function, by summing over all the
coefficients of the scalar bubble integrals associated with a given
$n$-point amplitude and subtracting off the contribution from IR 
divergences\footnote{In an earlier version we ignored the issue of IR
divergences, and therefore got the sign of the beta function wrong!  We thank
Lance Dixon for pointing out our mistake.} \cite{DixonNote}:
\begin{equation}
\sum_{{\rm bubbles}} C_2 - IR \propto b_0 M_{{\rm tree}}
\end{equation}
Since we now have an expression for the $C_2$'s directly in terms of
tree amplitudes, and since IR divergences can also be determined in terms of
tree amplitudes, this represents a highly non-trivial relationship
between tree amplitudes!

Let us see how this works in the case of pure Yang-Mills, which will
also serve as a consistency check on our expression for $C_2$.
Consider the amplitude $M(1^-,2^-,3^+,4^+)$ at 1-loop. We want to
compute the coefficient of the bubble integral $I_2(s_{14})$, which
by our formula for $C_2$ becomes
\begin{equation}
C = \int d {\rm LIPS} \int_{\cal C} \frac{dz}{z} \sum_{h=\pm}
M(\ell_2^{-h}(z),4^+,1^-,\ell_1^{h}(z))M(\ell_1^{-h}(z),2^-,3^+,\ell_2^{h}(z))
\end{equation}
Consider first the case with $h=+$; it is easy to see that $h=-$
will make the same contribution so we will multiply the result for
$h=+$ by a factor of 2.  The BCFW deformation is
$\lambda_{\ell_2}(z) = \lambda_{\ell_2}$ while $\lambda_{\ell_1}(z)
= \lambda_{\ell_1}+z\lambda_{\ell_2}$. Then the product of the
amplitudes is
\begin{equation}
\frac{\langle \ell_2~1\rangle^4}{\langle 1~\ell_1(z)\rangle\langle
\ell_1~\ell_2\rangle\langle \ell_2~4\rangle\langle 4~1\rangle}\times
\frac{\langle\ell_1(z)~2\rangle^3}{\langle2~3 \rangle\langle
3~\ell_2\rangle\langle \ell_2~\ell_1\rangle}
\end{equation}
The rational function of $z$ is then
\begin{equation}
\frac{(\langle \ell_1~2\rangle+z\langle \ell_2~2\rangle)^3}{(\langle
\ell_1~1\rangle+z\langle \ell_2~1\rangle)}
\end{equation}
Performing the $z$ integral is the same as extracting the constant
term in a Laurent series around $z=\infty$. Upon doing this, the
coefficient of the bubble becomes
\begin{eqnarray}
C_2 & =  & 2\int d{\rm LIPS}\frac{1}{\langle
\ell_1~\ell_2\rangle^2\langle \ell_2~4\rangle\langle
1~4\rangle\langle 2~3\rangle\langle 3~\ell_2\rangle}\times \nonumber \\
& & (3\langle \ell_1~2\rangle^2\langle \ell_2~2\rangle\langle
\ell_2~1 \rangle^3-3\langle 1~\ell_1 \rangle\langle \ell_1~2
\rangle\langle \ell_2~2 \rangle^2\langle \ell_2~1 \rangle^2+\langle
1~\ell_1 \rangle^2\langle \ell_2~2 \rangle^3\langle \ell_2~1
\rangle)
\end{eqnarray}
where we have included the factor of 2 coming from adding the $h=-$
term as well. There are three different $d$LIPS integrals to be
carried out here, and it is far from obvious that the result will be
proportional to $M_{{\rm tree}}$.
The computation of the integrals is described in appendix D. We indeed
do find that the three integrals give us $(1,1/2,1/3) \times M_{{\rm
tree}}$ respectively. The bubble coefficient is
\begin{equation}
C_2 = 2 \times (1+1/2+1/3) \times M_{{\rm tree}} = \frac{11}{3}
M_{{\rm tree}}
\end{equation}
This seems to be the beta function coefficient, but in fact the sign
is wrong.  The reason is that we have neglected the contributions from
massless bubble integrals attached to the external legs -- although these
diagrams are zero in dimensional regularization, they only vanish due
to a cancelation between UV and IR divergences.  

We can account for these diagrams by computing the IR divergence from 
the collinear limit of the five gluon amplitude
\be
M(i^-, j^-, ...) = \frac{\langle i j \rangle^4}
{\langle 1 2 \rangle \langle 23 \rangle \langle 34 \rangle \langle 45 \rangle \langle 51 \rangle}
\ee
and then subtracting them off.
The computation is a bit subtle because of an overlap between the soft 
and collinear divergences \cite{DixonNote}, but the final result is 
the usual beta function coefficient, precisely as required on general
grounds!

In this explicit computation, the correct result was arrived at in a
highly non-trivial way. It would be very interesting to see if there
is a transparent general reason, using purely on-shell reasoning,
for these relationships to hold. Certainly, the Wilsonian
understanding of effective field theory and the renormalization
group makes heavy use of off-shell information, so it would be
particularly interesting to try and understand these results from an
on-shell perspective.

Before closing this subsection, we state another observation without
proof here, which we will elaborate on elsewhere \cite{WIP}. In a
general theory, the behavior at infinity of the BCFW continued
product of the two tree amplitudes in the double cut $F_2(z)$, is
directly related to the full 1-loop structure of the theory, and not
just to the bubble coefficients. Indeed, consider a Laurent
expansion of $F_2(z)$ about infinity
\begin{eqnarray}
F_2(z) & = & a_0 + a_1 z + a_2 z^2 + \cdots \nonumber \\
&+& \frac{b}{z} \nonumber \\ &+& \frac{c_2}{z^2} + \frac{c_3}{z^3} +
\frac{c_4}{z^4} +  \cdots
\end{eqnarray}
Then, bubbles are present if and only if $a_0 \neq 0$, and triangles
are present if and only if $b \neq 0$. If only the $c_i \neq 0$, the
amplitudes only involve boxes. Indeed it is easy to motivate this
observation: after all, the double cut of any one box integral
contains two propagator factors and hence scales as $1/z^2$ at large
$z$, while the double cut of any one triangle integral scales as
$1/z$ as $z \to \infty$. In ${\cal N} = 4$ SYM and ${\cal N} = 8$
SUGRA, we will see that $F_2(z)$ scales as $1/z^{2s}$, giving us
another independent argument for the absence of triangles and
bubbles.

\subsection{Constructing $M^{{\rm 1-loop}}$ From Its Singularities}
In our discussion in this section, we assumed the decomposition of
the amplitude given by eqn. (\ref{unity}) as a given, beginning with
Feynman diagrams and invoking an integral reduction procedure. By
reversing the logic of this section, we can understand the physical
meaning of this decomposition more transparently. Indeed, suppose we
never knew directly about integral reduction procedures. We would
know that amplitudes have discontinuities which are associated with
putting particles on shell, and by counting equations we would know
that at most four particles can go on shell. If we want to make an
ansatz for an amplitude that reproduces this leading singularity, we
could take the quadruple cut of the amplitude and multiply it by an
object that has a non-vanishing quadruple cut; the simplest such
object is clearly a scalar box integral. Having reproduced the
quadruple cut, we turn to the triple cut. Our ansatz makes a
prediction for the triple cut since the scalar box has a
non-vanishing triple cut; if this agrees with the triple cut of the
amplitude we would be done. But as we have seen, the product of
three tree amplitudes in the triple cut can have a pole at infinity
that can not be reproduced with scalar boxes. Thus, if this pole at
infinity is non-zero, we have to correct our ansatz. Since we have
already accounted for the quadruple cut, we would like to correct it
with an object that has a non-vanishing triple cut but a vanishing
quadruple cut. Once again, the scalar triangle integral is the most
natural object with this property. We can continue in the same way
to see if our ansatz produces the correct answer for the double cut,
and the if product of two amplitudes has a pole at infinity, we have
to correct the ansatz by an object that has a non-vanishing double
cut but vanishing triple and quadruple cuts, which is naturally the
scalar bubble integral.

This leaves us with rational terms, which do not have branch cuts
and are invisible to these cutting procedures in four dimensions.
Ideally we would proceed in complete analogy with the previous
discussions, i.e. start with a single cut integral, study its phase
space integral and identify a contour integral which receives
contributions from poles at finite locations, corresponding to
double cuts, and from a pole at infinity, corresponding to the
rational piece. The procedure in this case is more subtle and
interesting, however, because the ``single" cut is not precisely a
physical amplitude, and poles at finite locations have to produce
complete phase space integrals of double cuts; furthermore in this
discussion the subtleties of dimensional regularization must be
properly dealt with. A related issue worthy of investigation is how
anomalies are interpreted in this on-shell language; indeed, there
are indications that rational terms can be interpreted as anomalies
in some sense. We postpone the analysis using single-cut integrals,
as well as a fuller exposition of the above discussion, for a future
study of general 1-loop amplitudes \cite{WIP}, where the whole
machinery can be used to compute rational terms in theories like QCD
(see \cite{forde:extensions} for a recent discussion along similar
lines).

\section{One-loop Amplitudes with Maximal SUSY}

It has been known for a long time that ${\cal N}=4$ SYM is special
in the sense that any 1-loop amplitude can be written solely in
terms of scalar boxes. This was proven using the Bern-Kosower rules
(string-based method) \cite{Bern:1990cu}. It has also been suspected
since the work of Bern et.al. \cite{Bern:1998sv} that the same
behavior might be true for ${\cal N}=8$ SUGRA. The first explicit
assumption that this be true was made in \cite{Bern:2005bb} and very
non-trivial evidence was presented in \cite{BohrNT}. Very recently,
it was shown how the same string-based method when applied to
theories without color ordering gives rise to cancelations not
present in SYM and imply in the case of ${\cal N}=8$ SUGRA that only
scalar boxes are present \cite{BV}.

In this section we provide a proof of the scalar box expansion
property which is identical for both maximally supersymmetric
theories. Since we have seen that the coefficients of triangles and
bubbles are given by the poles at infinity of the products of three
and two amplitudes appearing in triple and double cuts, in order to
prove the absence of triangles in any theory, it is sufficient to
prove that the species/helicity sum of the product of three tree
amplitudes $F_3(z)$ vanishes as $z \to \infty$, while to show the
absence of bubbles, it suffices to show that species/helicity sum of
the product of two tree amplitudes $F_2(z)$ vanishes at infinity.
This is not a necessary condition--it is in principle possible that
the these functions diverge at infinity with vanishing residue of
the pole at infinity; for instance if $F(z) = a_0 + a_1 z + \cdots$,
the pole at infinity is only given by $a_0$. But as we will show
shortly it is the stronger statement which is true for ${\cal N} =
4$ SYM and ${\cal N} = 8$ SUGRA. The strategy will precisely follow
our analysis of tree amplitudes. We will use SUSY to relate the sum
over the supermultiplet of a product of cut amplitude with some
fixed external states, to the same sum with different external
states but only the highest spin states running in the cut loop.
Then the excellent UV behavior of the highest spin amplitudes can be
used to prove that $F_2$ and $F_3$ vanish at infinity, establishing
the absence of triangles and bubbles.

We can see very clearly that SUSY is playing only a partial role in
this : it allows us to relate the sum over the supermultiplet
appearing in the cuts, to another amplitude with in general
different external states, but with only gluons/gravitons in the
loop. It is then the soft behavior of these tree amplitudes at
infinity, which is already present for non-supersymmetric pure
Yang-Mills and Gravity theories and is related to the enhanced
spin-Lorentz symmetry, that is responsible for the absence of
bubbles and triangles. The role of the excellent large $z$ scaling
of tree amplitudes in allowing ``non-SUSY" cancelations to take
place even in gravitational theories with no SUSY was observed in
\cite{Bern3}.

The absence of rational pieces is proven in a different manner. We
will in fact prove that in any theory where scalar triangles and
scalar bubbles are absent, rational pieces must satisfy collinear
and multi-particle factorization limits by themselves. In
the case of ${\cal N}=4$ SYM and ${\cal N}=8$ SUGRA, where rational
pieces can relatively easily be shown to be absence for $n=4$,
implies that they are absent for any $n$.

\subsection{Absence of Bubbles}

Let us begin with scalar bubble integrals, which are absent if
$F_2(z) \to 0$ as $z \to \infty$, where
\begin{eqnarray}
F_2(z) &=& \int d^{\cal N} \eta_1 d^{\cal N} \eta_2 \nonumber \\ &
&M_L(\{\eta_1,\lambda_1(z),\bar \lambda_1\},\{\eta_2,\lambda_2,\bar
\lambda_2(z)\},\eta^L_{i_L}) \nonumber \\
&\times & M_R( \{\eta_1,\lambda_1(z),- \bar \lambda_1
\},\{\eta_2,\lambda_2,- \bar \lambda_2(z)\},\eta^R_{i_R})
\end{eqnarray}

\begin{figure}[h]
\begin{center} \label{FigBubbleCut}
\includegraphics[width=6cm]{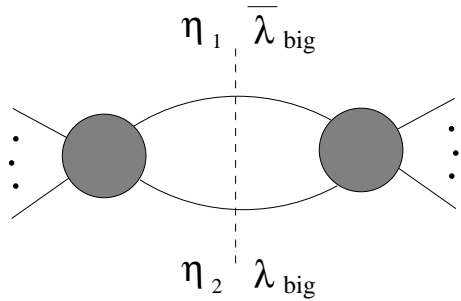}
\caption{\small{A double cut used to understand the contributions of
bubbles in ${\cal N} = 4$ SYM and ${\cal N} = 8$ Supergravity.  We
use supersymmetry to rotate $\eta_1, \eta_2 \to 0$, so that when we
analytically continue the cut loop momenta the double cut -- which
is just a product of two tree amplitudes -- vanishes as
$\frac{1}{z^{2s}}$ as $z \to \infty$.  This loop-level reflection of
the cancelations that make the BCFW Recursion Relations possible
proves the absence of bubbles in these theories.}}
\end{center}
\end{figure}

We follow our usual strategy, and use the $Q$ SUSY to translate
$\eta_{1,2}$ to zero; the required $\zeta$ is $z$-dependent and we
can write:
\begin{equation}
\zeta(z) = \frac{\lambda_1(z) \eta_2 - \lambda_2 \eta_1}{\langle
1(z) 2 \rangle} = \frac{\lambda_1 \eta_2 - \lambda_2
\eta_1(-z)}{\langle 1 2 \rangle}
\end{equation}
Note that $\eta_1(-z)$ now only enters through the shift of the
other $\eta_i$ via $\eta_i \to \eta_i + \langle \lambda_i \zeta(z)
\rangle = \eta_i^\prime(\eta_1(-z),\eta_2)$ where
\begin{equation}
\eta^\prime_i(\eta_1,\eta_2) = \eta_i + \frac{\langle i1
\rangle}{\langle 1 2 \rangle} \eta_2 - \frac{\langle i2
\rangle}{\langle 1 2 \rangle} \eta_1
\end{equation}
Note also that since $\eta_1(-z) = \eta_1 - z \eta_2$, $d^{\cal N}
\eta_1 d^{\cal N} \eta_2 = d^{\cal N} \eta_1(-z) d^{\cal N} \eta_2$.
Further re-defining $\eta_1(-z) \to \eta_1$ in the integral we have
\begin{eqnarray}
F_2(z) &=& \int d^{\cal N} \eta_1 d^{\cal N} \eta_2 \nonumber \\
& & M_L(\{0,\lambda_1(z),\bar \lambda_1\},\{0,\lambda_2,\bar
\lambda_2(z)\},\eta^{\prime R}_{i_L}(\eta_1,\eta_2)) \nonumber \\
&\times & M_R( \{0,\lambda_1(z),-\bar \lambda_1
\},\{0,\lambda_2,-\bar \lambda_2(z)\},\eta^{\prime
R}_{i_R}(\eta_1,\eta_2))
\end{eqnarray}
Now this is a sum over different soft backgrounds of the product of
two $--$ graviton amplitudes, with the only $z$ dependence coming in
the BCFW scaling of the two graviton momenta. But we know that this
amplitude dies as $\frac{1}{z^s}$, and hence
\begin{equation}
F_2(z) \to \frac{1}{z^{2s}} \, {\rm as} \,  z \to \infty
\end{equation}
and therefore there are no scalar bubble integrals at 1-loop. Note
that given our claim about the connection between the powers of $z$
appearing in the Laurent expansion $F_2(z)$ at infinity and the
presence of bubbles and triangles in a general theory, the fact that
$F_2(z)$ vanishes at least as fast as $\frac{1}{z^2}$ demonstrates
the absence of both triangles and bubbles. The fact that $F_2(z)$
vanishes as $\frac{1}{z^4}$ in ${\cal N} =8$ implies further
non-trivial cancellations at 1-loop that are absent for ${\cal N} =
4$. This is a mirror of the same phenomenon seen for the large $z$
scaling of tree amplitudes.

\subsection{Absence of Triangles}

To show the absence of triangles, we have to look at the behavior of
the product of three amplitudes at infinite momentum. It is
important to note that even though the momenta are becoming large,
the large $z$ limit is {\it not} exactly a BCFW deformation on each
of the amplitudes. In BCF, the momentum goes to infinity by having
the $\lambda$ of one line and the $\bar \lambda$ of the other line
become large. Now, for each loop momentum, as $z \to \infty$, we can
take either $\lambda$ or $\bar \lambda$ to be large. Explicitly, we
can write \begin{equation} (\sigma \cdot l)_i(z) = \left(
\begin{array}{cc} \alpha^+_i & \ell(z)
\\\bar \ell(z)& \alpha^-_i \end{array} \right) = \left(\begin{array}{c}
\alpha^+_i \\ \bar \ell(z) \end{array} \right)
\left(\begin{array}{cc}
1 & \ell(z)/\alpha^+_i \end{array}\right) = \left(\begin{array}{c} \ell(z)\\
\alpha^-_i \end{array} \right) \left(\begin{array}{cc} \bar
\ell(z)/\alpha^-_i & 1
\end{array}\right)
\end{equation}
where in the first form, as $z \to \infty$ and so $\ell \to \infty$,
it is $\bar \lambda$ that is becoming large, while in the second
form, it is $\lambda$ becoming large. It is possible to take e.g.
$\bar \lambda_3, \bar \lambda_1$ large and $\lambda_2$ large.

\begin{figure}[h]
\begin{center} \label{FigTripleCut}
\includegraphics[width=8cm]{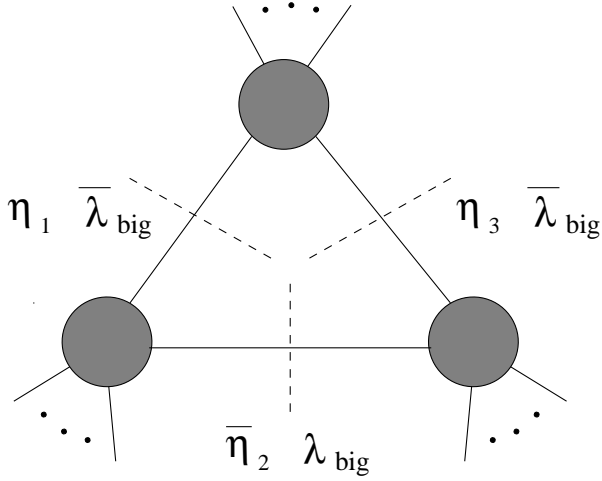}
\caption{\small{This is the structure of the triple cut in ${\cal N}
= 4$ SYM and ${\cal N} = 8$ Supergravity.  We use supersymmetry to
send $\eta_1, \bar \eta_2, \eta_3 \to 0$, so that we the particles
propagating in the loop are gluons or gravitons.  Then we
analytically continue the loop momentum in the way indicated in the
figure to show that this product of three tree amplitudes vanishes
as $\frac{1}{z^s}$ as $z \to \infty$, proving the absence of
triangles in these theories.}}
\end{center}
\end{figure}

This makes the large momentum scaling of the tree amplitudes
involving the external states $\{1\}$ and $\{2\}$ correspond to a
BCF deformation, and using the by now familiar SUSY methods we will
see that these do vanish very rapidly as $z \to \infty$. However,
the amplitude involving $\{3\}$ has two large momenta $l_1,l_3$,
where $\bar \lambda_1, \bar \lambda_3$ are becoming large, which is
not a familiar BCF deformation, so we must understand how to
characterize this. It is natural to label the $l_3,l_1$ lines by
$\eta_1,\eta_3$ and the $l_2$ line by $\bar \eta_2$. Then
there are no triangles if $F_3(z) \to \infty$, where
\begin{eqnarray}
F_3(z) &=& \int d^{\cal N} \eta_1 d^{\cal N} \eta_3 d^{\cal N} \bar \eta_2 \nonumber \\
& &M_3\left(\{\eta_3,\lambda_3(z),\bar
\lambda_3(z)\},\{\eta_1,\lambda_1(z),\bar
\lambda_1(z)\},\eta^{\{3\}}_{i_{\{3\}}}\right) \nonumber \\
&\times & M_1\left(\{\eta_3,\lambda_3(z),- \bar \lambda_3
(z)\},\{\bar \eta_2,\lambda_2(z),- \bar
\lambda_2(z)\},\eta^{\{1\}}_{i_{\{1\}}}\right) \nonumber \\
& \times & M_2\left(\{\bar \eta_2,\lambda_2(z),-\bar
\lambda_2(z)\},\{\eta_1,\lambda_1(z),- \bar \lambda_1(z)\},\bar
\eta^{\{2\}}_{i_{\{2\}}}\right)
\end{eqnarray}
Let us first get some intuition for the problem by considering the
simple case where the set $\{3\}$ has a single particle. Then the
momentum $P_3 \equiv p$ is null; in our parametrizaton we can choose
$E_- = 0$, so that \begin{equation} \lambda_{3,1}(z) =
\left(\begin{array}{c} 1 \\ 0 \end{array} \right), \, \bar
\lambda_1(z) = \left(\begin{array}{c} -1 \\ z \end{array} \right),
\bar \lambda_3(z)= \left(\begin{array}{c} -(1 + E_+) \\ z
\end{array} \right)
\end{equation}
Then, the three-point amplitude at the $\{3\}$ vertex is
\begin{eqnarray}
M_3\left(\{\eta_3,\lambda_3(z),\bar
\lambda_3(z)\},\{\eta_1,\lambda_1(z), \bar
\lambda_1(z)\},\eta_p\right) &=& \frac{1}{(E_+z)^{3s}} \delta^{{\cal
N}} \left(z(\eta_1 + \eta_3)\right) \delta^{\cal N}\left((E_+(\eta_1
+ \eta_p)\right) \nonumber \\
 &=& (E_+z)^s \delta(\eta_1 + \eta_3) \delta(\eta_1 +
\eta_p)
\end{eqnarray}
which does grow as $z^s$. We can use the delta functions to perform
the integrals over $\eta_{1,3}$. We can also do a $z$ independent
$\bar Q$ SUSY transformation to set $\bar \eta_2 \to 0$. Then, the
other two amplitudes $M_{1,2}$ are $(+s \, {\rm anything})$
amplitudes, which each vanish as $\frac{1}{z^s}$. Therefore, we
conclude in this case that
\begin{equation}
F_3(z) \to \frac{1}{z^s}
\end{equation}

In fact this result holds for the general case where $P_3$ is not
necessarily null. As usual, we can use the $Q$ SUSY to translate
$\eta_{1,3} \to 0$, with a $z$ dependent $\zeta$
\begin{equation}
\zeta(z) = \frac{\lambda_1(z) \eta_3 - \lambda_3(z) \eta_1}{\langle
1(z) 3(z) \rangle} = \frac{\lambda_1 \eta_3(z) - \lambda_3
\eta_1(z)}{\langle 1 3 \rangle}
\end{equation}
In doing the integral over $\eta_{1,3}$, we will change variables to
$\eta_{1,3}(z)$; it is easy to compute the Jacobian
\begin{equation}
d^{\cal N} \eta_1 d^{\cal N} \eta_3 = \left(\frac{\langle 1 3
\rangle}{\langle 1(z) 3(z) \rangle}\right)^{\cal N} d^{\cal N}
\eta_1(z) d^{\cal N}\eta_3(z) \to z^{4s} d^{\cal N} \eta_1(z)
d^{\cal N} \eta_3(z)
\end{equation}
where to get the first form we used the Schouten identity to
simplify the Jacobian, and for the large $z$ scaling we use the fact
that $\langle 1(z) 3(z) \rangle \to \frac{1}{z}$ for large $z$.

Having shifted $\eta_{1,3} \to 0$, the amplitude at the $\{3\}$
vertex has two $-s$ helicity particles, and we have to determine its
large $z$ scaling. As we reviewed in the introduction, the large
momentum behavior of the amplitude is determined by the spin Lorentz
invariance.  We choose $q$ to be the $O(z)$ piece of $\lambda_{1,3} \bar{\lambda}_{1,3}(z)$,
ie $q = (0,1,i,0)$.  Using spin lorentz invariance, we derive the ansatz
\begin{equation}
M_{s = 1}^{ab} = \left( c z \eta^{ab} + A^{ab} + q^a K^b + K^a q^b + \frac{B^{ab}}{z} +
\cdots \right)
\end{equation}
for a pair of helicity $\pm 1$ particles in $N=4$ SYM.
We then contract this ansatz with the appropriate polarization vectors.
Because the large $z$ scaling of the gravity amplitude is the square of
the large $z$ scaling of the gauge amplitude, it suffices to look at the
large $z$ scaling of the gauge amplitude. The explicit polarization
vectors are
\begin{eqnarray}
\epsilon_1^a & = &  \frac{\tilde{\mu} \sigma^a \lambda_1}{\left[\bar{\lambda}_1(z) \tilde{\mu}\right]}
= \frac{1}{z} (1,0,0,1) \\
\epsilon_3^b & = &  \frac{\tilde{\mu} \sigma^b \lambda_3}{\left[\bar{\lambda}_3(z) \tilde{\mu} \right]}
= \frac{1}{z} (1,0,0,1)
\end{eqnarray}
where we have chosen $\tilde{\mu} = (1 \ 0)$. When we contract these
polarization vectors with $z \eta^{ab}$, the anti-symmetric matrix
$A^{ab}$, and the $q^a K^b + K^a q^b$ term, the result vanishes, so
we are left with only the $\frac{B^{ab}}{z}$ term from the ansatz.
Thus this amplitude scales as $\frac{1}{z^{3s}}$. We can then
translate $\bar{\eta}_2 \to 0$ to conclude that each of the other
two amplitudes scale as $\frac{1}{z^s}$. We then find that in total
\begin{equation}
F_3(z) \to z^{4s} \times \frac{1}{z^{3s}} \times \frac{1}{z^s}
\times \frac{1}{z^s} \to \frac{1}{z^s}
\end{equation}
This completes the proof of the absence of triangles.

\subsection{Absence of Rational Terms}

Having proven that ${\cal N}=8$ SUGRA 1-loop amplitudes do not
contain scalar triangles or scalar bubbles, the only missing
ingredient is to prove the absence of rational pieces. As mentioned
at the end of our general discussion of 1-loop amplitudes, we should
most naturally proceed by studying the phase space integral
associated with a  ``single cut" and identify a contour integral
which receives contributions from poles at finite locations,
corresponding to double cuts, and from a pole at infinity,
corresponding to the rational piece. This procedure has subtleties
that we postpone to a future discussion \cite{WIP}. In this section
we instead present a very different sort of argument, of a sort
which is not usually encountered in the physics literature, which
not only has the advantage of being very compact, but highlights a
remarkable mathematical difference between rational terms and boxes,
that certainly deserves further exploration. The argument is based
on the number theoretic properties of the scalar box functions and
rational terms when the kinematical invariants are taken to be
algebraic numbers \footnote{We remind the reader that an algebraic
number is a complex number that is a root of a polynomial with
rational (equivalently integer) coefficients. A transcendental
number is a number that is not algebraic. The sum, difference,
product and ratio of algebraic numbers yield other algebraic
numbers, and so they form a field denoted by $\mathbb{\bar Q}$.}:
rational terms become algebraic while box functions are
transcendental \footnote{The latter is based on the mathematical
expectation that dilogarithms of algebraic numbers are
transcendental \cite{Lewin}. We could relax this and simply take the
kinematical invariants to be in $\mathbb{Q}$ (rational) which would
imply that rational terms and boxes becomes rational and irrational
respectively. This follows from the results of Chudnovsky
\cite{irrational} and ${\rm Li}_2$ identities.}. The assertion that
rational terms give rise to algebraic numbers when kinematical
invariants are algebraic is not obvious and requires a proof which
we give in appendix E.

The strategy is to use the fact that the behavior of boxes and
rational terms at algebraic kinematical points gives a sharp
separation among them. This separation persists even when special
kinematical points corresponding to collinear or multiparticle
singularities are approached following a sequence of algebraic
numbers. This is clear for the rational terms but it is a nontrivial
condition for the box functions which we prove in appendix E.

Actually, more needs to be said. Since the algebraic numbers form a
field while transcendental numbers do not, we also need the fact
that linear combinations of dilogarithms, ${\rm log}^2$'s and
$\pi^2$'s with algebraic coefficients cannot give rise to (non-zero)
algebraic numbers. This is the number theoretic version of the fact
proven in section 6 of the first reference in \cite{BDK} where it
was shown that no linear combination of scalar box integrals with
rational functions of the kinematical invariants as coefficients can
give rise to a rational function. The reason this is not enough in
our case is that in the singular limits numerical terms like
$st\times\pi^2$ are produced and these are rational functions of the
kinematical invariants. We therefore need a sharp separation that
continues to hold in all the singular limits of interest. The
transcendentality argument is such a separation as shown in appendix
E.2.

Knowing that boxes and rational terms do not mix even when expanded
around one of the physical singularities we proceed to study the
known singular behaviors in gravity in collinear and
multi-particle factorization channels~\cite{Bern:1998sv}.

The structure of the soft (also given in~\cite{Bern:1998sv}) and
collinear limits give additional evidence for the extra simplicity
of gravity compared to gauge theory: remarkably, the soft factors
and collinear splitting functions are not renormalized beyond tree
level, as a direct consequence of the dimensionful coupling constant
in gravity! This is not true for Yang-Mills theories, and so while
our argument for the absence of triangles and bubbles worked for
both ${\cal N} = 4$ SYM and ${\cal N} = 8$ SUGRA, the argument we
give below for the absence of rational terms will be restricted to
${\cal N} = 8$ SUGRA.

We write the soft limit explicitly, although it is not used in the argument, only to illustrate
the point about the extra simplicity in gravity,
\begin{equation}
M_{n}(\ldots,a, s^\pm,b,\ldots) \stackrel{p_s\to 0}{\longrightarrow} {\cal S}(s^\pm)\times M_{n-1}(\ldots ,a,b,\ldots)
\end{equation}
The collinear limit, $p_a\to z P$ and $p_b\to (1-z) P$, is
\begin{equation}
M_n(\ldots, a^{h_a},b^{h_b},\ldots)\stackrel{a||b}{\longrightarrow} \sum_h {\rm Split}_{-h}\left(z,a^{h_a},b^{h_b}\right) \times M_{n-1}(\ldots, P^h,\ldots).
\end{equation}
These equations are true order by order in perturbation theory. In
our argument only the tree-level and 1-loop cases will be relevant.
The universal soft function ${\cal S}(s^\pm)$ is Weinberg's soft
factor reviewed in section 4.2 while the split function ${\rm
Split}_{-h}$ is a simple rational function of $z$, $[ab]$ and
$\langle ab\rangle$ with rational coefficients (in $\mathbb{Q}$)
given in eq. 5.4 of \cite{Bern:1998sv}.

The multi-particle singular limit, $P^2 = (p_i+\ldots + p_j)^2\to
0$, is more complicated~\cite{ChS,BohrNT} and it is schematically
given as follows
\begin{equation}
\label{multi}
M_n^{1\hbox{-}{\rm loop}}\stackrel{P^2\to 0}{\longrightarrow} M^{1\hbox{-}{\rm loop}}_L \frac{1}{P^2}
M_R^{\rm tree} + M_L^{\rm tree}\frac{1}{P^2}M^{1\hbox{-}{\rm loop}}_R + M_L^{\rm tree}\frac{1}{P^2}M^{\rm tree}_R\times {\cal F}
\end{equation}
where ${\cal F}$ is a helicity independent factorization function
which depends in the so-called discontinuity functions $d_1$ and
$d_2$ introduced in~\cite{boxfunctions}. Now it is clear that on
the right hand side of all the limits, the separation we mentioned
of the boxes and rational terms is preserved provided $z$ is
algebraic (which is part of the original assumption) and provided
the discontinuity functions $d_1$ and $d_2$ do not mix terms. In
appendix E.2 we show that $d_1$ and $d_2$ produce, to order
$\epsilon^0$, only transcendental numbers of the form ${\rm Li}_2$
and $log^2$ of algebraic numbers! This means that rational terms do
not participate in the third term of~(\ref{multi}). If there had
been scalar bubbles this would not have been the case as a different
discontinuity function appears which can mix with the rational
terms.

Given that in the collinear and multi-particle factorization limits the separation between
rational terms and boxes holds on both sides of the equations it
must be the case that boxes and rational terms satisfy the limit
independently! In other words, if we write a general 1-loop
amplitude as
\begin{equation}
M^{\rm 1\hbox{-}loop}_n = \sum_i C^i_4 I_4^i + {\cal R}_n
\end{equation}
where ${\cal R}_n$ are the rational terms, then in the collinear
limit
\begin{equation}
\label{dos}
{\cal R}_n(\ldots, a^{h_a},b^{h_b},\ldots)\stackrel{a||b}{\longrightarrow} \sum_h {\rm Split}_{-h}\left(z,a^{h_a},b^{h_b}\right) \times {\cal R}_{n-1}(\ldots, P^h,\ldots)
\end{equation}
and in the multi-particle factorization limit one has
\begin{equation}
\label{tres}
{\cal R}_n\stackrel{P^2\to 0}{\longrightarrow} {\cal R}_L \frac{1}{P^2}
M_R^{\rm tree} + M_L^{\rm tree}\frac{1}{P^2}{\cal R}_R.
\end{equation}

Now we are ready to prove that ${\cal R}_{n}=0$ for all $n$.
Consider the function
\begin{equation}
H_n=\frac{{\cal R}_n}{M^{\rm tree}_n}.
\end{equation}
We now proceed by induction by assuming that all rational terms
${\cal R}_m$ with $m<n$ are zero. Note that it has been known for a
long time that for $n=4$ there are not rational terms
\cite{Green:1982sw} (more recently the same result has been found
for $n=5$ in \cite{Bern:1998sv} and $n=6$ in \cite{BohrNT}).

Using that $M_n^{\rm tree}$ is singular in the collinear and
multi-particle factorization limits, we can quickly see from
(\ref{dos}) and (\ref{tres}) that $H_n$ vanishes in all
the limits. Obviously, $H_n$ is a rational function of the
kinematical invariants $\langle i~j\rangle$ and $[i~j]$ which is
annihilated by the helicity operator
\begin{equation}
\lambda^i\frac{\partial}{\partial \lambda^i} -\tilde\lambda^i\frac{\partial}{\partial \tilde\lambda^i}
\end{equation}
for all $i$.

The absence of singularities (including collinear limits where
$\langle i~j\rangle$ and $[i~j]$ can be taken to zero independently
in order to be sensitive to phase singularities) implies that $H_n$
can only be a polynomial in $\langle i~j\rangle$ and $[i~j]$.
Dimensional analysis reveals that such a polynomial must have degree
two in the variables ($\langle i,j\rangle$, $[i,j]$). The
requirement that it be annihilated by the helicity operator implies
that the only possible combinations are $\langle i,j\rangle [i,j] =
s_{ij}$. Therefore we have reduced the possibilities to
\begin{equation}
H_n = \sum_{i<j} A_{ij} s_{ij}
\end{equation}
where $A_{ij}$ are numerical coefficients. Taking collinear limits
one can show that the only combination which vanishes in all limits is $A_{ij}=0$.

In making this argument, it was important to have already proven the
absence of triangles and bubbles. This is because e.g. one can show
that the three-mass triangle functions {\it do} produce algebraic
(and in fact rational) numbers in the singular limits under
consideration, and thus could in principle mix with rational terms.
Also, since the argument is inductive, it is important that the
rational pieces vanish for $n=4$. These conditions are violated for
other gravitational theories including pure gravity, so our argument
for the absence of rational terms only works for ${\cal N} = 8$
SUGRA.

This concludes the proof of the ``no-triangle" hypothesis, in other
words, that all 1-loop amplitudes in ${\cal N}=8$ SUGRA can be
expanded (up to order ${\cal O}(\epsilon)$) in terms of scalar boxes
defined in $D=4-2\epsilon$ with coefficients which are $\epsilon$
independent and which are rational functions of the kinematical
invariants.

\subsection{Explicit Solution for One Loop Amplitudes}

We have shown that 1-loop amplitudes in theories with maximal
supersymmetry only receive contributions from box integrals. Thus it
is possible to write down the explicit formula
\be M^{1\hbox{-}{\rm loop}}_n(\{\eta_i,\lambda_i,\tilde\lambda_i\})
= \!\!\!\!\! \sum_{L_t, R_t, L_b, R_b} \!\!\! C_{L_b, R_b}^{L_t,
R_t} \left(\{ \eta_1, \lambda_1, \bar \lambda_1 \}, \ldots, \{
\eta_n, \lambda_n, \bar \lambda_n \} \right) I(P_{L_t}, P_{R_t},
P_{L_b}, P_{R_b}) \ee for 1-loop amplitudes in these theories, where
the coefficients $C_{L_b, R_b}^{L_t, R_t}$ are rational functions of
the external momenta labeled by the sets of momenta that form the
corners of the box, and the $I$'s are standard box integrals given
in Appendix C. The sum runs over all partitions of the external
states into the four possible corners of the box. The $C$'s can be
written as products of four tree amplitudes
\begin{eqnarray}
&& C_{L_b, R_b}^{L_t, R_t} \left(\{ \eta_1, \lambda_1,
\bar \lambda_1 \}, \ldots, \{ \eta_n, \lambda_n, \bar \lambda_n \}
\right) =
 \\ && \sum_{\ell_*} \int \Pi_i d^{\mathcal{N}} \eta_i M_{L_t}(\ell_1^*,
-\ell_2^*;\eta_1,\eta_2)M_{R_t}(\ell_2^*, -\ell_3^*;\eta_2,\eta_3)
M_{R_b}(\ell_3^*; -\ell_4^*,\eta_3,\eta_4)M_{L_b}(\ell_4^*,
-\ell_1^*;\eta_4,\eta_1) \nonumber
\end{eqnarray}
where the $\ell_i^*$ are the complex on-shell loop momenta from the
quadruple cut and the sum refers to the (in general) two solutions
for the frozen momenta.

\begin{figure}[h]
\begin{center} \label{FigQuadrupleCut}
\includegraphics[width=8 cm]{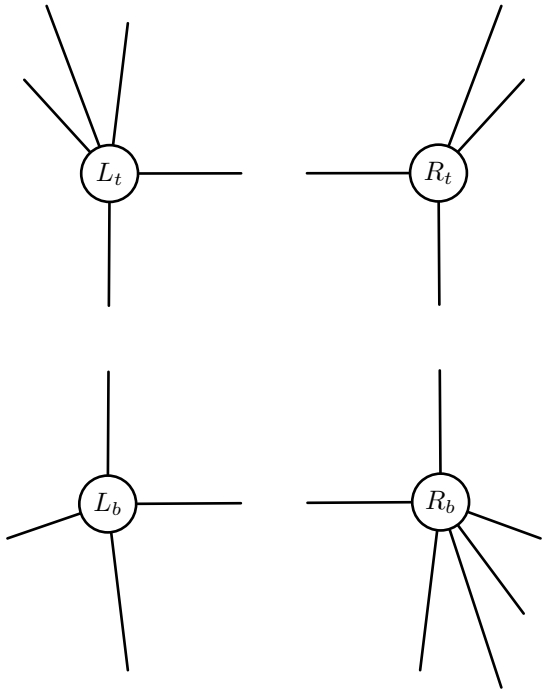}
\caption{\small{One-loop amplitudes in theories with maximal
supersymmetry can be represented entirely in terms of box integrals.
Here we show a quadruple-cut box integral, emphasizing the fact that
in these theories, one-loop amplitudes are completely determined by
tree amplitudes.  As discussed in the text, we also find further
relations among the coefficients of these box integrals in ${\cal N}
= 8$ SUGRA.}} \end{center} \end{figure}

Furthermore, we now know that {\it all} tree amplitudes in ${\cal N}
= 4$ SYM and ${\cal N} = 8$ SUGRA can be computed using recursion
relations, reducing all the way down to the three-point amplitudes
which are fixed by SUSY and Poincare invariance! We have therefore
arrived at a systematic, algebraic  procedure for determining all
1-loop amplitudes in maximally supersymmetric theories. This is a
striking concrete illustration of the great simplicity in these
theories; no such expression is available for all 1-loop amplitudes
in any other theory we are aware of.

These formulas hold for both ${\cal N} = 4$ SYM and ${\cal N} = 8$
SUGRA, but in the ${\cal N} = 8$ theory there are further relations
between the 1-loop box coefficients. As we have seen a number of
times, one difference is that some color-orderings in Yang-Mills are
different than others, obstructing the nicer behavior that is seen
in gravity amplitudes. For instance, the well-known IR behavior of
gravitational amplitudes
\begin{equation}
\left. M^{1 \hbox{-} \mathrm{loop}}_{IR} \right. =
-\frac{1}{\epsilon^2}\sum_{i,j} (-s_{ij})^{1-\epsilon} M^{\rm tree}.
\label{known}
\end{equation}
leads to relations among the box coefficients once we match the IR
divergences on both sides. Unlike in Yang-Mills, the leading
$1/\epsilon^2$ term cancels since $\sum_{i,j} s_{ij} = 0$; this
simple fact already yields a relation between box coefficients that
can arise due to the lack of color-ordering for ${\cal N} = 8$,
which does not exist in ${\cal N} =4$. But there are many more such
relations. For instance, as alluded to in our discussion of the
supersymmetric generalization of the BCFW recursion relations, for
${\cal N} = 4$ SYM, the PT symmetric form of the recursion relations
are a direct consequence of the IR relations. But in ${\cal N} = 8$
SUGRA, we have found through explicit analysis of the case with five
external particles that the IR equations are independent of the
recursion relations, meaning that the box coefficients of the SUGRA
theory are far more constrained.


Some of these extra relations follow from the fact that in ${\cal N}
= 8$ theory, when we BCFW deform a tree-level amplitude to ${\cal
M}(z)$, we have ${\cal M}(z) \propto \frac{1}{z^2}$ as $z \to
\infty$, rather than the $\frac{1}{z}$ scaling expected in ${\cal N}
=4$ SYM when the BCF deformed legs are adjacent in color. This
implies further relations between tree amplitudes given in eqn.
(\ref{relations}) that we reproduce here:
\begin{equation} 0 = \sum_{L} \int d^8 \eta M_L(\{p_1(z_P),
\eta_1(z_P) \},\{-P(z_P), \eta \}, L) \frac{z_P}{P^2}
M_R(\{p_2(z_P), \eta_2\}, \{P(z_P), \eta\}, R) \end{equation}
which allows us to derive a relation among tree amplitudes from
every tree amplitude in the theory.  Since we have determined the
1-loop amplitudes explicitly in terms of the tree amplitudes, these
equations imply many non-trivial relations among box coefficients.

For instance, if we take the triple cut of a 1-loop amplitude, we
get the product of three tree amplitudes that we will suggestively
represent by
$$M_t(\{\eta_1, \ell_1 \}, \{ \eta_2, \ell_2 \}),~ M_{L_b}(\{ \eta_2, -\ell_2 \}, \{ \eta_3, \ell_3 \}),~
{\rm and} ~
M_{R_b}(\{ \eta_3, -\ell_3 \}, \{ \eta_1, -\ell_1 \}).$$
If we consider the BCFW decomposition of $M_t$ as we analytically
continue the loop momenta as $\ell_1(z) = \ell_1 + zq$, $\ell_2(z) =
\ell_2 - zq$, then the relation above becomes \be 0 = \sum_{L_t}
\int d^8 \eta M_{L_t}(\{\ell_1(z_P), \eta(z_P) \},...)
\frac{z_P}{P^2} M_{R_t}(\{\ell_2(z_P), \eta_2\},...) \ee where we
have separated the external momenta in $M_t$ into sets $L_t$ and
$R_t$. Unfortunately, for general external momenta, an arbitrary
BCFW deformation will not keep all of the $\ell_i$ on-shell -- it
would lead to $\ell_1(z_P)$ and $\ell_2(z_P)$ that force $\ell_3^2
\neq 0$.  However, as we saw in section 5.4 when we analyzed the
triple cut in detail, it is possible to deform the momenta $\ell_i$
in such a way that $\ell_i \to z q + O(1)$ as $z \to \infty$.  We
must choose the BCFW $q$ so that $q \cdot P_{L_b} = q \cdot P_{R_b}
= 0$ in order to keep the $\ell_i$ on-shell in this limit, and this
uniquely determines $q$ and therefore the values of $z_P$.

Thus if we take the above relation with this choice of $q$, multiply
by $M_{L_b}$ and $M_{R_b}$, and set the propagator $\frac{1}{P^2}$
on-shell to get a quadruple cut, then we obtain the equation \be 0 =
\sum_{L_t, R_t } \int d^8 \eta \frac{C_{L_b, R_b}^{L_t, R_t}
\left(\{ \eta_1, \lambda_1, \bar \lambda_1 \}, \ldots, \{ \eta_n,
\lambda_n, \bar \lambda_n \} \right)}{q \cdot P_{L_t}} \ee where
$L_b$ and $R_b$ are held fixed.  The 1-loop four-point amplitude
provides a very simple example -- in that case we find \be 0 =
\frac{C^{1,2}_{3,4}}{q \cdot p_1} + \frac{C^{2,1}_{3,4}}{q \cdot
p_2} \ee Since we must have $q \cdot p_3 = q \cdot p_4 = 0$, $q
\cdot p_1 = - q \cdot p_2$ and so this relation is a very simple one
\begin{equation}
\label{wika}
C^{1,2}_{3,4} = C^{2,1}_{3,4}
\end{equation}
Despite first impressions this is a non-trivial fact! The only
symmetry properties of the Box integrals correspond to the rigid
motions of the square, and do not include a (12) flip. The new relation (\ref{wika})
implies that the coefficients do have such a symmetry and hence are invariant
under the full permutation group of four elements. An explicit
evaluation reveals that this is indeed the case since
\begin{equation}
C^{1,2}_{3,4} = stu M^{\rm tree}_4
\end{equation}
which is invariant under the permutation group given the absence of
any color ordering in $M^{\rm tree}_4$. For higher-point amplitudes
there are many more non-trivial relations. We leave a systematic
exploration of their structure to future work \cite{WIP}.

Another question to explore is the nature of the $E_{7(7)}$
invariance of the theory at loop level. This should be
straightforward at one loop, given that we have determined the box
coefficients in terms of tree amplitudes, and that we know the
behavior of the single- and double- soft scalar emission for trees.
We very naturally expect that the amplitude for single soft emission
continues to vanish and the amplitude for double soft emission has
the same form as eqn.(\ref{E7}). These results are {\it almost}
obvious. Consider single scalar emission; the cofficient of any box
is $M_1 M_2 M_3 M_4$, and the $M$ for whichever corner contains the
scalar vanishes in the soft limit. For generic boxes, the box
integral itself will be regular in the soft limit, so these
contributions vanish. One then only has to investigate the special
case where the soft scalar is alone in one corner of a box, where
singularities in the box integral can occur in the soft limit.
Similarly, for generic boxes the double scalar emission is easily
seen to be of precisely the form eqn.(\ref{E7}), leaving only boxes
whose associated integrals can possibly become singular in the
double soft limit. We will return to this issue in \cite{WIP}.

\section{Leading Singularity Conjecture for ${\cal N}$ = 8 SUGRA}

The discussion in this section is mainly speculative and the purpose
is to motivate further research in this direction since its
consequences might be far-reaching. As mentioned in the introduction
to the BCFW construction in section 1.1, one of its most surprising
aspects is that tree-level scattering amplitudes can be completely
constructed by using only a very small subset of the poles from
non-vanishing factorization channels. As we discussed in detail this
is always true in ${\cal N}=4$ SYM and in ${\cal N}=8$ SUGRA.

One might wonder if this simplicity continues at loop level. At higher orders in perturbation theory, scattering amplitudes develop, in addition to poles, branch cuts. At $L$-loop order we expect functions at least as complicated as polylogarithms of $(L+1)^{\rm th}$ order, i.e. ${\rm Li}_{L+1}(z)$ and ${\rm log}^{L+1}(z)$. Here we have written the polylog as a function of a single complex variable but in an actual amplitude the argument is a rational function of various kinematical invariants. These functions possess branch cuts with discontinuities which are given in terms of lower polylogs. This means that the discontinuities are not single valued functions and have branch cuts themselves. The structure of multiple branch cuts in the multi complex dimensional space of kinematical invariants makes the problem of determining the amplitudes quite formidable.

In the 60's, detailed studies of the singularity structure of
Feynman integrals were carried out~\cite{Smatrix}. Each Feynman
integral by itself provides a good laboratory for understanding the
intricate structure of branch cuts of multi-loop amplitudes. The
location of branch cuts and the corresponding discontinuities are
nicely encoded in the Landau equations. Discontinuities across the
branch cuts were shown to be computable in terms of the original
Feynman integral by cutting propagators in the sense defined in the
section 5. In other words, each cut integral can be interpreted as
the discontinuity across a given singularity. Using this connection
it is clear that a discontinuity computed by cutting two propagators
will possess higher codimension branch cuts if there are more
propagators left uncut in the integral. Cutting more propagators
gives the corresponding discontinuities. This process cannot
continue indefinitely as it is bounded by the dimension of spacetime
and the number of propagators in the integral under
consideration\footnote{One can go even further than the dimension of
spacetime by imposing polynomial relations among the external
momenta (these discontinuities are related to poles not to branch
cuts).}. The end point of this process computed the discontinuity
across what was called the {\it Leading Singularity} of the
corresponding integral.

Our interest in this section is the study of singularities of
amplitudes, not individual Feynman diagrams. Therefore, we make the
following definition: the Leading Singularity of an amplitude at a
given loop level is the highest codimension singularity allowed by
the spacetime dimension. Here and in the following we will abuse
terminology and refer to both the singularity and its discontinuity
as ``the leading singularity".

At 1-loop in four dimensions the leading singularity corresponds to
cutting four propagators (a quadruple cut). Clearly, in a general
QFT the knowledge of the leading singularity is necessary but not
sufficient to fix a 1-loop amplitude. One might naively think that
it is impossible to determine a full amplitude by only computing
such a restrictive set of singularities. Note that the same naive
expectation could have been stated at tree level! In fact, much more
is true. We already have two theories for which this surprising
property holds, i.e. 1-loop amplitudes in ${\cal N}=4$ SYM and in
${\cal N}=8$ SUGRA are determined entirely by their leading
singularities.

Let us amplify this point. In section 4 we gave the basic steps to
determine to which degree a general QFT is determined by its leading
singularities. An amplitude is determined by its leading
singularities if all its subleading singularities are. Recall that a
triple cut can be computed in terms of quadruple cuts if and only if
there is no contribution at infinity. If there is a contribution at
infinity them more information is needed. If the contribution at
infinity vanishes then one goes to the double cut and repeats the
analysis. Absence of all contributions at infinity is the statement
that any 1-loop amplitude can be expressed as a linear combination
of only scalar boxes. Since each scalar box possesses a particular
leading singularity not shared by the others it follows that the
amplitude can be determined only by computing quadruple cuts.

At higher loops, one would naively think that if the number of
external particles is not large enough the leading singularity will
always vanish. For example, a four-point two-loop amplitude does not
have eight propagators to cut. However, as shown in~\cite{BC} there
are hidden singularities which always account for the missing
propagators. Another problem might be that since the structure at
higher loops is more complicated (e.g. no simple basis is known)
then the leading singularity might not be enough to fix it. This is
actually the case for the leading singularity as defined in the
60's~\cite{Smatrix} and used \footnote{In [57], Bern et.al. have
developed a systematic and efficient technique where lower
codimension singularities are also used in order to get sufficient
information to determine the amplitudes. This is called the
``maximal cut method", and has been used to determined the
four-particle five-loop integrand in ${\cal N}=4$ SYM.}
in~\cite{quadruple, moresingular}, i.e. by replacing propagators
$1/P^2$ by $\delta^+(P^2)$. Recently~\cite{sharpleading}, it was
shown that one can roughly double the amount of information (per
loop) if one considers isolated ``complex leading singularities".
This has been shown to provide enough information to determine
two-loop five-point, two-loop six-point and three-loop five-point
MHV amplitudes in ${\cal N}=4$ SYM~\cite{sharpleading, sixleading,
threeloop}.

The natural question is whether all amplitudes in ${\cal N}=4$ SYM
and in ${\cal N}=8$ SUGRA are determined by their leading
singularities. There is a long way to go to answer this question. At
the moment the way the complex leading singularities have been used
is manifestly four dimensional and has no information about the way
the theory has been regulated. It is clear that in order to discuss
issues such as the UV behavior of a theory it is important to have a
consistent regularization procedure in place. Therefore one has to
extend the leading singularity technique to, e.g., $D=4-2\epsilon$
in the FDH scheme.

These problems are very interesting but beyond the scope of our
discussion here. Let us simply assume that all generalizations are
indeed possible and furthermore that at every given loop order all
scattering amplitudes in ${\cal N}=4$ SYM and in ${\cal N}=8$ SUGRA
are determined in terms of the leading singularities. Let us explore
some consequences of this assertion in order to motivate further
study. In a general non-asymptotically free field theory,
renormalizble or not, the presence of UV divergences in perturbation
theory means that the classical theory is only an effective
description at low energies and that higher derivative terms appear
in the low-energy theory, reflecting the nature of the underlying UV
completion. As mentioned in section 2.3 the two-loop divergence in
pure gravity implies the presence of a counterterm that changes the
structure of perturbation theory by introducing three-particle
vertices which were not present in the two-derivative lagrangian. A
consequence of this is that at higher orders in perturbation theory,
scattering amplitudes cannot possibly be determined by only the
information of the original lagrangian since the new vertices
generated at two loops need to be included.

Thus these UV divergences make it impossible to determine the
perturbative expansion of the S-matrix of a theory in terms of the
original lagrangian. On the other hand, if the full S-matrix of a
theory is completely determined by its leading singularities, which
are themselves computed only in terms of the tree-level S-matrix,
then UV divergences cannot be present and the theory must be finite.
Therefore, it is tempting to make the statement that the well-known
UV finiteness of ${\cal N}=4$ SYM as well as the hypothetical
finiteness of ${\cal N}=8$ SUGRA both follow from their
corresponding S-matrices being determined in terms of their leading
singularities. Note that this ``leading singularity conjecture" is
much stronger than finiteness; for instance, one can easily imagine
theories which are finite (such as non-supersymmetric Banks-Zacks
CFT's \cite{Banks}) which, already at 1-loop, have triangles and
bubbles and rational pieces in their amplitudes. However, since
every amplitude must have a leading singularity, the nicest possible
analytic structure is for {\it all} the subleading singularities to
be determined by it. If ${\cal N} = 4$ SYM and ${\cal N} = 8$ SUGRA
are indeed the simplest QFT's, they are the best candidates to have
this amazing property. Of course, this means that if UV divergences
are eventually found in the increasingly heroic explicit higher-loop
calculations of ${\cal N} = 8$ SUGRA amplitudes, the much stronger
leading singularity conjecture is manifestly false!

It goes without saying that the perturbative finiteness of ${\cal N}
= 8$ SUGRA is {\it by itself} not particularly important for
physics. Apart from the well-known phenomenological difficulties
with using this theory to describe the real world, there are
fundamental reasons to believe that the real difficulties with
quantum gravity have little to do with the technical issue of
non-renormalizable divergences in Feynman graphs; rather, they have
to do with the breakdown of local field theory, most obviously at
short distances, but also macroscopically in certain situations.
Indeed, even if the theory has no perturbative divergences, the
expansion parameter of perturbation theory is $(E/M_{Pl})$, which is
large at super-Planckian energies and so the finiteness of the
perturbation theory is useless; related to this, since the expansion
is asymptotic, the un-determinable $e^{-(M_{Pl}/E)^p}$ corrections
to the amplitudes are also large. Furthermore, as has been pointed
out by many authors, the continuous $E_{7(7)}$ symmetry of the
theory must be broken to a discrete subgroup to reflect Dirac
quantization of electrically and magnetically charged black hole
solutions. This dovetails nicely with the observation that it is
impossible to decouple wrapped brane states from the spectrum in
attempting to arrive at ${\cal N}= 8$ from compactifications of
M-theory \cite{Green:2007zzb}. Finally, everything we have learned
about the holographic properties of quantum gravity, from the
qualitative lessons of black hole complementarity \cite{BHC} to the
sharp lessons of the AdS/CFT correspondence \cite{juan} tells us
that because of non-perturbative gravitational effects, quantum
gravity can not be an ordinary quantum field theory.

The {\it real} interest in the possible finiteness of ${\cal N} = 8$
SUGRA is that it is telling us about unsuspected structures in the
theory, perhaps a new sort of dual formulation. We end with further
speculations along these lines.

\section{Towards the Holographic Theory of Flat Space}

Why does the conventional formulation of field theory shroud  the
remarkable structures in gauge and gravity amplitudes? Why does it
mislead us into thinking that the theories with the simplest
amplitudes are most complicated and vice-versa? The culprit is
locality. The conventional formulation of field theory, based on a
local Lagrangian, makes locality manifest. Already even for scalar
field theories, this involves a degree of redundancy -- field
refinitions can change the Lagrangian while leaving scattering amplitudes
invariant. But for describing particles of higher spin,
holding onto manifest locality forces the introduction of
increasingly tortured amounts of redundancy, from gauge invariance
for spin 1 to diffeomorphism invariance for spin 2. This leads
directly to the enormous complexity of perturbation theory. Perhaps
the most complicated theories from this point of view are ${\cal N}
= 4$ SYM for spin 1 and ${\cal N} = 8$ SUGRA for spin 2; the
difficulty in formulating an off-shell superspace makes the extra
SUSY a hindrance rather than an aid in explicit computations. The
fact that maximally supersymmetric theories seem to nonetheless have
the simplest scattering amplitudes strongly suggests the existence
of a ``weak-weak" dual formulation of QFT, which will not be
obviously local, but will make manifest other properties, such as
the simplicity of amplitudes. It stands to reason that the simplest
versions of such a theory should describe ${\cal N}=4$ SYM and
${\cal N} = 8$ SUGRA.

One of the things such a theory should clarify is an {\it a priori}
understanding of what the analytic structure of scattering
amplitudes should be. This is a major missing piece in the
``bottom-up" approach to uncovering the properties of amplitudes
that have been pursued so far. The analytic structure is very easy
to guess at tree-level, but already at 1-loop the rules are not
clear. In the usual view, the analytic structure of the amplitudes
reflects causality and unitarity in some way. However, given that
the structure of the amplitudes are most naturally given without any
past-future distinction and for complex momenta, it seems likely
that, from the dual point of view, causality and unitarity will be
derived consequences of more fundamental rules, when specialized to
in- an out- going real momenta with positive energies.

Apart from the clear patterns in the structure of scattering
amplitudes calling out for an explanation, there is a perhaps deeper
reason for suspecting that a dual formulation of field theory
exists. While Lorentz invariance is very likely an exact property of
Nature, non-perturbative gravitational effects make it impossible to
define local observables sharply, and therefore locality can not be
an exact property including gravity \cite{nonloc}. The usual
reaction to this fact is to formulate a holographic theory living on
the boundary of space-time, where precise observables can be
defined. The most successful formulation of quantum gravity we have
-- the AdS/CFT correspondence -- certainly has this character. But
it has not yet given us a way of understanding approximately local
observables in the bulk, for instance, the fate of an observer
falling into a black hole. Perhaps the fact that there are no sharp
observables associated with such observers tells us that we
shouldn't ask such questions, however, the situation becomes more
dire in cosmology, where we clearly don't live at the boundary of
space-time and are much more analogous to observers falling through
a black hole horizon. Since the ``local" geometry in any space-time
is flat space, an understanding of approximately local observables
is intimately tied up with understanding quantum gravity in flat
space.

One might think that we already have an approximately local
description of gravitational physics -- the long-distance effective
theory of GR coupled to matter! However we know that this
description can be drastically misleading. For instance the
long-distance theory unambigously and incorrectly predicts
information loss or remnants in the process of black hole formation
and evaporation; in an eternally inflating spacetime, the infinite
volumes of spatial slices are described with infinitely large
Hilbert spaces that seem clearly unphysical. What we are after is
another way of describing approximately local observables that is
also applicable to these situations.

There is a perhaps useful analogy to made here, between our attempts
to jettison locality to describe gravitational physics, and the
jettisoning of causality in the transition from classical to quantum
mechanics. The usual formulation of classical mechanics in terms of
Newton's laws or the equivalent differential equations in phase
space, are manifestly deterministic. One can formulate a rough guide
to where quantum effects might be important, invoking a fuzziness in
phase space set by $\hbar$; but this is a very approximate notion
missing fundamental aspects of quantum kinematics; the idea of a
wavefunction, interference and entanglement. This formulation of
classical mechanics does not allow a deformation with a small
parameter $\hbar$ to quantum mechanics. On the other hand, the least
action principle is a reformulation of classical mechanics in a way
that is not {\it manifestly} deterministic. This makes it a much
more convenient jumping off point to quantum mechanics via the path
integral.

In the analogy, our usual (local) approximation of the correct
underlying gravitational theory by effective field theory breaking
down in the UV at the Planck scale is like the crude approximation
to quantum mechanics afforded by Newton's laws with a fuzzy phase
space. What we are looking for is a reformulation of QFT that
is  not {\it manifestly} local, just as the least action principle
formulation of classical mechanics is not manifestly deterministic.
Much like the least action principle made other properties of
classical mechanics manifest (such as conservation laws), so the
desired reformulation of QFT should make the counter-intuitive
simplicity of scattering amplitudes manifest. And much like the
least action principle could be naturally ``deformed" to quantum
mechanics, thereby destroying determinism, so the hope is that this
reformulation of QFT can be more naturally deformed to a theory
incorporating the non-perturbative gravitational effects that
destroy locality.

It is very plausible that the theory we are after should be thought
of as the holographic theory of flat space. This is vacuously true
in the sense that it should compute the S-matrix, which is the only
boundary data available in flat space, but there is a more specific
sense in which thinking in this way could be useful. The best
attempt we currently have at a dual formulation of QFT is the
twistor string theory for ${\cal N} = 4$ SYM \cite{Twistor}. It is
notable that the twistor variables are associated with light-like
lines in Minkowski space and are therefore holographic co-ordinates
for its null boundary--indeed, Penrose's orignial motivations for
introducing twistor variables \cite{Penrose:1972ia} were essentially
holographic.

Over the past decade, we have learned a huge number of fascinating
things about quantum field theories in general and ${\cal N} = 4$
SYM in particular, stimulated by the AdS/CFT correspondence
\cite{juan}. It seems very hard to imagine that any theory could be
nicer than ${\cal N} = 4$ SYM in the planar limit, given its amazing
properties, from integrability determining the spectrum of operator
dimensions \cite{integrability} to the growing understanding of dual
conformal invariance in scattering amplitudes
\cite{Malda,perturbation}. But in this paper, we have suggested that
${\cal N} = 8$ SUGRA has the simplest S-matrix of any theory. There
are hints at tree-level and 1-loop for extra simplicity in ${\cal N}
= 8$ compared to ${\cal N} = 4$--the gravitational amplitudes die as
$\frac{1}{z^2}$ rather than $\frac{1}{z}$ at infinity under BCFW
deformation, and thereby satisfy further relations that are absent
in ${\cal N} = 4$. This tree-level fact extends to 1-loop level,
since both theories only have scalar box integrals and their
coefficients are determined by the tree-amplitudes, so the 1-loop
amplitudes in ${\cal N} = 8$ also satisfy further relations that are
absent for ${\cal N} = 4$. There is the observation that the
massless $S-$ matrix is defined in all the ${\cal N} = 8$ vacua,
transforming non-trivially under an $E_{7(7)}$ symmetry, while the
massless S-matrix is only defined at the origin of moduli space for
${\cal N} = 4$ SYM. Perhaps this compensates in simplicity for the
fact that ${\cal N} = 4$ is superconformal and the planar limit
appears to enjoy a dual superconformal invariance!

There is a final moral reason why ${\cal N} = 8$ should have the
simplest scattering amplitudes. As a non-gravitational theory,
${\cal N} = 4$ SYM has a wealth of physical off-shell information in
addition to the S-matrix. However as a gravitational theory, ${\cal
N} = 8$ SUGRA has the S-matrix as its only observable, and most
deserves the nicest S-matrix! On a less jocular note along these
lines, the twistor string theory fails beyond tree-level, since it
includes conformal supergravity that can't be decoupled at loop
level \footnote{More recent twistor-space theories purporting to
describe YM theories have been constructed \cite{oxford} but their
status at loop level is not yet clear}.This could be a hint that any
such dual theory is gravitational, which would be very natural for
the reason just mentioned, that it is gravity that renders the
S-matrix the only observable in flat space. Clearly a new physical
idea is needed to find a sensible version of such a theory for
${\cal N} = 8$ SUGRA.

\section*{Acknowledgments}

We are grateful to Zvi Bern, Emil Bjerrum-Bohr, Lance Dixon, David
Kosower, Juan Maldacena, Nathan Seiberg, Pierre Vanhove, Jay Wacker
and Edward Witten for interesting discussions. F.C. is also grateful
to the Institute for Advanced Study for hospitality during the
origination of this work. N-A.H. and J.K. also thank the Perimeter
Institute for its hospitality during an enjoyable visit.  N.-A.H.
would like to thank the organizers of the ``Wonders in Gauge Theory
and Gravity" workshop in Paris, for putting together a stimulating
meeting. N.A.-H. is supported by the DOE under grant
DE-FG02-91ER40654, F.C. was supported in part by the NSERC of Canada
and MEDT of Ontario, and J.K. is supported by a Hertz foundation
fellowship and an NSF fellowship.

\begin{appendix}

\section{One Loop IR Divergences and BCF Recursion Relations}

We closed section 3 by commenting on the relation between a manifestly PT invariant,
supersymmetric form of BCF
recursion relations and the IR singular structure of 1-loop
amplitudes in ${\cal N}=4$ SYM and in ${\cal N}=8$ SUGRA. In this
appendix we explain and derive the claims made as well as discuss
further issues related to the extra relations among coefficients
which appear in ${\cal N}=8$ SUGRA and are not present in ${\cal
N}=4$ SYM.

As discussed in section 3, the IR divergent part of any 1-loop
amplitude in ${\cal N}=4$ SYM has the form
\begin{equation}
\left. M^{\rm 1 \hbox{-} loop}_{IR} \right. =
-\frac{1}{\epsilon^2}\sum_{i=1}^n (-s_{i,i+1})^{-\epsilon} M^{\rm
tree}. \label{known}
\end{equation}
At 1-loop, ${\cal N}=4$ super Yang-Mills amplitudes can be
conveniently represented in terms of scalar box integrals whose
definition is not relevant for the present discussion. The
coefficients of the box integrals are determined in terms of
products of tree amplitudes using the quadruple cut technique.
Scalar box integrals contain IR singularities and one can show that
by combining the different constraints from the known IR behavior
(\ref{known}) the following must hold
\begin{eqnarray}
\label{compi}
&& M^{\rm tree}_n =  \\ && \frac{1}{2}\sum_{j=i+2}^{n+i-2} \sum_{{\cal S}, h}\! \frac{M_1(\ell_2,i+2,\ldots ,
j,\ell_3)M_2(\ell_3,j+1,\ldots
,i-1,\ell_4)M_3(\ell_4,i,\ell_1)M_4(\ell_1,i+1,\ell_2)}{(k_i+k_{i+1})^2
(k_{i+1}+k_{i+2}+\ldots + k_j)^2} \nonumber
\end{eqnarray}
for arbitrary but fixed $i\in \{ 1,\ldots , n\}$ and where the sum over $h$ means a sum over all states in the ${\cal N}=4$ supermultiplet for each internal line and the sum over ${\cal S}$ is over the solutions to the equations $\ell_i^2=0$ with
\begin{equation}
\ell_1:=\ell,\quad \ell_2 := \ell-k_{i+1},\quad \ell_3 := \ell +k_{i+1}+k_{i+2}\ldots +k_{j}, \quad \ell_4:=\ell+k_i.
\end{equation}
There are two solutions to these equations. These are
\begin{equation}
\ell= z\lambda_i\tilde\lambda_{i+1}, \qquad \ell=z\lambda_{i+1}\tilde\lambda_i
\end{equation}
with $z$ determined from the condition $\ell_3^2=0$. This equation
was first presented in~\cite{RSV}. In appendix B, we provide a proof
of this relation starting from the set of IR equations.

It is clear that the main complication in (\ref{compi}) is the sum
over all particles in the multiplet since traditional approaches
treat this computation state by state. Using the $\eta$
representation for all the amplitudes, one can easily compute the
product of the two three-particle amplitudes in (\ref{compi}) after
all the sums over multiplets in lines $\ell_4$,$\ell_1$ and $\ell_2$
are done. It turns out that each solution in ${\cal S}$ gives rise to one of
the supersymmetric recursion relations that enter in the equality (\ref{equality}).
Knowing that each form gives independently the full tree amplitude,
the factor of half is there to cancel the
factor of two coming from adding the two solutions.

We now show explicitly that (\ref{compi}) is the manifestly PT
invariant supersymmetric recursion relation. We choose to write all
external states in the $\eta$ representation. Our goal is to
explicitly compute the product of the two three-particle amplitudes
in (\ref{compi}) after all the sums over multiplets in lines
$\ell_4$,$\ell_1$ and $\ell_2$ are done.

Consider first the solution given by $\ell = z
\lambda_i\tilde\lambda_{i+1}$. We choose to write the two
three-particle amplitudes in the $\eta$ representation. The momentum
configuration for $M_3(\ell_4,i,\ell_1)$ is such that all
$\lambda$'s are proportional. This means that only the anti-holomorphic term
contributes and gives
\begin{equation}
M_3(\{\ell_4,\eta_{4}\},\{i,\eta_i\},\{\ell_1,\eta_1\}) =\int
d^4\bar\eta_id^4\bar\eta_4d^4\bar\eta_1 \frac{
\delta^8(\tilde\lambda_i(\bar\eta_i-\bar\eta_4)-z\tilde\lambda_{i+1}(\bar\eta_1+\bar\eta_4))}{[i,\ell_4][\ell_4,\ell_1]
[\ell_1,i]}e^{\bar\eta_1 \eta_1+\bar\eta_4 \eta_4+\bar\eta_i
\eta_i}.
\end{equation}
where we used that $\tilde\lambda_{\ell_1}=z\tilde\lambda_{i+1}$ and
$\tilde\lambda_{\ell_4}=z\tilde\lambda_{i+1}-\tilde\lambda_i$.

Since by assumption $[i,i+1]\neq 0$, the delta function localizes
$\bar\eta_i=\bar\eta_4$ and $\bar\eta_1=-\bar\eta_4$. The integrals
over $\bar\eta_1$ and $\bar\eta_i$ can be done at the expense of a
Jacobian. Recalling that these are Grassmann integrations on finds
that the Jacobian is a factor of $z^4[i,i+1]^4$. The answer is then
\begin{equation}
M_3(\{\ell_4,\eta_{4}\},\{i,\eta_i\},\{\ell_1,\eta_1\})
=z[i,i+1]\int d^4\bar\eta_4 e^{\bar\eta_4(\eta_4+\eta_i-\eta_1)}
\end{equation}
where we used that on the solution $[i,\ell_4][\ell_4,\ell_1]
[\ell_1,i]=z^3[i,i+1]^3$

Consider now the second three-particle amplitude. In the solution
under consideration only the holomorphic term contributes
and gives
\begin{equation}
M_3(\{\ell_1,\eta_1\},\{i+1,\eta_{i+1}\},\{\ell_2,\eta_2\}) =
\frac{\delta^8(\lambda_i(\eta_1+z\eta_{2})+\lambda_{i+1}(\eta_{i+1}+\eta_{2}))}{\langle
i+1,\ell_1\rangle\langle \ell_1,\ell_2\rangle\langle
\ell_2,i+1\rangle}. \label{asica}
\end{equation}

In the original formula  (\ref{compi}) the sum over all helicity
states is achieved by integrations over $\eta_i$ with $i=1,2,3,4$.
Using the delta function in (\ref{asica}) one can do the $\eta_1$
and $\eta_2$ integrals. Once again, since $\langle i, i+1\rangle
\neq 0$, the delta function localizes $\eta_1=-z\eta_{2}$ and
$\eta_{2}=-\eta_{i+1}$. A Jacobian is generated which is of the form
$\langle i, i+1\rangle^4$. Also note that $\langle
i+1,\ell_1\rangle\langle \ell_1,\ell_2\rangle\langle
\ell_2,i+1\rangle= z\langle i, i+1\rangle^3$.

Writing this explicitly one finds
\begin{eqnarray}
\int \prod_{i=1}^4 d^4\eta_i M_1(\ell_2,i+2,\ldots , j,\ell_3)M_2(\ell_3,j+1,\ldots ,i-1,\ell_4)M_3(\ell_4,i,\ell_1)M_4(\ell_1,i+1,\ell_2) =  & & \nonumber \\
\!\! \int \! d^4\eta_3 M_1(\ell_2,i+2,\ldots , j,\ell_3)\! \int\! d^4\eta_4
M_2(\ell_3,j+1,\ldots ,i-1,\{ \ell_4,\eta_4\})\! \int \! d^4\bar\eta_4
e^{\bar\eta_4(\eta_4+\eta_i+z\eta_{i+1})}
\end{eqnarray}
On the right hand side we have omitted the factor $\langle i,
i+1\rangle[i,i+1]$ which will cancel with a similar factor in the
denominator of (\ref{compi}). Also worth mentioning is the fact that
the $z$ dependence completely canceled out.

Performing the $\eta_4$ integral one goes from the $\eta$ to the
$\bar\eta$ representation
\begin{eqnarray}
&& \int d^4\bar\eta_4 e^{\bar\eta_4(\eta_i+z\eta_{i+1})}\int
d^4\eta_4 e^{\bar\eta_4\eta_4}M_2(\ell_3,j+1,\ldots ,i-1,\{
\ell_4,\eta_4\}) = \int d^4\bar\eta_4
e^{\bar\eta_4(\eta_i+z\eta_{i+1})}\qquad \qquad \nonumber \\ &&  \times
M_2(\ell_3,j+1,\ldots ,i-1,\{ \ell_4,\bar\eta_4\}) =
M_2(\ell_3,j+1,\ldots ,i-1,\{ \ell_4,\eta_i+z\eta_{i+1}\})
\end{eqnarray}
where in the last equation we performed the $\bar\eta_4$ integration
to go back to the $\eta$ representation but in terms of the new
variable $\eta_i+z\eta_{i+1}$. Combining all these terms we find
the contribution from the first solution to be
\begin{equation}
\frac{1}{2}\sum_{j=i+2}^{n+i-2} \int d^4\eta M_1(\ell_2,\eta_{i+1}\},i+2,\ldots , j,\{\ell_3,\eta\})\frac{1}{P^2_j} M_2(\{\ell_3,\eta\},j+1,\ldots ,i-1,\{ \ell_4,\eta_i+z\eta_{i+1}\})
\end{equation}
with $P_j^2 = (k_{i+1}+k_{i+2}+\ldots +k_{i-1})^2$. Note that the
factor $P_j^2$ is the original one in (\ref{compi}) while the factor
of $(k_i+k_{i+1})^2$ in the denominator is canceled by the $\langle
i, i+1\rangle[i,i+1]$ factor left over from the Jacobians.

A similar computation can be done for the second solution in $\cal
S$, {\it i.e.}, $\ell=z\lambda_{i+1}\tilde\lambda_{i}$. This would
give us the ``other" form of the recursion relation, with the roles
of $\eta_i$,$\eta_{i+1}$ reversed. Adding the two contributions
gives the manifestly PT invariant supersymmetric form of the
recursion relations. It is important to mention that nothing in this
argument can explain the identity (\ref{equality}), i.e., that the
two terms are equal to each other an hence equal to the original
tree amplitude. As mentioned in section 3, this is a remarkable fact
which is a consequence of the large $z$ behavior and not of PT
invariance. This shows that the proposal made in section 5.1
of~\cite{BCF} to prove the recursion relations for gluons, i.e., to
use supersymmetric Ward identities to derive (\ref{equality}),
cannot be carried out.

To conclude this appendix, we show some details on the way to prove that under a PT transformation the contributions from the
two solutions get exchanged. This is also a nice consistency condition for the supersymmetric BCF recursion relations presented in section 3. In fact the computation we present now uses the notation in section 3, more explicitly, we use the forms given in (\ref{equality}).

Consider the second form of the recursion relations in (\ref{equality})
\begin{equation}
\sum_{L,R}\int d^{4} \eta
M_L(\{\eta_1,\lambda_1,\bar\lambda_1(z_{P_R}) \}, \eta, \eta_L) \, \frac{1}{P_R^2}
M_R(\{\eta_2(z_{P_R}),\lambda_2(z_{P_R}),\bar\lambda_2\},\eta,\eta_R)
\end{equation}

PT invariance of each of the individual physical amplitudes gives
\begin{equation}
M_L(\{\eta_1,\lambda_1,\bar\lambda_1-z_{P_R}\bar\lambda_2 \}) = \int d^4\bar\eta_1 e^{\bar\eta_1\eta_1}M_L(\{\bar\eta_1,\bar\lambda_1-z_{P_R}\bar\lambda_2,\lambda_1 \})
\end{equation}
and
\begin{equation}
M_R(\{\eta_2+z_{P_R}\eta_1,\lambda_2+z_{P_R}\lambda_1,\bar\lambda_2\},\eta,\eta_R) = \int d^4\bar\eta_2 e^{\bar\eta_2(\eta_2+z_{P_R}\eta_1)}
M_R(\{ \bar\eta_2,\bar\lambda_2,\lambda_2+z_{P_R}\lambda_1\}).
\end{equation}
Note that in both expressions we have suppressed the transformation on the remaining particles since it is straightforward how to write down the full expression.

Collecting the terms in the exponential that are proportional to $\eta_1$ we find $(\bar\eta_1+z_{P_R}\bar\eta_2)\eta_1$. This suggest that it is natural to perform the following change of variables, $\bar\eta \to \bar\eta_1-z_{P_R}\bar\eta_2$, in the $\bar\eta_1$ integration. The jacobian is unity and we find
\begin{equation}
\label{fasi}
\sum_{L,R}\int d^{\cal N} \eta \int d^4\bar\eta_1d^4\bar\eta_2 e^{\bar\eta_1\eta_1+\bar\eta_2\eta_2}M_L(\{\bar\eta_1-z_{P_R}\bar\eta_2,\bar\lambda_1-z_{P_R}\bar\lambda_2,\lambda_1 \})M_R(\{ \bar\eta_2,\bar\lambda_2,\lambda_2+z_{P_R}\lambda_1\})
\end{equation}
Now it is important to use that the first form of the recursion relations by itself is equal to the full physical amplitude, $M_n$, i.e,
\begin{eqnarray}
\label{oles}
& M(\{\eta_1,\lambda_1,\bar\lambda_1\},\{\eta_2,\lambda_2,\bar\lambda_2\}) = &  \\ & \sum_{L,R}\int d^{\cal N} \eta M_L(\{\eta_1+z_{P_L}\eta_2,\lambda_1+z_{P_L}\lambda_2,\bar\lambda_1\}, \eta,
\eta_L) \, \frac{1}{P_L^2}
M_R(\{\eta_2,\lambda_2,\bar\lambda_2-z_{P_L}\bar\lambda_1\},\eta,\eta_R) \nonumber
\end{eqnarray}
so that by writing
\begin{equation}
M(\{\eta_1,\lambda_1,\bar\lambda_1\},\{\eta_2,\lambda_2,\bar\lambda_2\})  = \int d^4\bar\eta_1 d^4\bar\eta_2 e^{\bar\eta_1\eta_1+\bar\eta_2\eta_2} M(\{\bar\eta_1,\bar\lambda_1,\lambda_1\},\{\bar\eta_2,\bar\lambda_2,\lambda_2\})
\end{equation}
which is true for any physical amplitude in ${\cal N}=4$, and using (\ref{oles}) for the amplitude on the RHS one immediately finds (\ref{fasi}). In making the identification one has to realize that $z_{P_R}(\lambda,\bar\lambda) = -z_{P_L}(\bar\lambda,\lambda)$ since $P_R = -P_L$.

\subsection{${\cal N}=8$ SUGRA}

The IR behavior of gravitational amplitudes is also well-known:
\begin{equation}
\left. M^{\rm 1 \hbox{-} loop}_{IR} \right. =
-\frac{1}{\epsilon^2}\sum_{i,j} (-s_{ij})^{1-\epsilon} M^{\rm tree}.
\label{known}
\end{equation}
Note that unlike Yang-Mills, the leading $1/\epsilon^2$ term
cancels since $\sum_{i,j} s_{ij} = 0$.

Just as in SYM, given that any one-loop amplitude can be written in
terms of only scalar box integrals, matching the IR divergences
provides equations among the coefficients of the scalar boxes.

We have studied the case of $n=5$ external particles explicitly, to
see if some combination of the ${\cal N} = 8$ IR equations is
equivalent to our recursion relation. We have found that no formula
analogous to (\ref{compi}) can be derived from the IR equations.
However, from our calculation in SYM and knowing that there exists
recursion relations for ${\cal N}=8$ supergravity it is natural to
start with the following anzats which is a relation among
coefficients of one-mass and two-mass hard scalar box integrals
\begin{equation}
M^{\rm tree}_n = \frac{1}{2}\sum_{U,D} \sum_{\cal S}\sum_{\rm
Multiplet}\frac{M_1(\ell_2,U,\ell_3)M_2(\ell_3,D,\ell_4)M_3(\ell_4,i,\ell_1)M_4(\ell_1,j,\ell_2)}{s_{ij}^2
t_{U,i}} \label{cogra}
\end{equation}
where $U,D$ are two non-empty sets such that their union together with $\{i,j\}$ gives all external particle labels.
$s_{ij}=(k_i+k_j)^2$ and $t_{U,i} = (k_i+\sum_{m\in U}k_m)^2$.%

Following the same argument as before but this time using the
formulas for the three-particle amplitudes in gravity
one finds that the first three particle amplitude gives a factor of $z^2[i,j]^2$ while the
second one gives $\langle i,j\rangle^2/z^2$. Combining the two
factors gives $s_{i,j}^2$ which cancels the explicit factor in the
denominator. The rest follows in complete analogy with the
Yang-Mills case and leads to manifestly PT invariant form of the recursion relation.

From our study of $n=5$ we conclude that the relation among
coefficients implied by the recursion relations is in fact a new set
of relations! For $n=5$, there are 30 independent scalar box
coefficients possible. We have used the ten IR equations to
determine ten out of the thirty independent coefficients and then
checked that the recursion relations provide nine independent
equations, allowing the computation of $19$ coefficients. This
difference between SYM and SUGRA may not be surprising; recall that
in SUGRA amplitudes fall off as $\frac{1}{z^2}$ which means that
there are further relations among coefficients. We have found that
these extra relations further fix 18 of the remaining 19
coefficients, leaving only one free! We will return to examine these
very interesting relations at greater length elsewhere.

\section{Proof of IR Equation}

In appendix A, we showed how the PT symmetric form of the BCF
recursion relations can be derived from the known IR behavior of
1-loop amplitudes in ${\cal N}=4$ super Yang-Mills. An important
ingredient in the proof is an equation derived from
\begin{equation}
\left. M^{\rm 1 \hbox{-} loop} \right|_{IR} =
-\frac{1}{\epsilon^2}\sum_{i=1}^n (-s_{i,i+1})^{-\epsilon} M^{\rm
tree}. \label{api}
\end{equation}
by using that $M^{\rm 1 \hbox{-} loop}$ is written as a linear
combination of scalar box integrals whose coefficients are
determined using quadruple cuts as the product of tree amplitudes.

Each box integral has a particular set of IR divergences and by
imposing that the sum matches (\ref{api}) one finds many equations
relating the coefficients of the boxes to the tree level of the full
amplitude. In~\cite{RSV} it was stated that by taking a particular
linear combination of the equations one can find an equation which
relates $A^{\rm tree}$ to a sum over only one-mass and two-mass-hard
box integrals, all of them having an $s$-channel like singularity of
the form $(-(k_i+k_{i+1})^2)^{-\epsilon}$.

In this appendix we present explicitly the particular linear combination which gives rise to the equation and the proof that indeed does so for any $n$.

Let us explain how the equations are obtained and then we turn to
the particular linear combination of interest. Denote each box
integral by $I(K_1,K_2,K_3,K_4)$ and the main kinematical invariants
by $s=(K_1+K_2)^2$ and $t=(K_2+K_3)^2$. Each $K_i$ is the sum of
consecutive external momenta. If $K_i$ is equal to the momentum of a
single particle then it is null, if it is the sum of two or more
then we say that it is massive. We denote null momenta by lower case
letters $(p,q,r)$ while massive momenta by capital letters
$(P,Q,R)$. Clearly, integrals with all four massive legs are
completely finite and will not appear in our discussion. Integrals
with at least one massless leg are usually classified into four
classes. These are known as: One-mass, two-mass-easy, two-mass-hard
and three-mass. This is according to the number of massive legs and
their relative location. Using our notation these are respectively
given by: $I(p,q,r,P)$, $I(p,P,q,Q)$, $I(p,q,P,Q)$, and
$I(p,P,R,Q)$.

We now list the IR divergent structure of each of the four classes:
\begin{eqnarray}
\label{rhs}
I(p,q,r,P) & = & -\frac{1}{\epsilon^2}\left( (-s)^{-\epsilon} + (-t)^{-\epsilon} - (-P^2)^{-\epsilon} \right) \cr
I(p,P,q,Q) & = & -\frac{1}{\epsilon^2}\left( (-s)^{-\epsilon} + (-t)^{-\epsilon} - (-P^2)^{-\epsilon} - (-Q^2)^{-\epsilon}\right) \cr
I(p,q,P,Q) & = & -\frac{1}{\epsilon^2}\left( \frac{1}{2}(-s)^{-\epsilon}+(-t)^{-\epsilon}-\frac{1}{2}(-P^2)^{-\epsilon}-\frac{1}{2}(-Q^2)^{-\epsilon}\right)\cr
I(p,P,R,Q) & = &  -\frac{1}{\epsilon^2}\left( \frac{1}{2}(-s)^{-\epsilon} + \frac{1}{2}(-t)^{-\epsilon} - \frac{1}{2}(-P^2)^{-\epsilon} - \frac{1}{2}(-Q^2)^{-\epsilon}\right).
\end{eqnarray}

Let us denote the coefficients of each integral $I$ by $B$. For
example, the coefficient of $I(p,q,r,P)$ is $B(p,q,r,P)$. Now we use
the fact that any 1-loop amplitude can be written as
\begin{equation}
\label{sumi} M^{\rm 1 \hbox{-} loop} = \sum_{i} B_i I_i
\end{equation}
where $i$ runs over the set of all distinct scalar boxes that can be constructed with $n$ momenta and respecting the color ordering.

Imposing the condition (\ref{api}) on (\ref{sumi}) and using (\ref{rhs}) one finds linear equations among the coefficients $B_i$.

Consider for example the equation that comes from collecting all terms with $ -\frac{1}{\epsilon^2}(-(k_1+k_2)^2)^{-\epsilon}$. Here we denote the momenta of external particles by $k_i$. The equation one gets is of the form
\begin{eqnarray}
\label{quqi}
M^{\rm tree} & = & B(k_1,k_2,k_3,k_4+\ldots +k_n)+B(k_n,k_1,k_2,k_3+\ldots +k_{n-1}) \nonumber \\ && - B(k_n,k_1+k_2,k_3,k_4+\ldots k_{n-1}) + \ldots.
\end{eqnarray}
The first two coefficients are from the only two one-mass integrals which can have the corresponding singularity while the third term is the only two-mass-easy which does. The ellipses represent the many two-mass-hard and three-mass integrals that contribute.

One more example is the equation coming from $-\frac{1}{\epsilon^2}(-(k_1+k_2+k_3)^2)^{-\epsilon}$. This time the lhs is zero,
\begin{equation}
0 = -B(k_1,k_2,k_3,k_4+\ldots +k_n) + B(k_n,k_1+k_2,k_3,k_4+\ldots k_{n-1}) + \ldots.
\end{equation}
Note that here two of the coefficients which appeared in the first equation also appear here but with different signs. This is due to the fact that in (\ref{rhs}) singular terms come with different prefactors.

Now we are ready to state the result we want to prove.

Denote by $[k_i+k_{i+1}+\ldots +k_{j}]$ the equation obtained by collecting all terms proportional to $ -\frac{1}{\epsilon^2}(-(k_i+k_{i+1}+\ldots +k_{j})^2)^{-\epsilon}$. Then
\begin{equation}
\label{solu}
\sum_{j=i+1}^{n+(i-2)}[k_i+k_{i+1}+\ldots +k_{j}]
\end{equation}
gives
\begin{equation}
\label{gala}
2M^{\rm tree} = \sum_{j=i+1}^{n+(i-2)} B(k_i,k_{i+1},k_{i+2}+\ldots k_{j},k_{j+1}+\ldots + k_{i-1}).
\end{equation}
Clearly the coefficients appearing here are those of one-mass and two-mass-hard integrals only.

The proof of (\ref{gala}) is extremely complicated if one tries to directly write down the explicit equations entering in (\ref{solu}) and then find a pattern for adding them up as our attempt at writing explicit equations in (\ref{quqi}) demonstrated. There is something simple we can learn though from straightforwardly adding the equations. This is the lhs of (\ref{gala}). Note that out of all the equations entering (\ref{solu}) only two come from invariants in two-particle channels, {\it i.e}, $[k_i,k_{i+1}]$ and $[k_i+\ldots +k_{n+(i-2)}] = [k_{i-2}+k_{i-1}]$ (where in the last equality we used momentum conservation). According to (\ref{api}) each of these two equations gives $M^{\rm tree}$ while the rest contributes zero to the lhs of (\ref{gala}) and hence the $2M^{\rm tree}$ term.

The strategy we will follow to prove the rhs of (\ref{gala}) is very simple. We will prove that each two-mass-easy that can possibly contribute to one of the equations in (\ref{solu}) also appears in a second equation but with a different sign. This directly means that no two-mass-easy coefficient can appear in (\ref{gala}). The same will hold for three-mass coefficients. The proof of these statements will show that they drop out since their IR behavior is ``complete" in a sense we will make clear below. One the other hand, one-mass and two-mass-hard integrals are ``incomplete" and this is why they contribute.

In order to keep the notation simple we set $i=1$ without loss of generality since all arguments are cyclically symmetric.

\subsection{Absence of Two-Mass-Easy Coefficients}

Consider a two-mass-easy coefficient $B(p,P,q,Q)$. One way it can appear in (\ref{solu}) is if $P=k_1+k_2+\ldots$. Then it gives a $-1$ contribution to the equation $[P]$ while a $+1$ contribution to $[P+q]$ and hence they cancel. The second way it can appear is if $p=k_1$. Then it gives a $+1$ contribution to $[p+P]$ while a $-1$ contribution to $[p+P+q]$. This completes the proof.

\subsection{Absence of Three-Mass Coefficients}

Consider a three-mass coefficient $B(p,P,R,Q)$.
The first way it can appear is if $P=k_1+k_2+\ldots$.
Then it gives $-1/2$ to $[P]$ while $+1/2$ to $[P+R]$ and hence
 it cancels. The second way is if $p=k_1$. Then it gives $+1/2$ to $[p+P]$ while $-1/2$ to $[p+P+R]$. A third way is if $Q=k_1+k_2+\ldots$. Then it gives $-1/2$ to $[Q]$ while $+1/2$ to $[Q+p]$. Finally, there is a fourth way; this is if $R=k_1+k_2+\ldots$. Then it gives $+1/2$ to $[R+Q]$ while $-1/2$ to $R+Q+p$. Therefore there is no contribution to (\ref{solu}).

\subsection{Contribution from One-Mass Integrals}

Consider a one-mass coefficient $B(p,q,r,P)$. The first way it can appear is if $p=k_1$. Then it gives $+1$ to $[p+q]$ and $-1$ to $[p+q+r]$ and hence it cancels. The second way is if $q=k_1$. Then it gives $+1$ in $[q+r]$. Clearly this integral cannot contribute to any other equation and hence it is the first time we get a non-zero contribution. The third way it can appear is if $r=k_1$. Then it gives $+1$ to $[r+P]$. This is also a non-zero contribution since it cannot appear in any other equation. Finally, there is a fourth way it can appear. This is if $P=k_1+k_2+\ldots$ Then it gives $-1$ to $[P]$ while $+1$ to $[P+p]$ and hence it cancels.

Summarizing the one-mass contribution we have
$$B(k_n,k_1,k_2,k_3+\ldots k_{n-1}) + B(k_{n-2},k_{n-1},k_1,k_2+\ldots +k_{n-3}).$$

These are precisely the contributions of one-mass integrals in (\ref{gala}).

\subsection{Contributions from Two-Mass-Hard Integrals}

Consider a two-mass-hard coefficient $B(p,q,P,Q)$. The first way it can appear is if $p=k_1$. Then it gives $+1/2$ in $[p+q]$ and $-1/2$ in $[p+q+P]$ and hence it cancels. The second way is if $q=k_1$. This gives $+1$ to $[q+P]$. This integrals do not contribute to any other equations and hence give a nonzero contribution. The third case is if $P=k_1+k_2+\ldots$. Then it gives $-1/2$ to $[P]$ and $1/2$ to $[P+Q]$ and hence it cancels. Finally, there is fourth way and this is if $Q=k_1+k_2+\ldots$. Then it gives $-1/2$ to $[Q]$, $+1$ to $[Q+p]$ and $-1/2$ to $[Q+p+q]$ and hence it cancels.

Summarizing the two-mass contributions we have
\begin{equation}
\sum_{j=3}^{n-3}B(k_m,k_1,k_2+\ldots +k_j,k_{j+1}+\ldots + k_{n-1})
\end{equation}
and this is the final piece that gives (\ref{gala}).

This concludes the proof.

\section{Scalar Integrals}

Throughout the paper we have referred to scalar box, triangle and
bubble integrals. In this appendix we list their expansion in
$\epsilon$ for the reader's convenience. The expansions are through
${\cal O}(\epsilon^0)$ and are taken from appendix IV of
\cite{boxfunctions} with some rewriting of $\rm log$ terms.

Also, in section 4.2.3, we argued the absence of rational terms by using that scalar boxes possess a particular property when expanded in $\epsilon$. All those properties can be checked by inspection of the formulas below.

In general a scalar integral is defined by
\begin{equation}
I_n = (-1)^{n+1}i (4\pi)^{2-\epsilon}\int \frac{d^{4-\epsilon}L}{(2\pi)^{4-2\epsilon}}\frac{1}{L^2\prod_i^{n-1}(L-P_i)^2}
\end{equation}
with $P_i$ the sums of external momenta. If we label the momenta
going out of the vertex where the $1/L^2$ propagator ends by $K_1$
and the next by $K_2$ and so on then $P_i = K_1+\ldots + K_i$.

Let us start giving the explicit form of the integrals in terms of logarithms and polylogarithms. We start with the scalar bubble
\begin{equation}
I_2(K^2) = \frac{r_\Gamma}{\epsilon(1-2\epsilon)}(-K^2)^{-\epsilon},
\end{equation}
where
\begin{equation}
r_\Gamma = \frac{\Gamma(1+\epsilon)\Gamma^2(1-\epsilon)}{\Gamma(1-2\epsilon)}.
\end{equation}

Triangle integrals are given by
\begin{eqnarray}
&& I_3^{\rm 3m}(K_1^2,K_2^2,K_3^3)  = \frac{i}{\Delta}\sum_{i=j}^3\left[{\rm Li}_2\left(-\frac{1+i\delta_j}{1-i\delta_j}\right) - {\rm Li}_2\left(-\frac{1-i\delta_j}{1+i\delta_j}\right)\right], \\
&&
I^{\rm 2m}_{3}(K_1^2,K_2^2) = \frac{r_\Gamma}{\epsilon^2}\frac{(-K_1^2)^{-\epsilon}-(-K_2^2)^{-\epsilon}}
{(-K_1^2)-(-K_2^2)}, \\
&&
I^{\rm 1m}_3(K^2) = \frac{r_\Gamma}{\epsilon^2}(-K^2)^{-1-\epsilon}
\end{eqnarray}
where
\begin{equation}
\delta_1 = \frac{K_1^2-K_2^2-K_3^2}{\Delta},\quad \delta_2 = \frac{-K_1^2+K_2^2-K_3^2}{\Delta}, \quad \delta_3 = \frac{-K_1^2-K_2^2+K_3^2}{\Delta},
\end{equation}
and
\begin{equation}
\Delta = -(K_1^2)^2 -(K_2^2)^2-(K_3^2)^2 + 2K_1^2K_2^2+2K_2^2K_3^2+2K_3^2K_1^2.
\end{equation}

Finally we have the box integrals. Since our main concern in the paper is with the branch cuts and structure of logarithms and dilogarithms we have chosen to write scalar box {\it functions}, $F$, which are related to the actual box integrals, $I_4$, by a factor which turns out to be the jacobian of the change of variables to evaluate a quadruple cut. In other words,
\begin{equation}
I_4 = -\frac{r_\Gamma}{2\sqrt{{\rm det} S}} F_4
\end{equation}
where $S$ us a symmetric $4\times 4$ matrix with components
\begin{equation}
S_{ij} = -\frac{1}{2}(K_i+\ldots + K_{j-1})^2, ~i\neq j; ~~ S_{ii} = 0.
\end{equation}
The box functions are then given by
\begin{eqnarray}
\label{boxie}
F^{4m}(K_1,K_2,K_3,K_4) & = & \frac{1}{2}\left( -{\rm Li}_2((1-\lambda_1+\lambda_2+\rho)/2)+{\rm Li}_2((1-\lambda_1+\lambda_2-\rho)/2)\right. \nonumber \\
& & -{\rm Li}_2(-(1-\lambda_1-\lambda_2-\rho)/(2\lambda_1))+{\rm Li}_2(-(1-\lambda_1-\lambda_2+\rho)/(2\lambda_1)) \nonumber \\
& & \left. -\frac{1}{2}\ln \left(\frac{\lambda_1}{\lambda_2^2}\right)\ln\left( \frac{1+\lambda_1-\lambda_2+\rho}{1+\lambda_1-\lambda_2-\rho} \right) \right) , \nonumber \\
F^{3m}(k_1,K_2,K_3,K_4) & = & -\frac{1}{2\epsilon^2}\left( (-s)^{-\epsilon}+ (-t)^{-\epsilon}-(-K_2^2)^{-\epsilon}-(-K_4^2)^{-\epsilon}\right) \nonumber \\
& & + {\rm Li}_2\left( 1-\frac{K_2^2}{s}\right)+ {\rm Li}_2\left( 1-\frac{K_4^2}{t}\right) - {\rm Li}_2\left( 1-\frac{K_2^2K_4^2}{st}\right) \nonumber \\ & & + \frac{1}{2}\ln^2\left(\frac{s}{t}\right) -  \frac{1}{2}\ln\left(\frac{K_4^2}{s}\right)\ln\left(\frac{K_3^2}{s}\right)-
 \frac{1}{2}\ln\left(\frac{K_2^2}{t}\right)\ln\left(\frac{K_3^2}{t}\right), \nonumber \\
F^{2m\, h}(k_1,k_2,K_3,K_4) & = & -\frac{1}{2\epsilon^2}\left( (-s)^{-\epsilon}+ 2(-t)^{-\epsilon}-(-K_3^2)^{-\epsilon}-(-K_4^2)^{-\epsilon}\right) \nonumber \\
& & + {\rm Li}_2\left( 1-\frac{K_3^2}{t}\right)+ {\rm Li}_2\left( 1-\frac{K_4^2}{t}\right)+ \frac{1}{2}\ln^2\left(\frac{s}{t}\right)- \frac{1}{2}\ln\left(\frac{K_4^2}{s}\right)\ln\left(\frac{K_3^2}{s}\right), \nonumber \\
F^{2m\, e}(k_1,K_2,k_3,K_4) & = & -\frac{1}{\epsilon^2}\left( (-s)^{-\epsilon}+ (-t)^{-\epsilon}-(-K_2^2)^{-\epsilon}-(-K_4^2)^{-\epsilon}\right) \nonumber \\ & & + {\rm Li}_2\left( 1-\frac{K_2^2}{s}\right) + {\rm Li}_2\left( 1-\frac{K_2^2}{t}\right)+ {\rm Li}_2\left( 1-\frac{K_4^2}{s}\right)+ {\rm Li}_2\left( 1-\frac{K_4^2}{t}\right)\nonumber \\
& & - {\rm Li}_2\left( 1-\frac{K_2^2K_4^2}{st}\right)+\frac{1}{2}\ln^2\left(\frac{s}{t}\right) , \nonumber\\
F^{1m}(k_1,k_2,k_3,K_4) & = & -\frac{1}{\epsilon^2}\left( (-s)^{-\epsilon}+ (-t)^{-\epsilon}-(-K_4^2)^{-\epsilon}\right) \nonumber \\ & &
+{\rm Li}_2\left( 1-\frac{K_4^2}{s}\right)+{\rm Li}_2\left( 1-\frac{K_4^2}{t}\right)+\frac{1}{2}\ln^2\left(\frac{s}{t}\right)+\frac{\pi^2}{6}.
\end{eqnarray}
where $k_i$ denotes the on-shell momenta and the $K_i$ off-shell momenta. The definition of the functions appearing in four-mass case is
\begin{equation}
\rho = \sqrt{1-2(\lambda_1+\lambda_2)+(\lambda_1-\lambda_2)^2} \quad {\rm with}\quad \lambda_1 = \frac{K_1^2K_3^2}{st}, \;\; {\rm and} \;\; \lambda_2 = \frac{K_2^2K_4^2}{st}.
\end{equation}

For completeness let us write two equivalent definitions of the dilogarithm function
\begin{equation}
{\rm Li}_2(z) = \sum_{k=1}^{\infty}\frac{z^k}{k^2}, \qquad {\rm Li}_2(z) = -\int_0^z \frac{dt}{t} \ln (1-t).
\end{equation}

\section{Explicit Computation of a Bubble Coefficient}

In section 5.5, we gave a general formula for the computation of
bubble coefficients in any theory. We also discussed an example in
pure Yang-Mills where the coefficient of the bubble was related to
the beta function of the theory. In this appendix we give the
details the computation that led to the result.

Consider the four-gluon amplitude $M(1^-,2^-,3^+,4^+)$. We are
interested in the coefficient of the bubble integral $I_2(s_{14})$.
Using the general formula for the coefficient of a bubble in this
case we find
\begin{equation}
C_2 = \int d{\rm LIPS} \int_{\cal C} \frac{dz}{z} \sum_{h=\pm}
M(\ell_2^{-h}(z),4^+,1^-,\ell_1^{h}(z))M(\ell_1^{-h}(z),2^-,3^+,\ell_2^{h}(z))
\end{equation}
where the sum if over the helicity configurations of the internal
gluons which give non-vanishing contributions. We will take the BCFW
deformation with $\lambda_{\ell_2}(z) = \lambda_{\ell_2}$ while
$\lambda_{\ell_1}(z) = \lambda_{\ell_1}+z\lambda_{\ell_2}$.

Let us denote the contribution from $h=+$ ($h=-$) by $C_2^+$
($C_2^-$). Let us consider first $C^+$. It is straightforward to
compute the amplitudes to get
\begin{equation}
C_2^+ = \int d{\rm LIPS} \int_{\cal C} \frac{dz}{z} \frac{\langle
\ell_2~1\rangle^4}{\langle 1~\ell_1(z)\rangle\langle
\ell_1~\ell_2\rangle\langle \ell_2~4\rangle\langle 4~1\rangle}\times
\frac{\langle\ell_1(z)~2\rangle^3}{\langle2~3 \rangle\langle
3~\ell_2\rangle\langle \ell_2~\ell_1(z)\rangle}
\end{equation}
Performing the $z$ integral is the same as extracting the constant
term in a Laurent series around $z=\infty$ of the rational function
of $z$ coming from the product of the amplitudes. The rational
function of $z$ is explicitly,
\begin{equation}
\frac{(\langle \ell_1~2\rangle+z\langle \ell_2~2\rangle)^3}{(\langle
\ell_1~1\rangle+z\langle \ell_2~1\rangle)}
\end{equation}
Extracting the constant term gives
\begin{eqnarray}
C_2^+ & =  & \int d{\rm LIPS}\frac{1}{\langle
\ell_1~\ell_2\rangle^2\langle \ell_2~4\rangle\langle
1~4\rangle\langle 2~3\rangle\langle 3~\ell_2\rangle}\times \nonumber
\\ & & (3\langle \ell_1~2\rangle^2\langle \ell_2~2\rangle\langle
\ell_2~1 \rangle^3-3\langle 1~\ell_1 \rangle\langle \ell_1~2
\rangle\langle \ell_2~2 \rangle^2\langle \ell_2~1 \rangle^2+\langle
1~\ell_1 \rangle^2\langle \ell_2~2 \rangle^3\langle \ell_2~1
\rangle).
\end{eqnarray}
In order to carry out the integration over $d{\rm LIPS}$ we will use
a representation for it as a contour integral in $\mathbb{R}^+\times
\mathbb{CP}^1\times \mathbb{CP}^1$ \cite{CSW,dLIPS1,dLIPS2}. This is
obtained by parameterizing $\ell_2 = t\lambda\tilde\lambda$. Here
$\lambda$ and $\tilde\lambda$ are homogenous coordinates of a
different $\mathbb{CP}^1$ and the contour of integration is the
diagonal $\mathbb{CP}^1$, i.e, $\tilde\lambda = \bar\lambda$. More
explicitly,
\begin{eqnarray}
&& \int d{\rm LIPS} = \int d^4\ell_1d^4\ell_2
\delta^+(\ell_1^2)\delta^+(\ell_2^2)\delta^{(4)}(\ell_1+\ell_2-P) =
\\ && \int_0^\infty tdt\int_{\tilde\lambda =
\bar\lambda}\langle\lambda ,d\lambda\rangle
[\tilde\lambda,d\tilde\lambda]\theta((t\lambda\tilde\lambda +
P)^0)\delta(t\langle \lambda|P|\tilde\lambda]-P^2) =
P^2\int_{\tilde\lambda = \bar\lambda} \frac{\langle\lambda
,d\lambda\rangle [\tilde\lambda,d\tilde\lambda]}{\langle
\lambda|P|\tilde\lambda]^2}\nonumber
\end{eqnarray}
where $P=p_1+p_4$ and in the last inequality the delta function was
used to perform the $t$ integration. This localizes $t=P^2/\langle
\lambda|P|\tilde\lambda]$. Using this representation is useful
because it allows the evaluation of the phase space integral as a
purely algebraic procedure. Using the new parametrization we find
that
\begin{eqnarray}
\label{cplus} C^+ & =  & \frac{[1~4]}{\langle 2~3\rangle}\int
\frac{\langle\lambda ,d\lambda\rangle
[\tilde\lambda,d\tilde\lambda]}{\langle
\lambda|P|\tilde\lambda]^4}\times \\ && \frac{\left(
3[\tilde\lambda~3]^2\langle \lambda~2\rangle\langle \lambda~1
\rangle^3\langle 3~2\rangle^2 +
3[\tilde\lambda~4][\tilde\lambda~3]\langle\lambda~2 \rangle^2\langle
\lambda~1\rangle^2\langle 4~1\rangle\langle 3~2\rangle
+[\tilde\lambda~4]^2\langle\lambda~2 \rangle^3\langle
\lambda~1\rangle\langle 4~1\rangle^2\right)}{\langle \lambda
~4\rangle \langle \lambda ~ 3\rangle} \nonumber
\end{eqnarray}

All three terms have the same form,
\begin{equation}
\label{jiko} I = \int \frac{\langle\lambda ,d\lambda\rangle
[\tilde\lambda,d\tilde\lambda]}{\langle
\lambda|P|\tilde\lambda]^4}[\tilde\lambda~\mu][\tilde\lambda~\rho]
g(\lambda)
\end{equation}
where $g(\lambda)$ is a rational function of $\lambda$ with only
simple poles. In our case the simple poles are $\langle \lambda
~4\rangle$ and $\langle \lambda ~ 3\rangle$. This integral can be
done by noting first that
\begin{equation}
\frac{[\tilde\lambda ,d\tilde\lambda]}{[\tilde P ~
\tilde\lambda]^2}g(\lambda) = -d\tilde\lambda^{\dot
c}\frac{\partial}{\partial \tilde\lambda^{\dot c}}\left(
\frac{[\tilde\lambda,\eta]}{[\tilde P~\tilde\lambda][\tilde
P~\eta]}g(\lambda) \right) \label{valo}\end{equation}
where $\eta$ is some auxiliary spinor and $\tilde{P}_{\dot{\alpha}}
= \lambda^\alpha P_{\alpha \dot{\alpha}}$. Using that
\begin{equation}
\frac{1}{6}\rho^{\dot c}\mu^{\dot d}\frac{\partial^2}{\partial
\tilde P^{\dot c}\partial \tilde P^{\dot d}}\left(\frac{1}{[\tilde P
~ \tilde\lambda]^2}\right) =
\frac{[\tilde\lambda~\mu][\tilde\lambda~\rho]}{[\tilde P ~
\tilde\lambda]^4}.
\end{equation}
one immediately finds that
\begin{eqnarray}
\label{gami}
& & \frac{[\tilde\lambda,d\tilde\lambda]}{\langle \lambda|P|\tilde\lambda]^4}[\tilde\lambda~\mu][\tilde\lambda~\rho] g(\lambda) = \\
& & -d\tilde\lambda^{\dot c}\frac{\partial}{\partial
\tilde\lambda^{\dot c}}\left[[\tilde\lambda ~\eta]\left(
\frac{1}{3}\frac{[\mu~\tilde\lambda][\rho~\tilde\lambda]}{[\tilde
P~\tilde\lambda]^3[\tilde
P~\eta]}+\frac{1}{6}\frac{[\mu~\tilde\lambda][\rho~\eta]}{[\tilde
P~\tilde\lambda]^2[\tilde P~\eta]^2}+
\frac{1}{6}\frac{[\mu~\eta][\rho~\tilde\lambda]}{[\tilde
P~\tilde\lambda]^2[\tilde
P~\eta]^2}+\frac{1}{3}\frac{[\mu~\eta][\rho~\eta]}{[\tilde
P~\tilde\lambda][\tilde P~\eta]^3}\right)g(\lambda)
\right].\nonumber
\end{eqnarray}

If this equation was valid along the whole contour of integration we
would find that the integral is zero. A closer look at the formulas
reveals that the only places where this formula fails to be true is
at the location of the poles of $g(\lambda)$. This means that the
integrals of the form (\ref{jiko}) localize on the poles.

To correct (\ref{gami}) one has to notice that if the pole is
located at $\langle\lambda,\zeta\rangle =0$ then
\begin{equation}
-d\tilde\lambda^{\dot c}\frac{\partial}{\partial \tilde\lambda^{\dot
c}}\frac{1}{\langle\lambda,\zeta\rangle} = 2\pi \bar\delta
(\langle\lambda,\zeta\rangle)
\end{equation}
where the delta function is defined such that
\begin{equation}
\int \langle\lambda,d\lambda\rangle \bar\delta
(\langle\lambda,\zeta\rangle) H(\lambda) = -iH(\zeta)
\end{equation}
Applying these equations to the computation of $C^+$ in
(\ref{cplus}) one finds that
\begin{equation}
\frac{[1~4]}{\langle 2~3\rangle}\int \frac{\langle\lambda
,d\lambda\rangle [\tilde\lambda,d\tilde\lambda]}{\langle
\lambda|P|\tilde\lambda]^4}\frac{\left( 3[\tilde\lambda~3]^2\langle
\lambda~2\rangle\langle \lambda~1 \rangle^3\langle
3~2\rangle^2\right)}{\langle \lambda ~4\rangle \langle \lambda ~
3\rangle} = M^{\rm tree}
\end{equation}
while
\begin{equation}
\frac{[1~4]}{\langle 2~3\rangle}\int \frac{\langle\lambda
,d\lambda\rangle [\tilde\lambda,d\tilde\lambda]}{\langle
\lambda|P|\tilde\lambda]^4}\frac{\left(3[\tilde\lambda~4][\tilde\lambda~3]\langle\lambda~2
\rangle^2\langle \lambda~1\rangle^2\langle 4~1\rangle\langle
3~2\rangle \right)}{\langle \lambda ~4\rangle \langle \lambda ~
3\rangle} = \frac{1}{2}M^{\rm tree}
\end{equation}
and finally
\begin{equation}
\frac{[1~4]}{\langle 2~3\rangle}\int \frac{\langle\lambda
,d\lambda\rangle [\tilde\lambda,d\tilde\lambda]}{\langle
\lambda|P|\tilde\lambda]^4}\frac{\left([\tilde\lambda~4]^2\langle\lambda~2
\rangle^3\langle \lambda~1\rangle\langle
4~1\rangle^2\right)}{\langle \lambda ~4\rangle \langle \lambda ~
3\rangle} = \frac{1}{3}M^{\rm tree}.
\end{equation}
where
\begin{equation}
M^{\rm tree}(1^-,2^-,3^+,4^+) = \frac{\langle 1~2\rangle^3}{\langle
2~3\rangle\langle 3~4\rangle\langle 4~1\rangle}.
\end{equation}
Combining all contributions we find that $C^+ = 11 M^{\rm tree}/6$.
Further noticing that $C^+$ and $C^-$ are related by a relabeling of
the external particles which is a symmetry of $M^{\rm tree}$ one
find that
\begin{equation}
C = C^+ + C^- = \frac{11}{3}M^{\rm tree}.
\end{equation}
\section{Algebraic vs. Transcendental}

In section 6 we discussed the general structure of 1-loop amplitudes
in theories with maximal SUSY. At the end of the section we argued
that rational terms are absence in 1-loop amplitudes in ${\cal N}=8$
SUGRA. The argument relies on the fact that rational terms become
algebraic numbers when the kinematical invariants $\langle
i~j\rangle$ and $[i~j]$ are algebraic and that box functions give
rise to transcendental numbers. We remind the reader that algebraic
numbers are complex numbers which are roots of polynomials with
rational coefficients. This means that the field of algebraic
numbers is the algebraic completion of $\mathbb{Q}$ and it is
denoted by $\bar{\mathbb{Q}}$. A complex number which is not in
$\bar{\mathbb{Q}}$ is called transcendental.

\subsection{Algebraic Property of Rational Terms}

In this appendix we prove the statement about the rational terms.
Note that being rational functions in the variables $\langle
i~j\rangle$ and $[i~j]$ does not means that if the variables are in
$\bar{\mathbb{Q}}$ so is the function. As simple counterexample is
$\pi [1~2]^3/\langle 4~7\rangle$.

The goal is to prove that all rational terms that are produced as a
consequence of reduction procedures have the form
\begin{equation}
\label{fomi}
{\cal R}_n = \frac{P_1}{Q_1} + \left( \frac{P_2}{Q_2} + \pi^2 \frac{P_2}{Q_2} \right)\epsilon + {\cal O}(\epsilon).
\end{equation}
where $P_i$ and $Q_i$ are in $\mathbb{Q}[\langle i~j\rangle,[i~j]]$,
i.e., are polynomials with rational coefficients.

Rational terms come from two sources. One is tensor bubbles and the
other is from the ``$\mu$-integrals". The definition of
$\mu$-integrals is given in the next appendix were we discuss a
particular reduction procedure. They are IR and UV finite and can
easily be shown to give rise to rational functions in the
kinematical invariants with numerical coefficients which are
rational numbers, i.e., belong to $\mathbb{Q}$ (see the next
appendix).

Let us discuss the tensor bubbles, which can be computed explicitly
\cite{smirnov} and are given by
\begin{equation}
e^{\epsilon \gamma}\int d^{4-2\epsilon}\ell \frac{\ell^{\mu_1}\ldots \ell^{\mu_n}}{\ell^2(\ell-q)^2} = \frac{e^{\epsilon \gamma}}{(q^2)^{\epsilon}}\sum_{r=0}^{[n/2]}A(n,r)\left(\frac{q^2}{2}\right)^r\{ [g]^r[q]^{n-2r}\}^{\mu_1\ldots\mu_n}
\end{equation}
with
\begin{equation}
A(n,r) = \frac{\Gamma(\epsilon-r)\Gamma(n+1-\epsilon-r)\Gamma(1-\epsilon+r)}{\Gamma(2+n-2\epsilon)}
\end{equation}
and $[g]^r[q]^{n-2r}$ represents tensor structures with $r$ metric
tensors and $n-2r$ factors of $q$ which are symmetric in all
indices\footnote{We are using conventions where an $L$-loop integral
is multiplied by $e^{L \epsilon \gamma}$ where $\gamma$ is Euler's
constant.}.

Note that for any $(n,r)$ there is a single source of divergences
when $\epsilon\to 0$, i.e, $\Gamma(\epsilon-r)$. Now it is clear
that this can be expanded as $\alpha (q^2) /\epsilon + \beta(q^2)
\log(q^2) +\rho(q^2) + {\cal O}(\epsilon)$. The key observation is
that $\alpha$, $\beta$ and $\rho$ are polynomials in $q^2$ with
rational number coefficients, i.e, the coefficients belong to
$\mathbb{Q}$. To see this note that by repeatedly using the identity
$\Gamma(1+z)=z \Gamma(z)$ one can show that
\begin{equation}
A(n,r) = R(n,r,\epsilon) \frac{\Gamma(\epsilon)\Gamma(1-\epsilon)^2}{\Gamma(2-2\epsilon)}
\end{equation}
where $R(n,r,\epsilon)$ is the ratio of two polynomials in
$\epsilon$ with coefficients in $\mathbb{Q}$. Finally, multiplying
by $e^{\epsilon \gamma}$ and expanding in $\epsilon$ one finds
\begin{equation}
e^{\epsilon \gamma}A(n,r) = R(n,r,\epsilon)\left( \frac{1}{\epsilon} + 2 + \left( 4-\frac{\pi^2}{12} \right)\epsilon + {\cal O}(\epsilon^2) \right).
\end{equation}
Note that the transcendental term, i.e. $\pi^2/12$ is pushed to
${\cal O}(\epsilon)$ since $R(n,r,\epsilon)$ does not have a pole at
$\epsilon =0$.

This means that the rational pieces in the amplitude which come from
combinations of tensor and scalar bubbles with coefficients which
are rational functions of $\langle i~j\rangle$ and $[i~j]$, arise
from combinations of such coefficients with rational numbers. The
${\rm log}(q^2)$ term has a branch cut and does not contribute to
the rational terms and can be treated effectively as coming from a
scalar bubble (the same applies to the divergence $1/\epsilon$).

Combining the results for the tensor bubbles with that of the
$\mu$-integral we conclude that the rational terms are of the
form~(\ref{fomi}) as claimed.

\subsection{Transcendental Properties of Box Functions}

Now we turn to the properties of the box functions. Of course, we
are not going to prove that if $x\in \bar{\mathbb{Q}}$ then ${\rm
Li}_2(x)$ is transcendental! This is surely a very challenging
mathematical problem. However, it is believed to be a
true statement~\cite{Lewin}. Luckily, we need a weaker form. All we need is
that there exist sequences of algebraic numbers for which ${\rm
Li}_2$'s are transcendental (which is the generic case~\cite{Lewin})
that lead to the singular kinematical points we are interested in.

As mentioned in the footnote in section 5.2.3, if we want to relax
this assumption, we can restrict our attention to kinematical
invariants that take values in $\mathbb{Q}$. Using that ${\rm
Li}_2(x)$ is irrational when $x\in \mathbb{Q}$ and $0<x<1/2$
\cite{irrational} and repeatedly using dilogarithm identities, like
Euler's formula ${\rm Li}_2(x)+{\rm Li}_2(1-x) = \pi^2/6-{\rm
log}(x){\rm log}(1-x)$ and Landen's formula which relates ${\rm
Li}_2(1-x)$ to ${\rm Li}_2(1-1/x)$, one can extend this result to a
range where the box functions can be shown to be irrational.

In this part of the appendix we fill up the final gap in the
argument of section 5.2.3. We have to show two facts: one is that in
the various limits when physical singularities are approached the
expansion of the box functions still preserves the property of being
transcendental and expressible in terms of ${\rm Li}_2$'s, ${\rm
log}^2$'s and $\pi^2$'s. This is crucial in order to be able to
compare both sides of the limits and conclude that the rational
terms must satisfy equations (\ref{dos}) and
(\ref{tres}). The second fact is that combinations of ${\rm
Li}_2$'s, ${\rm \log}^2$'s evaluated at algebraic numbers and
$\pi^2$'s cannot give rise to algebraic numbers.

Once again the statement we want to prove is by no means obvious
since any function of the form ${\rm log}^2(1-z)$ or ${\rm
Li}_2(z)$ in the limit when $z\to 0$ by taking algebraic values
turns into an algebraic number to leading order.

By inspection of the box functions in~(\ref{boxie}) it is easy to
realize that all functions with at least one massless leg do not
posses any of the dangerous functions mentioned above. In fact, in
the collinear and multi-particle factorization limits, one only finds
terms of the form ${\rm
log}(z)^2$, ${\rm Li}_2(1-z)$ and ${\rm Li}_2(1-1/z)$. The first two
terms are clearly transcendental to leading order; the logarithms is
always transcendental if $z$ is algebraic while the leading order
terms of the dilogarithm in the limit is $\pi^2/6$. The third term
can be related to the previous two by using Landen's identity
\begin{equation}
{\rm Li}_2(1-1/z)+{\rm Li}_2(1-z) = -\frac{1}{2}{\rm log}(z)^2.
\end{equation}

The four-mass box is more complicated to study. If a certain
singularity does not make any of the $K_i^2$ or $s$ or $t$ vanish
then there is nothing to show since the box remains in the same form
evaluated at algebraic values. If any of the $K_i^2\to 0$ then we
can use the formalism of discontinuity functions developed in
\cite{boxfunctions}. The idea is that all box integrals are related
to one another by certain limits provided one is allowed to add
certain discontinuity functions. These are given by
\begin{eqnarray}
&& d_1(P^2) = \frac{1}{\epsilon^2}(-K^2)^{-\epsilon} \\
&& d_2(P_1^2,P_2^2,P_3^2) = \frac{1}{2\epsilon^2}(-P_1^2)^{-\epsilon} - \frac{1}{2\epsilon^2}\frac{(-P_1^2)^{-\epsilon}(-P_2^2)^{-\epsilon}}{(-P_3^2)^{-\epsilon}}-{\rm Li}_2\left(1-\frac{P_2^2}{P_3^2}\right).
\end{eqnarray}
All relations needed are found in table 5 of \cite{boxfunctions}. We
give as an example the one which relates a four-mass box, $F^{4m}=F^{4m}(K_1,K_2,K_3,K_4)$ to a
three-mass integral,
\begin{eqnarray}
F^{4m}-d_2(K_1^2,K_2^2,(K_1+K_2)^2)-d_2(K_1^2,K_4^2,(K_4+K_1)^2) \stackrel{K_1^2\to 0}{\longrightarrow} F^{3m}(k_1,K_2^2,K_3^2,K_4^2).
\end{eqnarray}

Noting that the discontinuity functions are such that when expanded
up to finite order in $\epsilon$ one finds only functions which
preserve the transcendentality property we have shown what we wanted
to prove.

Note in passing that the discontinuity functions $d_1$ and $d_2$ are
also the ones that enter in the factorization function ${\cal F}$ in
section 5.2.3 and as we have already mentioned they satisfy the
properties needed to show the separation among rational terms and
box functions.

Finally, as mentioned in the text, the fact that no combination of
${\rm Li}_2$'s, ${\rm log}^2$'s evaluated at algebraic numbers and
$\pi^2$'s can give rise to algebraic numbers follows for combination
that arise from scalar box integrals from the result in section 6 of
the first reference in \cite{BDK}. These combinations are the only
ones of interest in our analysis. It might be worth mentioning that
even finding linear relations of ${\rm Li}_2$'s, $\log^2$'s and
$\pi^2$'s with rational coefficients which equal zero (called
dilogarithm ladders (see e.g.~\cite{Lewin})) and which are not
derivable from Landen or Abel's identities is a very challenging
problem. An example is
$$  {\rm Li}_2(\rho^{20}) - 2{\rm Li}_2(\rho^{10})-15{\rm Li}_2(\rho^4)+10{\rm Li}_2(\rho^2) - \frac{\pi^2}{15} = 0$$
with $\rho = (\sqrt{5}-1)/2$.

\section{Reduction Procedure}

In this section we briefly describe the reduction procedure we apply
to 1-loop Feynman integrals in ${\cal N}=8$ SUGRA. More generally it
can be applied in any massless theory in $D=4-2\epsilon$ and in the
FDH scheme. The four dimensional helicity scheme simply instructs us
to keep all external kinematic in four dimensions, including the
number of degrees of freedom. As we will now show, the FDH scheme
combined with the van Neerven-Vermaseren method \cite{reduction}
give us the bases needed in appendix E to prove the absence of
rational terms in maximally supersymmetric theories.

Consider a general tensor integral
\begin{equation}
I^{\mu_1\ldots \mu_m}_n = \int d^DL\frac{L^{\mu_1}\ldots L^{\mu_m}}{\prod_{i=1}^n (L-P_i)^2}.
\end{equation}
In a Feynman diagram, tensor integrals are contracted with some tensors made out of external kinematical objects and the metric. If any two indices are contracted with a metric tensor we write the corresponding $L^2$ term in the numerator as $(L-P_i)^2 + 2L\cdot P_i-P_i^2$ for some $i$. Canceling the first term with a propagator we conclude that all tensor structures can be taken as contracted with external objects. Note that the case $n=2$ is special, i.e. tensor bubbles. We choose to keep all tensor bubbles in our basis explicitly and they have been analyzed in detain the main text.

Having proven that all integrals can be written in terms of tensor integrals with contractions with external objects we use that such objects are kept in four dimensions. This means that we can decompose the loop integration into the four dimensional part $\ell$ and the $-2\epsilon$ dimensional part. Since nothing depends on the angular variables of the $-2\epsilon$ integral this can be performed and we are left with
\begin{equation}
I^{\mu_1\ldots \mu_m}_n = -\frac{i\epsilon}{(4\pi)^{2-\epsilon}}\frac{1}{\Gamma (1-\epsilon)}\int_0^{\infty} \frac{d\mu^2}{(\mu^2)^{1+\epsilon}}\int d^4\ell \frac{\ell^{\mu_1}\ldots \ell^{\mu_m}}{\prod_{i=1}^n ((\ell-P_i)^2-\mu^2)}.
\label{epsi}
\end{equation}

This is the moment where the procedure developed by van Neerven and Vermaseren becomes applicable straightforwardly since the integral
\begin{equation}
\int d^4\ell \frac{\ell^{\mu_1}\ldots \ell^{\mu_m}}{\prod_{i=1}^n ((\ell-P_i)^2-\mu^2)}
\label{fordy}
\end{equation}
is precisely their object of study. In their reduction technique $\mu^2$ is a fixed mass term and did not play a particularly special role. In our case, $\mu^2$ is very important as powers of it in the numerator, which appear naturally in the procedure, can be interpreted as shifting the dimension of the integral.

Before continuing with the general discussion let us illustrate the reduction with a simple example. Consider the reduction of a scalar pentagon in $D=4-2\epsilon$. The procedure of van Neerven and Vermaseren directly gives rise to (borrowing formula (19) of their paper in \cite{reduction})
\begin{eqnarray}
E_{01234}(w^2-4\Delta \mu^2) &=& D_{1234}(2\Delta - w\cdot(v_1+v_2+v_3+v_4))+ D_{0234} v_1\cdot w \nonumber \\ & & + D_{0134} v_2\cdot w + D_{0124} v_3\cdot w + D_{0123} v_4\cdot w.
\end{eqnarray}
where
\begin{equation}
E_{01234} = \int \frac{d^4 \ell}{N_0N_1N_2N_3N_4}, \quad D_{ijkl} = \int \frac{d^4 \ell}{N_iN_jN_kN_l}
\end{equation}
with $N_0 = \ell^2-\mu^2$, $N_1 = (\ell-K_1)^2-\mu^2$, $N_2 = (\ell-K_1-K_2)^2-\mu^2$ and so on. It is also convenient to define $p_i$ as the sum of the external momenta appearing in $N_i$, e.g. $p_2 = K_1+K_2$. $v_i^\mu$ are a basis of dual vectors
\begin{equation}
v_1^\mu = \epsilon^{\mu p_2p_3p_4}, \; v_2^\mu = \epsilon^{p_1\mu p_3p_4}, \; {\rm etc.,}
\end{equation}
and $w^\mu = \sum_{i=1}^4 p_i^2 v_i^\mu$. Finally, $\Delta$ is the Gram determinant of the system.

Using this decomposition of the $D=4$ pentagon integral in the formula for the $D=4-2\epsilon$ pentagon integral
\begin{equation}
I_5[1] = -\frac{i\epsilon}{(4\pi)^{2-\epsilon}}\frac{1}{\Gamma (1-\epsilon)}\int_0^{\infty} \frac{d\mu^2}{(\mu^2)^{1+\epsilon}}\int d^4\ell \frac{1}{(\ell^2-\mu^2)\prod_{i=1}^4 ((\ell-p_i)^2-\mu^2)}
\end{equation}
one finds a formula relating $I_5[1]$ (where $[1]$ means numerator $1$) to a sum of $D=4-2\epsilon$ boxes and one more ``mysterious" integral. This remaining integral comes from the term $E_{01234}\times \mu^2$ which ends up being $I_5[\mu^2]$. As mentioned before, this can also be interpreted as a pentagon in $D=6-2\epsilon$. This formula is equivalent to that found by using differential equations in the last reference in the list \cite{reduction}.

Now we go back to the general case. From the example it is clear that we should keep all factors of $\mu^2$ in the numerator as the procedure is applied. This is also natural from the point of view of a Passarino-Veltman reduction of (\ref{fordy}). The reason is that in our case, $\mu^2$ is really an integration variable!

Performing the reduction we end up with a basis on integrals which can be divided naturally into two groups. The first is that of scalar boxes, scalar triangles, and scalar bubbles with no powers of $\mu^2$ in the numerator. The second is that of scalar polygons with $(\mu^2)^{r}$ terms in the numerator with $r>0$.

Note that all integrals in the second group can be thought of as
being integrals in $D>5$ and therefore are IR finite. This means
that the only divergences can be UV divergences and manifest
themselves as simple poles in $\epsilon$. The pole terms are the
only ones relevant to the computation of (\ref{epsi}) since they
turn into ${\cal O}(\epsilon^0)$ terms due to the explicitly
$\epsilon$ factor coming from the $-2\epsilon$ angular integration.
All higher order terms become irrelevant for our purposes.

Now it is clear that these integrals only contribute to the
generation of rational terms. For the purpose of the argument in
appendix E.1 all we need is to compute the leading UV singular piece
of those integrals. This can be done, following \cite{UV}, by
setting to zero all the external momenta and introducing an
auxiliary mass term $\lambda^2$. Therefore all integrals become
\begin{equation}
I_n[\mu^{2r}] \rightarrow -\frac{i\epsilon}{(4\pi)^{2-\epsilon}}\frac{1}{\Gamma (1-\epsilon)}\int_0^{\infty}d\mu^2 \int_0^\infty d\ell^2 \frac{\ell^2(\mu^2)^{r-1-\epsilon}}{(\ell^2+\mu^2+\lambda^2)^n}
\end{equation}
where the angular integration over $\ell$ has been performed.

These integrals can easily be done and give
\begin{equation}
-\frac{i\epsilon}{(4\pi)^{2-\epsilon}}\frac{1}{\Gamma (1-\epsilon)}(\lambda^2)^{2-n+r-\epsilon}\frac{\Gamma (r-\epsilon)\Gamma (-2+n-r+\epsilon)}{\Gamma (n)}.
\end{equation}

From this form it is clear that the expansion in $\epsilon$ always gives something of the form
\begin{equation}
I_n[\mu^{2r}] \rightarrow -\frac{i\epsilon}{(4\pi)^{2-\epsilon}}\left( (\lambda^2)^{2-n+r}\frac{A}{\epsilon} + {\cal O}(\epsilon^0)\right)
\end{equation}
with $A=\Gamma(n-r-2)\Gamma(r)/\Gamma(n)$ a numerical factor which is a rational number, i.e., $A\in \mathbb{Q}$.

The presence of the factor $\lambda^2$ implies that a rational function of the kinematical invariant which can be formed from the momenta $P_i$ appearing in the propagators appears.

A well known example is the case of $n=4$ and $r=2$,
\begin{equation}
I_4[\mu^4] \rightarrow -\frac{i\epsilon}{(4\pi)^{2-\epsilon}}\left( -\frac{1}{6\epsilon} + {\cal O}(\epsilon^0)\right).
\end{equation}
Note that here we have not included the pre-factor $e^{\epsilon\gamma}$ since its effect is only sub-leading in $\epsilon$. Also worth mentioning is that the $(4\pi)^{2-\epsilon}$ factors are overall factors in the amplitude and do not play any role in the discussion as can be absorbed in the definition of the coupling constant.

As mentioned at the beginning of this appendix we keep all tensor bubbles explicitly in the basis. A detailed analysis of their contribution to the rational terms in the amplitude is given in appendix E.1. Once again we find that the contribution to the rational pieces is of the form required in~(\ref{fomi}).

\end{appendix}

\end{document}